\documentclass[superscriptaddress,twocolumn,prd,preprintnumbers,amsmath,amssymb,nofootinbib]{revtex4-1}

\usepackage{graphicx}
\usepackage{epstopdf}
\usepackage{dcolumn}
\usepackage{bm}
\usepackage{epsfig}
\usepackage{amssymb}
\usepackage{amsmath}
\usepackage{subfigure}
\usepackage{color}
\usepackage[dvipsnames]{xcolor}
\usepackage[utf8]{inputenc}
\usepackage{enumitem}
\usepackage{hyperref}
\usepackage{aas_macros}
\usepackage{placeins}

\newcommand{\beq}{\begin{equation}}
\newcommand{\eeq}{\end{equation}}

\newcommand{\mpl}{m_{\mathrm{pl}}}

\newcommand{\sigmav}{\ensuremath{\langle \sigma v \rangle}}
\newcommand{\diff}{\ensuremath{\mathrm{d}}}
\newcommand{\e}{\mathrm{e}}

\newcommand{\TKR}{\ensuremath{T_\mathrm{KR}}}
\newcommand{\kcut}{\ensuremath{k_\mathrm{cut}}}
\newcommand{\kKR}{\ensuremath{k_\mathrm{KR}}}

\newcommand{\MKR}{\ensuremath{M_\mathrm{KR}}}
\newcommand{\rhoc}{\ensuremath{\rho_\chi}}
\newcommand{\rhom}{\ensuremath{\rho_m}}
\newcommand{\rhobar}{\bar{\rho}_\chi}
\newcommand{\boost}{\ensuremath{\overline{\rhoc^2}/{\rhobar}^2}}
\newcommand{\boostrho}{\ensuremath{\overline{\rhoc^2}/{\rhobar}}}
\newcommand{\rmax}{\ensuremath{r_\mathrm{max}}}
\newcommand{\mmax}{\ensuremath{M_\mathrm{max}}}
\newcommand{\Tf}{\ensuremath{T_\mathrm{F}}}
\newcommand{\Tkd}{\ensuremath{T_\mathrm{kd}}}
\newcommand{\TkdS}{\ensuremath{T_\mathrm{kdS}}}

\newcommand{\Mmin}{\ensuremath{M_\mathrm{min}}}
\newcommand{\lfs}{\ensuremath{\lambda_\mathrm{fs}}}

\begin{document}

\title{How an Era of Kination Impacts Substructure and the Dark Matter Annihilation Rate}

\author{M. Sten Delos}
\email{sten@mpa-garching.mpg.de}
\affiliation{Max Planck Institute for Astrophysics, Karl-Schwarzschild-Str. 1, 85748 Garching, Germany}
\affiliation{Department of Physics and Astronomy, University of North Carolina at Chapel Hill, Phillips Hall CB3255, Chapel Hill, NC 27599 USA}
\author{Kayla Redmond}
\author{Adrienne L. Erickcek}
\affiliation{Department of Physics and Astronomy, University of North Carolina at Chapel Hill, Phillips Hall CB3255, Chapel Hill, NC 27599 USA}

\begin{abstract}

An era of kination occurs when the Universe's energy density is dominated by a fast-rolling scalar field.  Dark matter that is thermally produced during an era of kination requires larger-than-canonical annihilation cross sections to generate the observed dark matter relic abundance.  Furthermore, dark matter density perturbations that enter the horizon during an era of kination grow linearly with the scale factor prior to radiation domination.  We show how the resulting enhancement to the small-scale matter power spectrum increases the microhalo abundance and boosts the dark matter annihilation rate. We then use gamma-ray observations to constrain thermal dark matter production during kination.  
The annihilation boost factor depends on the minimum halo mass, which is determined by the small-scale cutoff in the matter power spectrum. Therefore, observational limits on the dark matter annihilation rate imply a minimum cutoff scale for a given dark matter particle mass and kination scenario. 
For dark matter that was once in thermal equilibrium with the Standard Model, this constraint establishes a maximum allowed kinetic decoupling temperature for the dark matter. This bound on the decoupling temperature implies that the growth of perturbations during kination cannot appreciably boost the dark matter annihilation rate if dark matter was once in thermal equilibrium with the Standard Model.

\end{abstract}

\maketitle

%%%%%%%%%%%%%%%
\section{Introduction}
\label{sec:Intro}
%%%%%%%%%%%%%%%%

The thermal history between the end of inflation and the beginning of Big Bang nucleosynthesis (BBN) remains uncertain \cite{Allahverdi:2020bys}.  The energy scale of inflation is generally assumed to be greater than ${10^{10} \, \mathrm{GeV}}$, but the abundances of light elements predicted by BBN and the effective neutrino density implied by the anisotropies in the cosmic microwave background only require that the Universe be radiation dominated at temperatures less than 8 MeV \cite{Kawasaki:1999, Kawasaki:2000, Hannestad:2004, Ichikawa:2005, Ichikawa:2006, Hasegawa:2019jsa, dBPM08, deSalas:2015glj}.   Future detections of gravitational waves \cite{Boyle:2005GW, Easther:2006GW, Easther:2008GW, Amin:2014, Giblin:2014GW, Figueroa:2019paj} or a network of cosmic strings \cite{Cui:2017GW} may be used fill in this gap in the cosmic record, but we currently have no direct observational probes of this era.  

Fortunately, the matter power spectrum provides an indirect way to probe of the evolution of the Universe prior to BBN.  For example, it has been shown that an early matter-dominated era (EMDE) enhances the small-scale matter power spectrum and leads to an abundance of early-forming, highly dense dark matter microhalos \cite{Erickcek:2011, Barenboim:2013, Fan:2014, Erickcek:2015}. These microhalos increase the dark matter annihilation rate to the point of bringing some EMDE scenarios into tension with Fermi-LAT observations of dwarf spheroidal galaxies \cite{Erickcek:2015,Erickcek:Sinha:Watson:2015} and the isotropic gamma-ray background \cite{Blanco:2019eij, StenDelos:2019xdk}.  An early era of cannibal domination, during which the Universe is dominated by massive particles that are self-heated by number-changing interactions, generates a similar enhancement to the small-scale matter power spectrum \cite{Erickcek:2020wzd,Erickcek:2021fsu}.

Another possibility is that between the end of inflation and the beginning of BBN there was a period of kination, during which the Universe was dominated by a fast-rolling scalar field (a kinaton) \cite{Spokoiny:1993, Joyce:1996, Ferreira:1997}.  Kination was initially proposed as a postinflationary scenario that allows the Universe to transition to radiation domination even if the inflaton does not decay into relativistic particles \cite{Spokoiny:1993}.
Postinflationary kination phases naturally arise in string theory (e.g.~\cite{Conlon:2022pnx,Apers:2022cyl}).
While BBN bounds on gravitational waves generally rule out scenarios that include no couplings between the inflaton and the Standard Model \cite{Figueroa:2018twl}, a period of kination can still serve to dilute the inflaton energy density after inflation.
Kination also facilitates baryogenesis \cite{Joyce:1996}, and the kinaton can
%mimic the effects of a cosmological constant if its potential energy becomes dominant at very late times 
transition into dark energy \cite{Ferreira:1997, Peebles:1998, Dimopoulos:2001, Dimopoulos:2002Curvaton, Chung:2007}
or dark matter \cite{Li:2013nal,Li:2016mmc,Li:2021htg}.

An early period of kination alters the evolution of dark matter density perturbations \cite{Redmond:2018}.  When a perturbation mode enters the horizon during an era of kination, the gravitational potential drops sharply and then oscillates with a decaying amplitude while the dark matter density perturbations grow linearly with the scale factor.  This linear growth leaves an imprint on the matter power spectrum.  Specifically, for modes that enter the horizon during kination, the matter power spectrum ${P_\delta(k)}$ is proportional to ${k^{n_s-3}}$, where $k$ is the wavenumber of the perturbation mode and $n_s$ is the scalar spectral index.  In comparison, ${P_\delta \propto k^{n_s-4} \, \mathrm{ln}^2[k/(8k_\mathrm{eq})]}$ for modes that enter the horizon during radiation domination, where $k_\mathrm{eq}$ is the wavenumber of the perturbation mode that enters the horizon at matter-radiation equality.  Therefore, a period of kination generates a small-scale enhancement to the matter power spectrum.  In this work, we explore how this enhancement affects the abundance of dark matter microhalos and the extent to which it strengthens observational constraints on dark matter production during kination.  

The expansion rate during kination is higher than the expansion rate in a radiation-dominated universe at the same temperature.  Consequently, a larger dark matter annihilation cross section is required to generate the observed dark matter abundance if dark matter is thermally produced during kination \cite{Profumo:2003, Pallis:2005, Pallis:2006, Gomez:2008, Lola:2009, Pallis:2009, DEramo:2017gpl, Redmond:2017, DEramo:2017ecx}, even if the dark matter abundance is diluted by kinaton decays \cite{Visinelli:2017qga}.  Therefore, an era of kination widens the field of dark matter candidates to include particles that have velocity-averaged annihilation cross sections that are larger than $3\times10^{-26}\,\mathrm{cm}^3\, \mathrm{s}^{-1}$, such as dominantly wino or higgsino neutralinos \cite{Profumo:2003}.  These large annihilation cross sections also imply that such scenarios are already tightly constrained by  observational limits on dark matter annihilation within dwarf spheroidal galaxies \cite{Fermi:Constraints} and the Galactic Center \cite{HESS:Constraints}.  If the dark matter freezes out from thermal equilibrium during an era of kination, these limits strongly restrict the dark matter mass and the temperature at which kinaton-radiation equality occurs \cite{Redmond:2017}.

Given these constraints on dark matter freeze-out during kination, any boost to the dark matter annihilation rate from enhanced dark matter structure could easily rule out these scenarios. 
This boost, quantified by $\overline{\rho_\chi^2}$/$\bar{\rho}_\chi^2$ where $\rho_\chi$ is the dark matter density, depends not only on the kination scenario but also on dark matter properties.
If the dark matter was once in kinetic equilibrium with Standard Model particles, then its thermal streaming motion suppresses the amplitudes of density variations below a cutoff scale determined by the temperature at which dark matter kinetically decoupled.  For each of the allowed kination parameter sets, we calculate the cutoff scale and hence the kinetic decoupling temperature $\Tkd$ required to rule out each scenario based on observations of the isotropic gamma-ray background, which provide the strongest bounds on the dark matter annihilation cross section in scenarios with enhanced small-scale structure \cite{StenDelos:2019xdk}.
In particular, we frame our results in terms of the temperature $\TkdS$ at which the dark matter would decouple within a standard (radiation-dominated) expansion history, since this parameter is a property of the dark matter particle alone, has been calculated for many dark matter models, and can be constrained by laboratory experiments \cite{Profumo:2006,Cornell:2013rza}.

We begin by determining the dark matter power spectrum in kination cosmologies in Section \ref{sec:TransferFunction}.  In Section \ref{sec:KDandFS}, we calculate the small-scale cutoff to the matter power spectrum if dark matter kinetically decouples during a period of kination.   In Section \ref{sec:PrimordialStructures}, we analyze the growth of structure following a period of kination and find that a period of kination triggers an earlier start to halo formation and enhances the abundance of sub-earth-mass microhalos.  In Section \ref{sec:SubstructureConstraints}, we calculate the annihilation boost from these early-forming microhalos following the procedure established in Ref.~\cite{StenDelos:2019xdk}, and
 we use the isotropic gamma-ray background to constrain 
scenarios where dark matter freezes out during kination.
In Section \ref{sec:Conclusion}, we summarize our results and discuss their implications.  Natural units $(\hbar = c=k_B=1)$ are used throughout this work.

%%%%%%%%%%%%%%%
\section{The Matter Power Spectrum After Kination}
\label{sec:DMEvolutionWavenumber}
%%%%%%%%%%%%%%%

A period of kination occurs when the pressure $P$ of the dominant component of the Universe equals its energy density $\rho$.  One way to realize this equation of state is with a scalar field $\phi$ whose kinetic energy greatly exceeds its potential energy: $(d\phi/dt)^2/2 \gg V(\phi)$.  The scalar field's equation of state is then 
\begin{equation}
w \equiv \frac{P_\phi}{\rho_\phi} = \frac{(d\phi/dt)^2/2 - V(\phi)}{(d\phi/dt)^2/2 + V(\phi)} \simeq 1.
\end{equation}
This equation of state implies that the energy density of the kinaton field drops as $\rho_\phi\propto a^{-6}$, where $a$ is the scale factor describing the expansion of the Universe.  

In addition to the kinaton, we assume that the postinflationary Universe contains relativistic Standard Model (SM) particles.  Since the kinaton's energy density decreases faster than the density of relativistic particles ($\rho_r \propto a^{-4}$), a period of kination will give way to a period of radiation domination even if the kinaton does not decay.  We characterize this transition by the radiation temperature $T_\mathrm{KR}$ at kinaton-radiation equality.  We also assume that dark matter is a thermal relic that freezes out from the SM radiation bath
and that the radiation's entropy is conserved between dark matter freeze-out and today.
This assumption allows us to use the methods of Ref.~\cite{Redmond:2017} to determine the dark matter annihilation cross section that yields the observed dark matter abundance as a function of $T_\mathrm{KR}$ and the dark matter mass $m_\chi$.

%%%%%%%%%%%%%%%
\subsection{The dark matter transfer function}
\label{sec:TransferFunction}
%%%%%%%%%%%%%%%

We first calculate the power spectrum of linear dark matter density perturbations during the matter epoch.  The transfer function $T(k,a)$ parametrizes how dark matter density perturbations $\delta_\mathrm{\chi} \equiv (\rho_\chi - \bar{\rho}_\chi)/\bar{\rho}_\chi$ vary with wavenumber $k$ during matter domination:
\begin{align}
\delta_\chi(k,a \gg a_\mathrm{eq}) = \frac{3}{5} \frac{k^2}{\Omega_M H^2_0} \Phi_\mathrm{p}(k) \, T(k,a),
\label{eq:TransferFunctionDefinition}
\end{align}
where $a_\mathrm{eq}$ is the scale factor at matter-radiation equality, $\Omega_M$ is the current matter density divided by the current critical density, $H_0$ is the present-day Hubble rate, and $\Phi_\mathrm{p}$ is the gravitational potential at superhorizon scales during radiation domination. 
To obtain $T(k,a)$, we use the evolution of $\delta_\chi$ during kination and the subsequent radiation-dominated era presented in Ref. \cite{Redmond:2018} to determine $\delta_\chi(k,a)$ during the matter-dominated era.

During kination, subhorizon dark matter density perturbations grow as $\delta_\chi\propto a$, even though the gravitational potential quickly decays upon horizon entry.
This growth rate arises because dark matter particles converge on regions that are initially overdense, and the comoving distance traversed by particles in the absence of peculiar gravitational forces is  proportional to $a$ during kination \cite{Redmond:2018}.
It follows that $\delta_\chi(a_\mathrm{KR}) \propto a_\mathrm{KR}/a_\mathrm{hor}$, where $a_\mathrm{KR}$ is the scale factor at kinaton-radiation equality and $a_\mathrm{hor}$ is the scale factor when the mode $k$ enters the horizon, i.e., $k=a_\mathrm{hor}H(a_\mathrm{hor})$.  Since $H\propto a^{-3}$ during kination, $a_\mathrm{hor} \propto k^{-1/2}$, which implies that $\delta_\chi \propto k^{1/2}$ for modes that enter the horizon during kination.  

During the radiation-dominated era that follows a period of kination, $\delta_\chi$ grows logarithmically with the scale factor while $k>aH$.
For modes that enter the horizon well before kinaton-radiation equality, Ref.~\cite{Redmond:2018} found that
\begin{align}
\delta_\chi(a) = 2.7 \, \Phi_0 \left(\frac{ k \sqrt{2}}{k_\mathrm{KR}} \right)^{1/2} \mathrm{ln}\left[\e\left(\frac{k_\mathrm{KR}}{k \sqrt{2}}\right)^{1/2} \frac{a}{a_\mathrm{hor}} \right]
\label{eq:TransferFunctionKD}
\end{align}
during radiation domination.
In this expression,  $k_\mathrm{KR}$ is the wavenumber of the mode that enters the horizon at kinaton-radiation equality, and ${\Phi_0= (9/8)\Phi_\mathrm{p}}$ is the gravitational potential on superhorizon scales during kination.  Equation (\ref{eq:TransferFunctionKD}) matches the numeric solution for the evolution of $\delta_\chi$ very well for modes with ${k/k_\mathrm{KR} \gtrsim 100}$.  As expected, $\delta_\chi \propto k^{1/2}$ for these modes.   Ref.~\cite{Redmond:2018} also provides a fitting function that describes the evolution of $\delta_\mathrm{\chi}$ for modes that enter the horizon during the transition between kination and radiation domination.
For modes with ${0.5 \lesssim k/k_{\mathrm{KR}}}$, 
\begin{align}
\delta_\chi(k,a) &= \frac{8}{9} \Phi_0 
%\left[
\,A(k) \, \mathrm{ln}\left(\frac{B(k) \, a}{a_\mathrm{hor}} \right)
%\right]
\nonumber \\
A(k) &= 2.29 \left[0.11\times 9.11^{2.64} + 2.9\left(\frac{k}{k_\mathrm{KR}}\right)^{1.32} \right]^{0.38} \nonumber \\
B(k) &= \left[0.594^{-6.59} + \e^{-6.59} \left(\frac{k}{k_\mathrm{KR}} \right)^{3.29} \right]^{-0.15}
\label{eq:FittedFunction}
\end{align}
during radiation domination.\footnote{This evolution of $\delta_\chi(k,a)$ assumes that dark matter freezes in, but it is still accurate to within $\sim\!\!10\%$ for modes that enter the horizon after dark matter freezes out \cite{Redmond:2018}.}. Modes with $k/k_{\mathrm{KR}} \lesssim 0.5$ follow the standard evolution of modes that enter the horizon during radiation domination, with $A = 9.11$ and $B=0.594$ \cite{Hu:1995Meszaros}.

Given the fitting functions $A(k)$ and $B(k)$ in Eq.~(\ref{eq:FittedFunction}), we solve for the evolution of $\delta_\chi(a)$ during matter domination for modes that enter the horizon during either an era of kination or radiation domination.  To do so, we use Eq.~(\ref{eq:FittedFunction}) as an initial condition for the Meszaros equation, which is valid when ${\rho_{\chi}\delta_\chi \gg \rho_r\delta_r}$ \cite{Meszaros:1974}.  Prior to baryon decoupling and well after matter-radiation equality, the resulting solution to the Meszaros equation is given by \cite{Hu:1995Meszaros}
\begin{align}
\delta_\chi(a) = \frac{3A}{2} f_1 \left(\frac{8}{9} \Phi_0 \right) \mathrm{ln}\left[ \left( \frac{4}{\e^3} \right)^{f_2/f_1} \frac{B \, a_\mathrm{eq}}{a_\mathrm{hor}} \right] \mathcal{D}(a),
\label{eq:Meszaros}
\end{align}
where $f_1$ and $f_2$ are determined by the baryon fraction ${f_b \equiv \rho_\mathrm{b}/(\rho_\mathrm{b}+\rho_\chi)}$,
\begin{align}
&f_1 = 1 - 0.568f_\mathrm{b} + 0.094 f_\mathrm{b}^2  \nonumber \\
&f_2 = 1 - 1.156 f_\mathrm{b} + 0.149 f_\mathrm{b}^2 - 0.074 f_\mathrm{b}^3, \nonumber
\end{align}
and $\mathcal{D}(a)$ is the growing solution to the Meszaros equation.  Prior to baryon decoupling, the baryons do not fall into the potential wells created by the dark matter density perturbations.  Accounting for the fact that the baryons do not participate in gravitational collapse \cite{Hu:1995Meszaros},
\begin{align}
\mathcal{D}(a) = \left( 1 + \frac{a}{a_\mathrm{eq}} \right)^{-\alpha} {}_2F_1 \left[ \alpha,\alpha + \frac{1}{2} ; 2\alpha + \frac{1}{2} ; \frac{a_\mathrm{eq}}{a+a_\mathrm{eq}} \right]    ,
\label{eq:GrowingMode}
\end{align}
where ${{}_2F_1 \left[ a,b;c;x \right]}$ is Gauss's hypergeometric function, and
\begin{align}
\alpha = \frac{1}{4} \left[ 1- \sqrt{1+24(1-f_\mathrm{b})} \right]  .
\label{eq:alpha}
\end{align}

To evaluate Eq.~(\ref{eq:Meszaros}), we first evaluate $a_\mathrm{eq}/a_\mathrm{hor}$ for modes that enter the horizon during an era of kination:
\begin{align}
\frac{a_\mathrm{eq}}{a_\mathrm{hor}} = 2^{1/4} \left(\frac{k}{k_\mathrm{KR}} \right)^{1/2} \frac{k_\mathrm{KR}}{k_\mathrm{eq}} \left( \frac{g_{*\mathrm{eq}}}{g_{*\mathrm{KR}}} \right)^{1/2} \left( \frac{g_{*s\mathrm{KR}}}{g_{*s\mathrm{eq}}} \right)^{2/3} ,
\label{eq:aeqOVERahor}
\end{align}
where $k_\mathrm{eq}$ is the wavenumber of the mode that enters the horizon at matter-radiation equality, $g_*(T)$ is the effective number of relativistic degrees of freedom at temperature $T$, and $g_{*s}(T)$ is the effective number of degrees of freedom that contribute to the entropy density at temperature $T$.  We define ${g_{*\mathrm{KR}} = g_*(T_\mathrm{KR})}$, ${g_{*\mathrm{eq}} = g_*(T_\mathrm{eq})}$, ${g_{*s\mathrm{KR}} = g_{*s}(T_\mathrm{KR})}$, and ${g_{*s\mathrm{eq}} = g_{*s}(T_\mathrm{eq})}$.  In comparison, if modes enter the horizon during radiation domination, then ${a_\mathrm{eq}/a_\mathrm{hor} = \sqrt{2}k/k_\mathrm{eq}}$.  The fitting function
\begin{align}
\frac{a_\mathrm{eq}}{a_\mathrm{hor}} = 1.98 \frac{k}{k_\mathrm{eq}} \left[ 1+1.4\left( \frac{k_\mathrm{KR}}{k} \right)^{-0.99} \right]^{-0.515}
\label{eq:aeqOVERahorFittedFunction}
\end{align}
is valid both for modes that enter the horizon during an era of kination and for those that enter the horizon during radiation domination.
Utilizing Eq.~(\ref{eq:aeqOVERahorFittedFunction}) and the fitting functions given in Eq.~(\ref{eq:FittedFunction}), we
can evaluate Eq.~(\ref{eq:Meszaros}) for the evolution of $\delta_\chi$ during matter domination for modes that enter the horizon during either an era of kination or radiation domination.  

When we analyze structure formation in Section~\ref{sec:PrimordialStructures}, we require a transfer function that is valid at all scales.  To construct this transfer function, we use the matter transfer functions computed by CAMB Sources \cite{Lewis:2007CAMB} and Eisenstein \& Hu (1998) \cite{Eisenstein:1997}, which do not take include deviations from radiation domination prior to matter-radiation equality.  For modes with ${k/k_\mathrm{eq} \leq 10^7}$, we use CAMB Sources \cite{Lewis:2007CAMB} to evaluate the matter transfer function $T_\mathrm{CAMB}(k)$ at redshift ${z = 50}$.  For modes with ${k/k_\mathrm{eq} > 10^7}$, we use the transfer function $(T_\mathrm{EH})$ computed by Eisenstein \& Hu (1998) \cite{Eisenstein:1997} since it provides the same scale dependence as that computed by CAMB.  Matching these two transfer functions at ${k/k_\mathrm{eq} = 10^7}$ allows us to extend the matter transfer function at $z=50$ to large $k$ values by taking
\begin{align}
T_{50}(k/k_\mathrm{eq} \geq 10^7) = T_\mathrm{EH}(k) \frac{T_\mathrm{CAMB}(k/k_\mathrm{eq} = 10^7)}{T_\mathrm{EH}(k/k_\mathrm{eq} = 10^7)}  .
\label{eq:TransferCombined}
\end{align}
When evaluating the transfer function we use the Planck 2018 parameters \cite{Planck:2018vyg}.

Even after they decouple from the photons, the baryons have nonzero pressure and do not participate in gravitational collapse on scales smaller than the baryon Jeans length  \cite{Bertschinger:2006}.  To account for this suppression, we define a scale-dependent growth function $D(k,z)$ such that $T(k,z) = T_{50}(k) D(k,z)$ and $D(k,z=50)$ mimics the scale dependence of $T_{\mathrm{CAMB}}/T_{\mathrm{EH}}$.   We choose to pin our transfer function at $z = 50$ because the first microhalos generally form around this redshift, as will be shown in Section \ref{sec:PrimordialStructures}.  For ${k/k_\mathrm{eq} \lesssim 10^4}$, ${T_{\mathrm{CAMB}}/T_{\mathrm{EH}} \simeq 1}$.  For ${k/k_\mathrm{eq} \gtrsim 10^5}$, ${T_{\mathrm{CAMB}}/T_{\mathrm{EH}} \simeq 0.789}$ at $z=50$ and continues to decrease at later redshifts.  The suppression of ${T_{\mathrm{CAMB}}/T_{\mathrm{EH}}}$ results from small-scale perturbations experiencing slower growth after recombination, which $T_\mathrm{EH}$ does not take into account.  To match this transition at ${z=50}$, we follow Ref.~\cite{Erickcek:2011} and take
\begin{align}
\frac{T_\mathrm{CAMB}}{T_\mathrm{EH}} \simeq D_s(k) \equiv \left[ \frac{\left(D_A - D_B \right)}{\left( 1 + \left(\frac{k/k_\mathrm{eq}}{48500} \right)^{2.1}  \right)} \right] + D_B ,
\label{eq:Ds}
\end{align}
where ${D_A = 1}$ and ${D_B = 0.789}$.

For ${3 \lesssim z \lesssim 500}$, modes with ${T_{\mathrm{CAMB}}/T_{\mathrm{EH}} \simeq 1}$ have ${\delta_\chi(a) \propto 2/3 + a/a_\mathrm{eq}}$.  This growth of $\delta_\chi$ comes from the growing mode of the Meszaros equation.  For modes with ${T_{\mathrm{CAMB}}/T_{\mathrm{EH}} \simeq 0.789}$, baryons are still pressure supported and ${\delta_\chi(a) \propto \mathcal{D}(a)}$, where $\mathcal{D}(a)$ is defined in Eq.~(\ref{eq:GrowingMode}).  Utilizing these two relations, the scale-dependent growth function ${D(k,z)}$ is obtained by reevaluating Eq.~(\ref{eq:Ds}) with $D_A$ and $D_B$ now taken to be functions of redshift \cite{Erickcek:2011}:
\begin{subequations}
\begin{align}
& D_A(z) = \frac{\frac{2}{3} + \frac{1+z_\mathrm{eq}}{1+z}}{\frac{2}{3} + \frac{1+z_\mathrm{eq}}{51}} , \\
& D_B(z) = \frac{\mathcal{D}(z)}{\mathcal{D}(z=50)} .
\end{align}
\label{eq:ScaleDependentFunctions}%
\end{subequations}

\begin{figure}
\centering\includegraphics[width=3.4in]{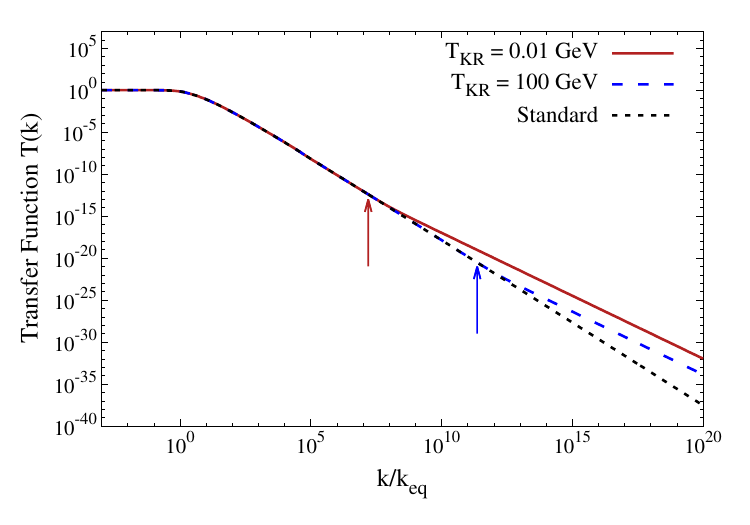}
\caption{The matter transfer function evaluated at ${z=50}$ including an era of kination with two different values of $T_\mathrm{KR}$: 0.01 GeV and 100 GeV.  The black line represents the standard matter transfer function without a period of kination.  The arrows represent the values of ${k_\mathrm{KR}/k_\mathrm{eq}}$ for the two kination cases, and thus represent where the transfer function begins to deviate from the standard case.}
\label{Fig:TransferFuctionz50}
\end{figure}

Finally, we account for an era of kination by multiplying the transfer function by the ratio ${R(k) = \delta_\chi(T_\mathrm{KR})/\delta_\chi(T_\mathrm{KR} = \infty)}$, where $\delta_\chi$ is evaluated after the era of kination and $T_\mathrm{KR}$ is the temperature at kinaton-radiation equality.  This ratio is evaluated by rescaling the solution to the Meszaros equation given by Eq.~(\ref{eq:Meszaros}) to account for an era of kination. The result is
\begin{align}
R(k) = \frac{A(k) \, \mathrm{ln}\left[ \left( \frac{4}{\e^3} \right)^{f_2/f_1} \frac{B(k) a_\mathrm{eq}}{a_\mathrm{hor}} \right]   }{9.11 \, \mathrm{ln} \left[ \left( \frac{4}{\e^3} \right)^{f_2/f_1} 0.594 \frac{\sqrt{2} k}{k_\mathrm{eq}} \right]} ,
\label{eq:RRatio}
\end{align}
where $A(k)$ and $B(k)$ are given by the fitting functions in Eq.~(\ref{eq:FittedFunction}) for ${k \gtrsim 0.5 \, k_\mathrm{KR}}$.  For modes with ${k \lesssim 0.5 \, k_\mathrm{KR}}$, $\delta_\chi$ is not affected by the period of kination and ${R(k \lesssim 0.5 \, k_\mathrm{KR}) = 1}$.

Figure \ref{Fig:TransferFuctionz50} shows matter transfer functions evaluated at ${z=50}$ that include an era of kination with various values of $T_\mathrm{KR}$ as well as the standard matter transfer function without an era of kination. The arrows represent the values of ${k_\mathrm{KR}/k_\mathrm{eq}}$ for the two kination cases, and they thus represent where the transfer function begins to deviate from the standard case.  Perturbation modes with ${k/k_\mathrm{eq}<1}$ enter the horizon during matter domination, and $T(k)$ is scale invariant there.  Perturbation modes with ${k_\mathrm{eq}< k < k_\mathrm{KR}}$ enter the horizon during radiation domination, and ${T(k) \propto \mathrm{ln}[k/(8k_\mathrm{eq})]/k^2}$ in this regime. Finally, modes with ${k> k_\mathrm{KR}}$ enter the horizon during an era of kination, and ${T(k) \propto k^{-3/2}}$ here.  
%It is clear from Eq.~(\ref{eq:TransferFunctionKD}) that for modes entering the horizon during an era of kination, ${\delta_\chi \propto k^{1/2}}$.  Combining this with Eq.~\ref{eq:TransferFunctionDefinition} shows that ${T(k) \propto k^{-3/2}}$ for modes with ${k> k_\mathrm{KR}}$.  
It follows that the power spectrum of density perturbations (${P_\delta \propto k^{n_s} \, T^2}$) is proportional to $k^{n_s-3}$ for modes with ${k> k_\mathrm{KR}}$.

The transfer function described above is not applicable to arbitrarily high wavenumbers; there is a cutoff imposed that suppresses power for modes with $k> k_\mathrm{cut}$.  To do this, we introduce a Gaussian exponential cutoff to the transfer function:
\begin{align}
T(k) = \mathrm{exp}\left[ -\frac{k^2}{2 k_\mathrm{cut}^2} \right]T_0(k),
\label{eq:Cutoff}
\end{align}
where $T_0(k)$ is the transfer function not taking into account a cutoff.  The cutoff accounts for the effects of interactions between dark matter particles and the SM particles and the free streaming of dark matter particles following their kinetic decoupling from the SM particles \cite{Green:2005fa, Loeb:2005pm, Bertschinger:2006}. We discuss these effects next.

%%%%%%%%%%%%%%%
\subsection{Kinetic decoupling and free streaming}
\label{sec:KDandFS}
%%%%%%%%%%%%%%%

The kinetic decoupling of dark matter occurs when the dark matter ceases to efficiently exchange momentum with relativistic particles.  The momentum exchange rate for non-relativistic dark matter with mass $m_\chi$ scattering off of relativistic particles with temperature $T$ and number density $n_{\mathrm{rel}}$ is
\begin{align}
\Gamma_{\mathrm{mt}} \simeq \sigma_{\mathrm{el}} \, n_{\mathrm{rel}} \frac{T}{m_\chi},
\label{eq:MomentumTransfer}
\end{align}
where ${\sigma_\mathrm{el}}$ is the dark matter scattering cross section and the factor of $T/m_\chi$ accounts for the fact that it takes $m_\chi/T$ collisions to significantly change the dark matter particle's momentum \cite{Bringmann:2006mu}.  The dark matter remains in kinetic equilibrium while $\Gamma_{\mathrm{mt}} > H$, where $H$ is the Hubble rate.  Once ${\Gamma_{\mathrm{mt}} = H}$, the dark matter kinetically decouples from the relativistic particles as the average time needed to change the dark matter particles' momenta becomes longer than the age of the Universe.  The kinetic decoupling temperature ${T_\mathrm{kd}}$ is defined from the relation ${\Gamma_{\mathrm{mt}}(T_\mathrm{kd}) = H(T_\mathrm{kd})}$.

The time of kinetic decoupling directly influences the growth of dark matter perturbations.
After a mode enters the horizon during radiation or kinaton domination, density perturbations in these species oscillate due to their pressure, but the gravitational potential that they source decays rapidly due to cosmic expansion.
Dark matter particles are kicked by this transient gravitational potential, and in the absence of collisions, their resulting motion yields the $k\gg k_\mathrm{eq}$ 
portion of the transfer function discussed in Sec.~\ref{sec:TransferFunction}. 
However, if the dark matter is still collisionally coupled to the radiation after horizon entry, it instead inherits the radiation's oscillations, with further damping induced if the coupling is imperfect.
Consequently, dark matter perturbations that enter the horizon prior to kinetic decoupling are modified and suppressed \cite{Green:2005fa, Loeb:2005pm, Bertschinger:2006}.
Additionally, the dark matter temperature is maintained at the temperature of the radiation up until kinetic decoupling, and the associated random motion of dark matter particles further suppresses perturbations, potentially even at scales that enter the horizon significantly after kinetic decoupling.

To calculate $T_\mathrm{kd}$, we first evaluate the Hubble rate during an era of kination under the assumption that the kinaton is not decaying.  At kinaton-radiation equality, ${H(T_\mathrm{KR}) = \sqrt{2 (4\pi^3/45m_\mathrm{pl}^2) \, g_{*\mathrm{KR}} \, T_\mathrm{KR}^4}}$.  While the kinaton dominates the energy density of the Universe,
\begin{align}
H(T) = \left(\frac{4\pi^3}{45} \right)^{1/2} \left(\frac{T^3}{T_\mathrm{KR} \,m_\mathrm{pl}} \right) \left( \frac{g_{*s}(T)}{g_{*s\mathrm{KR}}} \right) g_{*\mathrm{KR}}^{1/2}.
\label{eq:HubbleTemp}
\end{align}
For many dark matter candidates, ${\sigma_{\mathrm{el}} \propto (T/m_\chi)^2}$, and therefore ${\Gamma_{\mathrm{mt}} \propto T^6}$ \cite{Gelmini:2008KD}.  Using this relation for $\Gamma_{\mathrm{mt}}(T)$ and Eq.~(\ref{eq:HubbleTemp}), it follows that the kinetic decoupling temperature during an period of kination is given by
\begin{align}
T_\mathrm{kd} = \left( \frac{T_\mathrm{kdS}^4}{T_\mathrm{KR}} \right)^{1/3} \left(\frac{g_{*s}(T_\mathrm{kd})}{g_{*s\mathrm{KR}}} \right)^{1/3} \left( \frac{g_{*\mathrm{KR}}}{g_{*}(T_\mathrm{kdS})} \right)^{1/6},
\label{eq:Tkd}
\end{align}
where $T_\mathrm{kdS}$ is the temperature at which the dark matter would kinetically decouple during radiation domination:
\begin{align}
\Gamma_{\mathrm{mt}}(T_\mathrm{kdS}) \equiv \sqrt{\frac{8\pi}{3 m_\mathrm{pl}^2} \, \frac{\pi^2}{30} g_*(T_\mathrm{kdS})T_\mathrm{kdS}^4}.
\label{eq:DefinitionTkds}
\end{align}
The value of $T_\mathrm{kdS}$ has been calculated for many dark matter models and only depends on the dark matter microphysics that determines the elastic scattering rate \cite{Profumo:2006}.  The wavenumber of the mode that enters the horizon when $T = T_{\mathrm{kd}}$ is $k_\mathrm{kd}$:
 \begin{align}
\frac{k_\mathrm{kd}}{k_\mathrm{KR}} & \equiv \frac{a_\mathrm{kd}\,H(T_\mathrm{kd})}{a_\mathrm{KR}\,H(T_\mathrm{KR})} \nonumber \\
&= \frac{1}{\sqrt{2}} \left(\frac{T_\mathrm{kd}}{T_\mathrm{KR}} \right)^2 \left(\frac{g_{*s\mathrm{kd}}}{g_{*s\mathrm{KR}}} \right)^{2/3}  ,
\label{eq:kkdOVERkRH}
\end{align}
where $a_\mathrm{kd}$ is the scale factor value at kinetic decoupling, $a_\mathrm{KR}$ is the scale factor value at kinaton-radiation equality, and ${g_{*s\mathrm{kd}} = g_{*s}(T_{\mathrm{kd}})}$.

After kinetic decoupling, the dark matter velocity decreases as $1/a$.   The dark matter free-streaming length, $\lambda_{\mathrm{fs}}$, is the comoving distance covered by the dark matter from the time of kinetic decoupling to the present:
\begin{align}
\lambda_\mathrm{fs} = \int_{t_\mathrm{kd}}^{t_0} \frac{v}{a} \mathrm{d}t \simeq \sqrt{\frac{T_\mathrm{kd}}{m_\chi}} \, a_\mathrm{kd} \int_{a_\mathrm{kd}}^{1} \frac{\mathrm{d}a}{a^3 H(a)}.
\label{eq:FreeStreamingLength}
\end{align}
Since entropy is conserved both during and after kination, $T \propto g_{*s}(T)^{-1/3} \, a^{-1}$, and
\begin{align}
\lambda_\mathrm{fs} =& \sqrt{\frac{2T_\mathrm{kd}}{m_\chi}} \frac{a_\mathrm{kd}}{H(T_\mathrm{KR})} \nonumber\\
&\times\int_{a_\mathrm{kd}}^1\frac{\diff a}{a^3}\left[\left(\frac{a_\mathrm{KR}}{a}\right)^{-6}+\frac{g_*(a)g_{*\mathrm{KR}}^{1/3}}{g_{*S}^{4/3}(a)} \left( \frac{a_\mathrm{KR}}{a}\right)^4\right.\nonumber\\
&\quad\quad\quad\quad\quad\left.+ \frac{g_*(a_\mathrm{eq})g_{*\mathrm{KR}}^{1/3}}{g_{*S}^{4/3}(a_\mathrm{eq})}\left( \frac{a_\mathrm{KR}^4}{a^3 a_\mathrm{eq}}\right) 
\right]^{-1/2}.
\label{eq:FreeStreamingLength2}
\end{align}
Perturbations with $k\gtrsim \lambda_\mathrm{fs}^{-1}$ are suppressed by the free streaming of dark matter particles out of overdense regions and into underdense regions. 

\begin{figure}
\centering\includegraphics[width=\columnwidth,trim={1.4cm .9cm 1.4cm .5cm},clip]{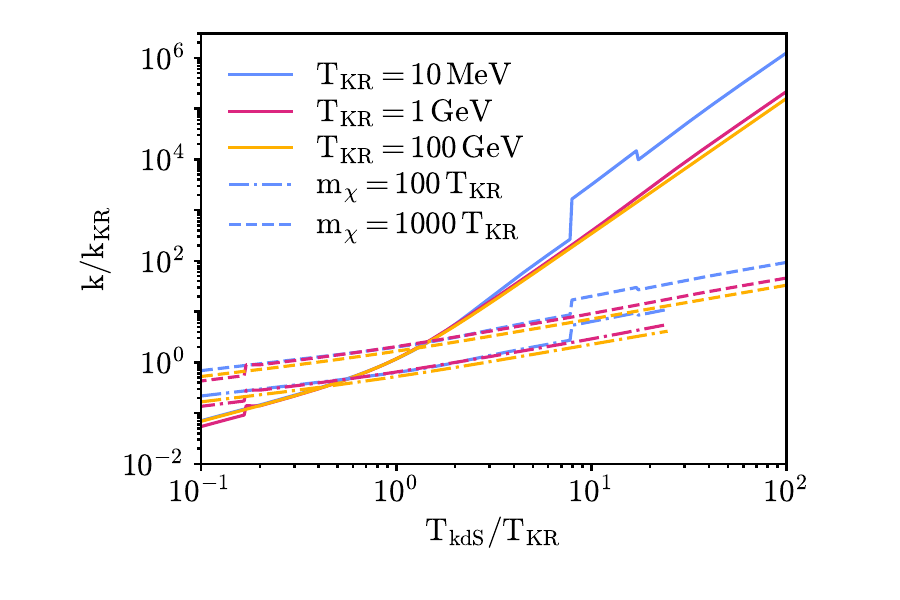}
\caption{The kinetic decoupling and free-streaming scales as a function of $T_\mathrm{kdS}/T_\mathrm{KR}$ for multiple values of $T_\mathrm{KR}$.  The solid lines represent ${k_\mathrm{kd}/k_\mathrm{KR}}$ and the dashed lines represent ${k_\mathrm{fs}/k_\mathrm{KR}}$.  The values of ${k_\mathrm{fs}/k_\mathrm{KR}}$ were calculated taking ${m_\chi = 100 \, T_\mathrm{KR}}$ and ${m_\chi= 1000 \, T_\mathrm{KR}}$.  Since our calculation of the free-streaming scale is only valid when \mbox{$T_\mathrm{kd} \lesssim m_\chi$}, the $k_\mathrm{fs}/k_\mathrm{KR}$ curves for ${m_\chi = 100 \, T_\mathrm{KR}}$ are restricted to \mbox{$T_\mathrm{kdS}\lesssim 25 T_\mathrm{KR}$}.}
\label{Fig:kOVERkRH}
\end{figure}

Figure \ref{Fig:kOVERkRH} shows the kinetic decoupling and free streaming scales ($k_\mathrm{fs} \equiv \sqrt{2}/\lambda_\mathrm{fs}$) as a function of $T_\mathrm{kdS}/T_\mathrm{KR}$ for multiple values of $T_\mathrm{KR}$.  The solid lines represent ${k_\mathrm{kd}/k_\mathrm{KR}}$ and the dashed lines represent ${k_\mathrm{fs}/k_\mathrm{KR}}$ for two different values of $m_\chi/T_\mathrm{KR}$.  Figure \ref{Fig:kOVERkRH} shows there is little variation in ${k_\mathrm{fs}/k_\mathrm{KR}}$ and ${k_\mathrm{kd}/k_\mathrm{KR}}$ for different values of $T_\mathrm{KR}$.  Figure  \ref{Fig:kOVERkRH} also shows that \mbox{$k_\mathrm{kd}$} is generally larger than $k_\mathrm{fs}$; the same efficient free streaming that is responsible for the rapid growth of dark matter perturbations during kination \cite{Redmond:2018} enhances the dark matter free-streaming length relative to the horizon size at kinetic decoupling.   Consequently, we expect free streaming to provide the dominant cutoff to the matter power spectrum when dark matter decouples during kination.

Ref.~\cite{Green:2005fa} found that free streaming suppresses $\delta_\chi$ by a factor of 
\begin{equation}
T_\mathrm{fs}(k) = \left[1-\frac23\left(\frac{k}{k_\mathrm{fs}}\right)^2\right]\exp\left[-\left(\frac{k}{k_\mathrm{fs}}\right)^2\right].
\end{equation}
It is customary, however, to apply a Gaussian cutoff in the matter power spectrum, $P(k) \propto \exp(-k^2/k_\mathrm{cut}^2)$ (e.g. \cite{Loeb:2005pm,Bertschinger:2006}), and we wish to maintain that convention.  For $T_\mathrm{fs}^2(k) > 0.2$, $\exp[-(10/3)k^2/k_\mathrm{fs}^2]$ matches $T_\mathrm{fs}^2(k)$ to within 10\%.  We therefore set $k_\mathrm{cut} = \sqrt{3/5}\lambda_\mathrm{fs}^{-1}$ when applying a \mbox{$P(k) \propto \exp[-k^2/k_\mathrm{cut}^2]$} cutoff to the matter power spectrum.  

\begin{figure}
\centering\includegraphics[width=\columnwidth,trim={1.0cm .9cm .9cm .5cm},clip]{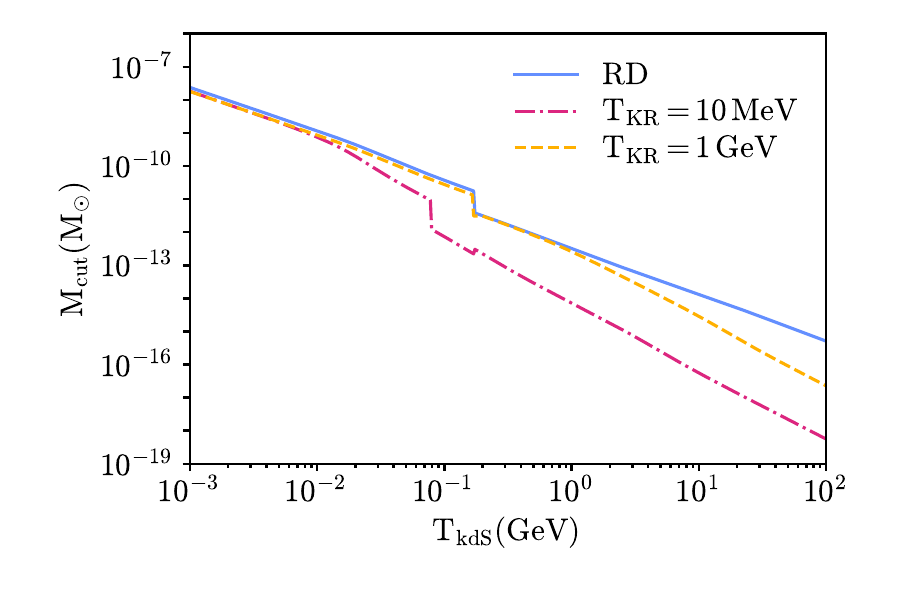}
\caption{The minimum halo mass as a function of $T_\mathrm{kdS}$ -- the temperature at which the dark matter would decouple in a radiation-dominated universe -- for different cosmological histories.  The solid curve shows the minimum halo mass if dark matter decouples while the universe is radiation dominated, while the dashed lines correspond to kination scenarios with different values of $T_\mathrm{KR}$.  Dark matter particles decouple earlier during a period of kination, leading to a reduction in the minimum halo mass for $T_\mathrm{kdS} < T_\mathrm{KR}$.}
\label{Fig:Mcut}
\end{figure}

The cutoff in the matter power spectrum sets the minimum size of dark matter halos: $M_\mathrm{cut} = 4\pi \rho_{m,0} k_\mathrm{cut}^{-3}$, where $\rho_{m,0}$ is the present-day matter density.  Figure \ref{Fig:Mcut} shows that a period of kination reduces $M_\mathrm{cut}$ for a given value of $T_\mathrm{kdS}$ in spite of the fact that free-streaming dark matter particles traverse more comoving distance during kination.  The reduction in $M_\mathrm{cut} $ arises because a period of kination effectively cools the dark matter by forcing it to decouple earlier: Eq.~(\ref{eq:Tkd}) implies that kination increases $T_\mathrm{kd}$ by roughly a factor of $(T_\mathrm{kdS}/T_\mathrm{KR})^{1/3}$.  Since the temperature of the dark matter particles falls as $1/a^2$ after decoupling, dark matter particles that decouple earlier are colder than particles that remain in kinetic equilibrium with relativistic particles longer.  As seen in Fig.~\ref{Fig:Mcut}, the cooling effect of the earlier decoupling more than compensates for the larger distance that dark matter particles can travel during kination, leading to a net reduction in $M_\mathrm{cut}$.  The existence of smaller dark matter halos implies that a period of kination will enhance the dark matter annihilation rate even if most of the perturbation modes that experience linear growth during kination are erased by free streaming, as Fig.~\ref{Fig:kOVERkRH} shows is the case if $m_\chi < 1000 T_\mathrm{KR}$.

It is also possible that dark matter particles do not interact with SM particles and instead exist as part of a hidden sector that is thermally decoupled from the visible sector \cite{KST85,PhysRevD.47.456,Chen:2006ni,Feng:2008mu,Berlin:2016First, Berlin:2016}.  
Since the temperature of particles in the hidden sector can differ from the SM temperature  \cite{Adshead:2016xxj},
the dark matter temperature when it decouples from the other hidden particles may not equal the SM temperature at that time.  If the particles in the hidden sector have a temperature $T_\mathrm{HS} = \epsilon T$ and the dark matter begins to free stream when the SM temperature is $T_\mathrm{kd}$, the free-streaming horizon is $\sqrt{\epsilon}$ times the expression given by Eq.~(\ref{eq:FreeStreamingLength2}).  Therefore, $\kcut/k_\mathrm{KR}$ can be much larger than the values shown in Fig.~\ref{Fig:kOVERkRH} if the hidden sector is colder than the visible sector.

%%%%%%%%%%%%%%%
\section{Kination's Impact on Structure}
\label{sec:PrimordialStructures}
%%%%%%%%%%%%%%%

\begin{figure}
\centering\includegraphics[width=3.4in]{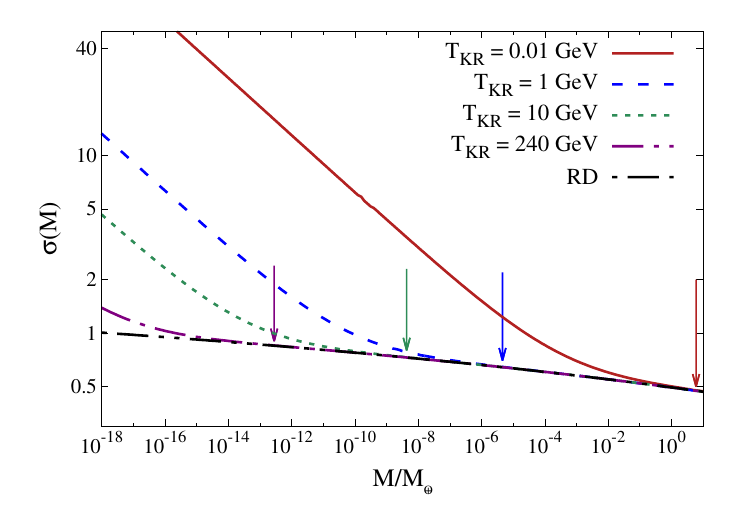}
\caption{The rms density perturbation $\sigma$ in a sphere of average mass $M$ evaluated at $z = 50$ for scenarios with $T_\mathrm{KR} = 0.01$~GeV, 1~GeV, 10~GeV, and 240~GeV.  Also included is the $\sigma(M)$ value calculated assuming uninterrupted radiation domination.  The arrows correspond to the values of $M_\mathrm{KR}/M_\oplus$ for each value of $T_\mathrm{KR}$.  For ${M<M_\mathrm{KR}}$, $\sigma(M)$ is larger than it would be if there were no period of kination. }
\label{Fig:Sigma}
\end{figure}

We use Press-Schechter halo mass functions \cite{Press:1973} to determine how the enhancement to $\delta_\chi$ during kination affects dark matter halo formation.  We first calculate the rms density perturbation in a sphere containing mass $M$ on average:
\begin{align}
\sigma^2(M,z) = \int \frac{\mathrm{d}^3k}{(2\pi)^3} [D(k,z) T_{50}(k)]^2 P_\mathrm{P}(k) F^2(kR) ,
\label{eq:Sigma}
\end{align}
where $D(k,z)$ is the scale-dependent growth function defined in Section \ref{sec:TransferFunction}; $T_{50}(k)$ is the transfer function defined in Eq.~(\ref{eq:TransferFunctionDefinition}) evaluated at $z=50$; $P_\mathrm{P}(k)$ is the present-day power spectrum of large-scale ($k \ll k_\mathrm{eq}$) matter density perturbations; and $F(kR)$ is a filter function that suppresses contributions from modes with ${k^{-1} \ll R = [3M/(4\pi \rho_{m,0})]^{1/3}}$.  If the matter power spectrum has a small-scale cutoff, using a sharp-$k$ filter to calculate $\sigma(M)$ generates more accurate Press-Schechter mass functions than a top hat filter \cite{Benson:2012}.  We use 
\begin{align}
F(kR) = \left\{
        \begin{array}{ll}
            1 & \quad kR \leq 1.85; \\
            0 & \quad kR > 1.85;
        \end{array}
    \right.
\label{eq:Filter}
\end{align}
the transition at $kR=1.85$ is chosen such that this sharp-$k$ filter gives the same value for $\sigma(M)$ as a top hat filter in the absence of a small-scale cutoff in the power spectrum.  Figure \ref{Fig:Sigma} shows $\sigma(M)$ evaluated at ${z=50}$ without a period of kination and for kination scenarios with different values of $T_\mathrm{KR}$.  It is evident that the growth of subhorizon density perturbations during kination enhances $\sigma(M)$ for small masses.  In addition, as $T_\mathrm{KR}$ decreases, $\sigma(M)$ deviates at larger values of $M$ from that predicted assuming radiation domination.

There is a characteristic mass scale $M_\mathrm{KR}$ defined by the mass enclosed in ${R(M_\mathrm{KR}) \equiv k^{-1}_{\mathrm{KR}}}$:
\begin{align}
M_\mathrm{KR} = 21.4 \, M_{\oplus}\, \left(\frac{g_{*s\mathrm{KR}}}{3.91}\right) \, \left(\frac{3.36}{g_{*\mathrm{KR}}}\right)^{3/2} \, \left( \frac{0.01 \, \mathrm{GeV}}{T_\mathrm{KR}} \right)^3,
\label{eq:MRH}
\end{align}
where $M_{\oplus}$ is the Earth mass. In obtaining Eq.~(\ref{eq:MRH}), we set $\rho_{m,0}$ equal to the density of dark matter alone (with $\Omega_\chi h^2 = 0.12$ \cite{Planck:2018vyg}) because only dark matter is expected to accrete onto such small halos \cite{Bertschinger:2006}.  The mass scale $M_\mathrm{KR}$ is the largest mass for which a period of kination enhances $\sigma(M)$.  In Figure \ref{Fig:Sigma}, the arrows indicate the values of ${M_\mathrm{KR}}$ for each value of $T_\mathrm{KR}$.  For $M > M_\mathrm{KR}$, $\sigma(M)$ is insensitive to $T_\mathrm{KR}$ because it only depends on modes with $k < k_\mathrm{KR}$.  For ${M \ll M_\mathrm{KR}}$,  ${\sigma(M)}$ is most sensitive to modes with ${k \gg k_\mathrm{KR}}$, which enter the horizon during an era of kination.  Since ${P_\delta(k) \propto k^{n_s-3}}$ for these modes, ${\sigma(M \lesssim M_\mathrm{KR})\propto M^{-n_s/6}}$.

\begin{figure}
\centering\includegraphics[width=3.4in]{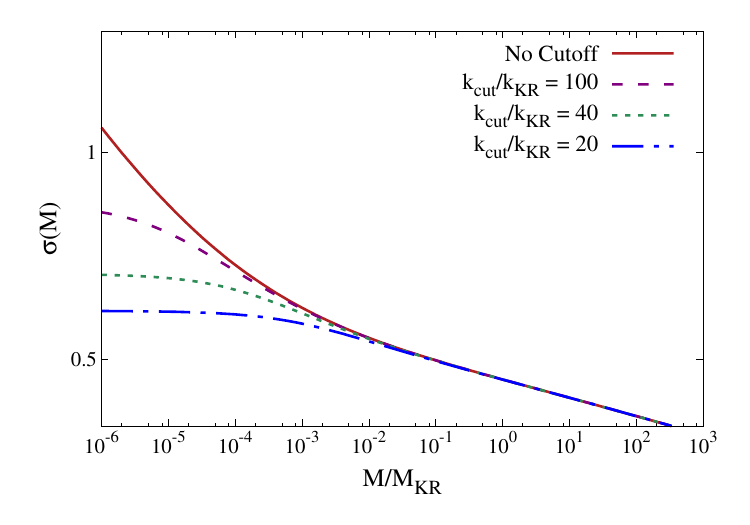}
\caption{The rms density perturbation $\sigma(M)$ in a sphere of average mass $M$ evaluated at $z = 50$ for scenarios with $T_\mathrm{KR} =$0.01 GeV and cutoffs to the matter power spectrum of ${k_\mathrm{cut}/k_\mathrm{KR} = 100,40,20}$.  The red line corresponds to $\sigma(M)$ with no cutoff to the matter power spectrum.}
\label{Fig:SigmaCutoff}
\end{figure}

\begin{figure*}[t]
 \centering
\begin{subfigure}
\centering
 \includegraphics[width=.45\textwidth]{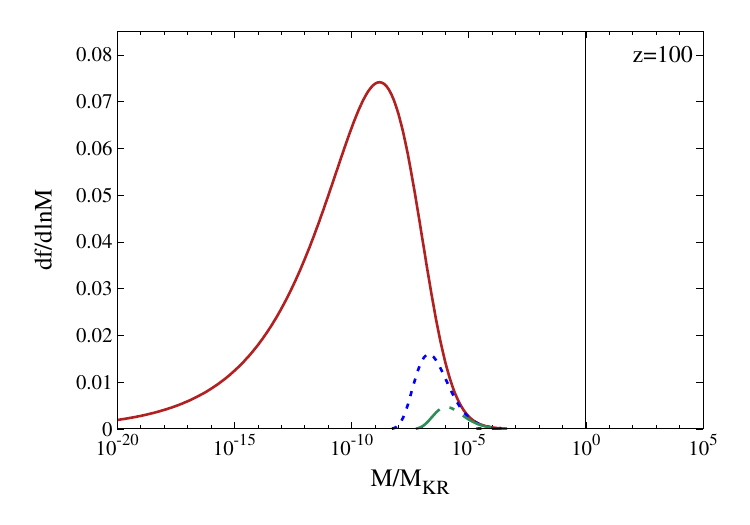}
 \end{subfigure}%
\begin{subfigure}
\centering
 \includegraphics[width=.45\textwidth]{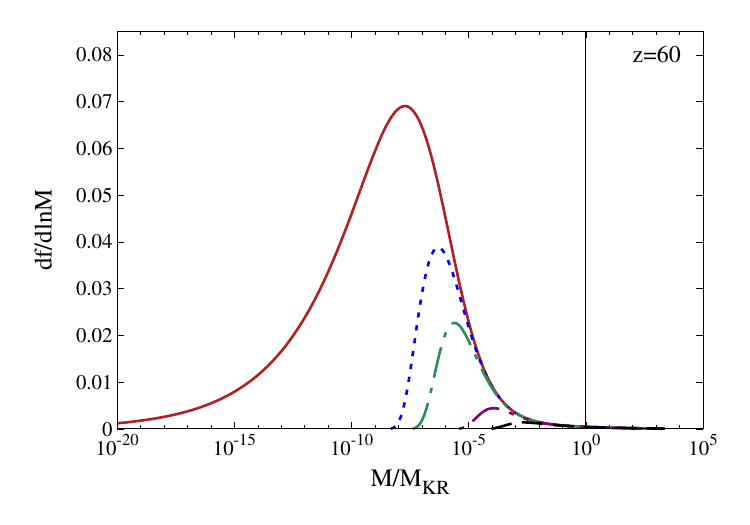}
 \end{subfigure}%
\begin{subfigure}
\centering
 \includegraphics[width=.45\textwidth]{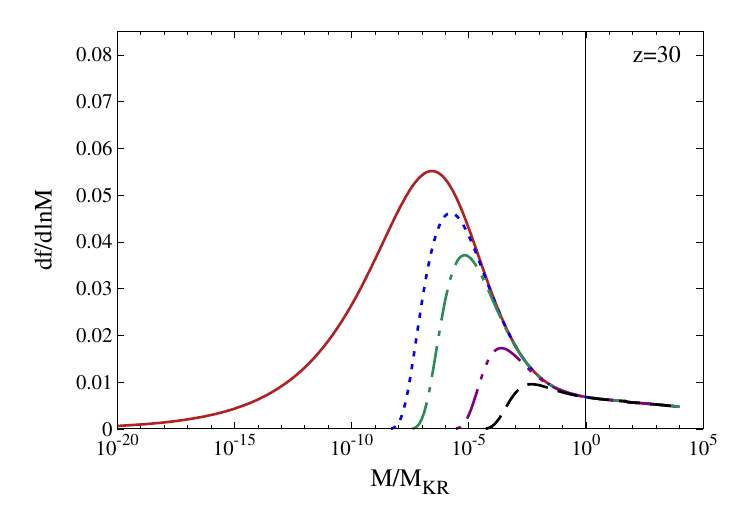}
 \end{subfigure}%
 \begin{subfigure}
\centering
 \includegraphics[width=.45\textwidth]{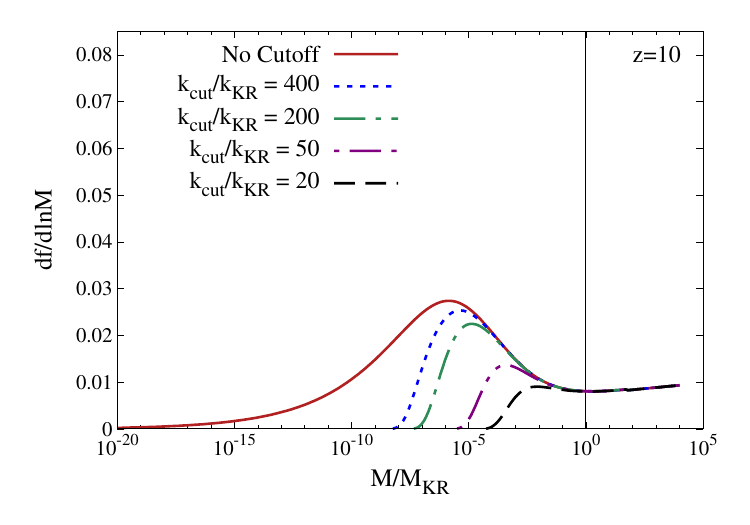}
 \end{subfigure}%

\caption{The evolution of $\diff f/\diff\mathrm{ln}M$ with $T_\mathrm{KR} = 1 \, \mathrm{GeV}$.  The different lines represent $df/d\mathrm{ln}M$ without a cutoff in the matter power spectrum and with cutoff values of  ${k_\mathrm{cut}/k_\mathrm{KR}=400,200,50}$, and 20 respectively.  The solid vertical line marks ${M = M_\mathrm{KR}}$. }
\label{Fig:dfdlnM}
\end{figure*}

To take into account the effects of free streaming, we impose a cutoff to the matter power spectrum that suppresses power for modes with $k> k_\mathrm{cut}$.
Figure \ref{Fig:SigmaCutoff} demonstrates the implications of imposing a cutoff on $\sigma(M)$; imposing a cutoff suppresses $\sigma(M)$ for small values of $M$.  As $k_\mathrm{cut}/k_\mathrm{KR}$ decreases, $\sigma(M)$ becomes more suppressed at smaller mass scales.  

The Press-Schechter formalism predicts that microhalos are common once $\sigma(M)$ exceeds the critical linear density contrast $\delta_c=1.686$.
With no cutoff, $\sigma(M)$ increases without limit as $M$ decreases, and microhalos form at arbitrarily high redshifts.  Imposing a cutoff limits the amplitude of $\sigma(M)$, and its maximum value is highly sensitive to ${k_\mathrm{cut}/k_\mathrm{KR}}$.  Therefore, the ratio ${k_\mathrm{cut}/k_\mathrm{KR}}$ largely determines when the first microhalos form.  The mass of these microhalos is determined by $k_\mathrm{cut}$.

After calculating ${\sigma(M)}$, we use the Press-Schechter formalism to calculate the differential comoving number density
\begin{align}
\frac{\diff n}{\diff\mathrm{ln}M} = \sqrt{\frac{2}{\pi}} \frac{\rho_{m,0}}{M} \left|\frac{\diff\mathrm{ln}\sigma}{\diff \mathrm{ln}M}\right| \frac{\delta_c}{\sigma(M,z)} \mathrm{exp}\left[ -\frac{\delta^2_c}{2 \sigma^2(M,z)} \right],
\label{eq:HaloMassFunction}
\end{align}
of halos with mass $M$ at redshift $z$. It follows that the differential fraction of mass contained in halos is
\begin{align}
\frac{\diff f}{\diff\mathrm{ln}M} = \frac{M}{\rho_{m,0}} \frac{\diff n}{\diff\mathrm{ln}M} .
\label{eq:DifferentialBoundFraction}
\end{align}
Figure \ref{Fig:dfdlnM} shows ${\diff f/\diff\mathrm{ln}M}$ as a function of $M$ for various cutoffs to the matter power spectrum with ${T_\mathrm{KR} = 1 \, \mathrm{GeV}}$; the solid vertical line marks ${M = M_\mathrm{KR}}$.  Figure \ref{Fig:dfdlnM} illustrates how a cutoff to the matter power spectrum influences the minimum halo mass.  For example, at a redshift of 60 with no cutoff to the matter power spectrum, halos with $10^{-17} M_\mathrm{KR} \lesssim M \lesssim 10^{-2} M_\mathrm{KR}$ are prevalent, whereas setting  $k_\mathrm{cut}/k_\mathrm{KR} = 400$ eliminates halos with $M\lesssim 10^{-8} M_\mathrm{KR}$.   As $k_\mathrm{cut}/k_\mathrm{KR}$ decreases, the minimum halo mass increases and microhalo formation is postponed to later times.  
%In addition, Figure \ref{Fig:dfdlnM} shows that the fraction of halos bound with masses greater than $M_\mathrm{KR}$ is unaffected by the cutoff to the matter power spectrum.

\begin{figure}
\centering\includegraphics[width=3.4in]{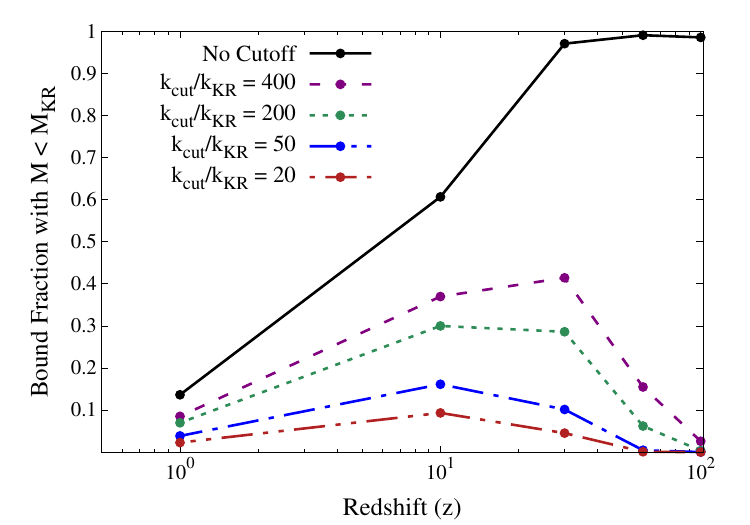}
\caption{The total bound fraction for ${M<M_\mathrm{KR}}$, evaluated with and without cutoffs to the matter power spectrum.  The total bound fraction was evaluated with $T_\mathrm{KR} = 0.01 \, \mathrm{GeV}$.  }
\label{Fig:TotalBoundFraction}
\end{figure}

Figure \ref{Fig:TotalBoundFraction} shows, as a function of redshift $z$, the fraction
\begin{equation}
    f_{<M_\mathrm{KR}}(z)\equiv 
    \int_{0}^{M_\mathrm{KR}} \frac{\diff M}{M} \left.\frac{\diff f}{\diff \mathrm{ln}M}\right|_z,
\end{equation}
of dark matter bound into halos with masses less than $M_\mathrm{KR}$
for scenarios with and without cutoffs to the matter power spectrum.  We fix ${T_\mathrm{KR} = 0.01 \, \mathrm{GeV}}$ here, but the total bound fraction is only weakly sensitive to $T_\mathrm{KR}$ because $\sigma(M)$ is nearly flat for $M>M_\mathrm{KR}$.  Figure \ref{Fig:TotalBoundFraction} confirms that as ${k_\mathrm{cut}/k_\mathrm{KR}}$ decreases, the formation of halos is delayed and the fraction of dark matter within the halos is decreased.  For example, by a redshift of $30$, at least a third of the dark matter is bound into halos with masses smaller than $M_\mathrm{KR}$ for ${k_\mathrm{cut}/k_\mathrm{KR} \gtrsim 200}$, yet at the same redshift only $10\%$ of the dark matter is bound into halos with $M<M_\mathrm{KR}$ for ${k_\mathrm{cut}/k_\mathrm{KR} = 50}$.

The ${M<M_\mathrm{KR}}$ bound fraction in Fig.~\ref{Fig:TotalBoundFraction} decreases at late times due to mergers, which produce more halos with larger masses and decrease the abundance of smaller-mass field halos. However, many of the smaller-mass halos are likely to survive as subhalos inside larger hosts; Press-Schechter calculations do not account for subhalos. Since subhalos are an important contributor to the dark matter annihilation rate, this consideration (among others) motivates the use of a different approach to evaluate the dark matter annihilation rate in kination scenarios, as we discuss next.

%%%%%%%%%%%%%%%
\section{Limits on dark matter annihilation in kination scenarios}
\label{sec:SubstructureConstraints}
%%%%%%%%%%%%%%%

If dark matter is a thermal relic, the abundance of dark matter microstructure induced by kination scenarios could substantially boost the dark matter annihilation rate today. In this section, we explore this boost and the extent to which it improves upon the constraints on kination scenarios derived in Ref.~\cite{Redmond:2017}.

An annihilation signal originating from unresolved microhalos traces those halos' spatial distribution, which follows the (smoothed) dark matter distribution. Consequently, the signal from this annihilation scenario closely resembles that of dark matter decay, and we follow the procedure in Refs.~\cite{Blanco:2019eij,StenDelos:2019xdk} to convert published bounds on the dark matter lifetime into constraints on annihilation within unresolved microhalos. In particular, we employ the limits on the dark matter lifetime derived in Ref.~\cite{Blanco:2018esa}. These constraints use the Fermi Collaboration's measurement of the isotropic gamma-ray background (IGRB) \cite{Fermi-LAT:2014ryh}. Reference~\cite{StenDelos:2019xdk} found that for microhalo-dominated annihilation scenarios, the IGRB yields constraints that are substantially stronger than those that employ gamma rays from dwarf spheroidal galaxies.

To convert the dark matter lifetime constraints in Ref.~\cite{Blanco:2018esa} into limits on the annihilation cross section, we equate the annihilation rate per mass, $\Gamma/M$, of particles with mass $m_\chi$ and cross section $\sigmav$, to the decay rate per mass of particles with mass $2m_\chi$ and effective lifetime $\tau_\mathrm{eff}$. That is,
\begin{equation}\label{decay}
\frac{\Gamma}{M_\chi}=\frac{\sigmav}{2m_\chi^2}\frac{\overline{\rhoc^2}}{\rhobar} = \frac{1}{2m_\chi\tau_\mathrm{eff}},
\end{equation}
where $\rhobar$ and $\overline{\rhoc^2}$ are the mean and mean squared dark matter density, respectively. This equation converts a lower bound on $\tau_\mathrm{eff}$ for particles with masses equal to $2m_\chi$ into an upper bound on $\sigmav$ for particles with masses equal to $m_\chi$. It connects to the dark matter halo population through the overall annihilation boost factor $\boost$.

%%%%%%%%%%%%%%%
\subsection{Predicting the annihilation boost}
\label{sec:boost}
%%%%%%%%%%%%%%%

To predict the annihilation boost factor $\boost$, we must know not only the population of dark matter halos but also their internal mass distributions, characterized by each halo's radial density profile $\rho(r)$. Moreover, we seek not only the population of field halos, discussed in Sec.~\ref{sec:PrimordialStructures}, but also that of subhalos that reside inside larger halos. While Figs.~\ref{Fig:dfdlnM}~and~\ref{Fig:TotalBoundFraction} show that the mass in field halos smaller than $\MKR$ decreases dramatically by the present day, we expect many of these microhalos to persist as subhalos.

The common approach to the problem of predicting $\boost$, exemplified in Ref.~\cite{Fermi-LAT:2015qzw}, is as follows.
\begin{enumerate}[label={(\arabic*)}]
	\item Use a Press-Schechter-like mass function $\diff n/\diff\ln M$ to characterize the field halo population.
	\item Describe the subhalo population using a subhalo mass function, which is typically modeled as a power law $\diff N/\diff M\propto M^{-\alpha}$ with $\alpha\sim 2$; $N$ and $M$ here are the subhalo count and mass, respectively.
	\item Use a concentration-mass relation to map each halo's mass to the typical density profile of halos of that mass.
\end{enumerate}
However, the subhalo mass functions and concentration-mass relations are tuned to the results of cosmological simulations carried out using a conventional cold dark matter power spectrum. We cannot expect them to remain accurate in a kination scenario.

Instead, we utilize the connection developed in Refs.~\cite{Delos:2019mxl,Delos:2022yhn} between the properties of local maxima in the linear density field and the density profiles of the halos that form when they collapse. This Peak-to-Halo (P2H) method associates each peak in the primordial (linear) density field to a collapsed halo and predicts the halo's density profile from the properties of the peak. 
Importantly, this mapping was shown to remain accurate for vastly different power spectra, making it suitable for the study of kination and other nonstandard cosmologies (e.g. Refs.~\cite{StenDelos:2019xdk,Delos:2021rqs}). Using the prescription in Ref.~\cite{Delos:2019mxl}, we exploit the Gaussian statistics of the linear density field to sample $10^6$ peaks for each kination power spectrum. We assume that each peak collapses to form a prompt $\rho=A r^{-3/2}$ cusp \cite{White:2022yoc,Delos:2022bhp}, and we use the P2H model to predict the cusp's coefficient $A$. Since baryonic matter does not cluster at mass scales below about $10^5 M_\odot$ \cite{Bertschinger:2006}, the growth rate of dark matter density perturbations below this scale is suppressed, an effect not accounted for in Ref.~\cite{Delos:2019mxl}. Thus, we use the modification presented in Ref.~\cite{StenDelos:2019xdk} that accounts for this slower growth when computing $A$.\footnote{\textsc{Python} code that implements these calculations is publicly available at \url{https://github.com/delos/microhalo-models}.}

The P2H method yields the initial population of microhalos, which is altered by subsequent hierarchical clustering processes. We consider two estimates of the impact of microhalo mergers.
\begin{enumerate}[label={(\arabic*)}]
	\item For a conservative estimate, we assume that interactions between microhalos soften the initial $\rho\propto r^{-3/2}$ cusps, transforming them to the \mbox{$\rho\propto r^{-1}$} cusps associated with Navarro-Frenk-White (NFW) profiles \cite{1996ApJ...462..563N,1997ApJ...490..493N}. This outcome has been suggested by Refs.~\cite{2016MNRAS.461.3385O,2017PhRvD..96l3519G,2017MNRAS.471.4687A,2018MNRAS.473.4339O,Delos:2019mxl,2020MNRAS.492.3662I}.
    \item For an optimistic estimate, we follow Ref.~\cite{Delos:2022bhp} and assume, based on the most recent simulation results \cite{Delos:2022yhn}, that all halos retain their initial $\rho\propto r^{-3/2}$ density cusps in their central regions.  Since this prompt cusp dominates the annihilation signal, we neglect annihilation beyond its extent.
    \end{enumerate}

As we will see in Sec.~\ref{sec:Constraints}, the optimistic assumption so severely limits possible dark matter parameters that scenarios in which kination affects halo structure are ruled out. Consequently, for that case, we may use results from Ref.~\cite{Delos:2022bhp}, which
found that the annihilation boost obtained from the P2H method assuming prompt-cusp survival is fit well by the expression
\begin{equation}\label{promptcusps}
    \frac{\overline{\rhoc^2}}{\rhobar^2} = 0.08\left[\log\left(\frac{m_\chi}{\mathrm{GeV}}\frac{\Tkd}{\mathrm{GeV}}\right)+36\right]^5
\end{equation}
for standard thermal histories (without kination). Appendix~\ref{sec:halomodel} shows that this annihilation boost factor is about an order of magnitude larger than the results of standard halo-based computations (e.g.~\cite{Sanchez-Conde:2013yxa}).

For the conservative case, we use the same procedure as Ref.~\cite{StenDelos:2019xdk}. We assume that mergers relax the microhalos' density profiles to the NFW form, \cite{1996ApJ...462..563N,1997ApJ...490..493N}
\begin{equation}\label{NFW}
\rho(r) = \frac{\rho_s}{(r/r_s)(1+r/r_s)^{2}},
\end{equation}
and that the scale radius $r_s$ and density $\rho_s$ of this profile are set by $\rho_s^2 r_s^3=\omega A^2$, where $\omega$ is an undetermined proportionality constant. The $J$-factor for each halo, $J\equiv \int \rho^2 \diff V$, is then
\begin{equation}\label{JA}
J=(4\pi/3) \omega A^2.
\end{equation}
Based on the arguments in Ref.~\cite{StenDelos:2019xdk}, we conservatively set $\omega=(4/3)^2$.

The other main impact of mergers is to reduce the number of halos while increasing the surviving halos' sizes. Reference~\cite{Delos:2019mxl} found that the sum $\sum A^2$ over all halos remains close to its value for the initial peak population even after mergers have taken place. As Ref.~\cite{StenDelos:2019xdk} discusses, the notion that $\sum A^2$ is approximately conserved during mergers is also consistent with the idealized merger studies of Ref.~\cite{Drakos:2018fgl}. Mergers between identical halos in those studies generated a halo with nearly the same characteristic density $\rho_s$ as the progenitor halos. Since $\sum M$ is approximately conserved during the merger, the sum over $A^2\propto \rho_s M$ must also be preserved. 

\begin{figure}
	\centering
	\includegraphics[width=\columnwidth]{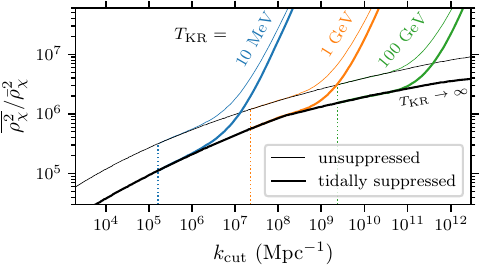}
	\caption{\label{fig:kinboost} The annihilation boost under the conservative assumption that prompt cusps do not survive. We plot the overall boost factor (relative to a homogeneous universe) for several kination scenarios (colors) as a function of the cutoff scale $\kcut$. In each case, $\kKR$, the wavenumber entering the horizon at kinaton-radiation equality, is shown with a dotted line. When $\kcut\gtrsim 10 \kKR$, kination's impact on small-scale power can boost the dark matter annihilation rate considerably compared to a scenario without a kination boost (black). We plot the boost both without tidal suppression of microhalos (thin curve) and with such tidal suppression (thick curve). As discussed in Appendix~\ref{sec:tidal}, we conservatively scale the boost by the smaller of the factors corresponding to tidal evolution of extragalactic and Galactic microhalos.}
\end{figure}

Assuming that $\sum A^2$ remains constant as the halo population evolves, the cosmologically averaged squared dark matter density predicted by the P2H model is
\begin{equation}\label{rho2}
\overline{\rhoc^2} = \frac{\bar n}{N}\sum_{i=1}^N J_i,
\end{equation}
where $\bar n$ is the number density of peaks \cite{1986ApJ...304...15B}, and we sum over the $N=10^6$ sampled peaks, computing $J_i$ for each peak using Eq.~(\ref{JA}). The boost factor $\boost$ computed using this procedure is depicted as the thin solid lines in Fig.~\ref{fig:kinboost} for a range of kination scenarios. This figure shows how kination's boost to the small-scale power spectrum results in a boost to the dark matter annihilation rate as long as $\kcut\gtrsim 10 \kKR$.

Beyond mergers between microhalos, the microhalo population is also influenced by accretion onto larger, later-forming dark matter structures such as galactic halos. The tidal influence of these larger halos gradually disperses and strips material from their subhalos.
We use the tidal evolution model developed in Ref.~\cite{Delos:2019lik} to estimate how this effect suppresses the annihilation rate within subhalos. In Appendix~\ref{sec:tidal}, we detail this calculation and show that tidal effects suppress the annihilation rate by a factor of about 3 for scenarios without significant kination-boosted structure. The suppression is weaker when kination's boost to structure is significant. Figure~\ref{fig:kinboost} shows the tidally suppressed annihilation boost.

\begin{figure}
	\centering
	\includegraphics[width=\columnwidth]{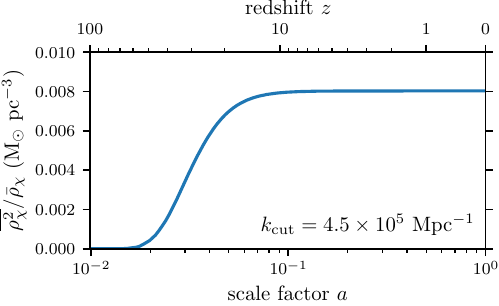}
	\caption{\label{fig:boostevo} Evolution of $\boostrho$, which is proportional to the annihilation rate per dark matter mass. We consider a standard thermal history (no kination), and tidal suppression is neglected. $\boostrho$ is essentially time independent for $z<20$. We assume here that prompt cusps do not survive, but Ref.~\cite{Delos:2022bhp} presents a similar conclusion under the assumption that they do persist.}
\end{figure}

The mapping in Eq.~(\ref{decay}) from decay to annihilation bounds requires that the annihilation rate per dark matter mass, proportional to $\boostrho$, be time-independent over the range of redshifts relevant to the IGRB. In Fig.~\ref{fig:boostevo}, we plot the time evolution of $\boost$ for one particular $\kcut$ and no kination boost. For simplicity, we neglect tidal suppression here.  Evidently, $\boostrho$ has negligible time dependence for the redshifts $z<20$ relevant to the IGRB.

While conventional halo model methods cannot account for kination-induced changes to the power spectrum, they have been applied to scenarios without these effects. It is reasonable to wonder how the predictions of our P2H model compare to halo-model predictions in these scenarios. We show in Appendix~\ref{sec:halomodel} that under the conservative assumption that prompt cusps do not survive, the P2H calculation of the annihilation boost matches the predictions of a halo-model calculation to within a factor of about 2. Moreover, for most parameters, the P2H results are bracketed by the halo-model predictions obtained under two common assumptions about the subhalo mass function. Note that in this comparison, the peak-based calculation is altered to neglect the fact that baryons do not cluster at small scales, since that matches the assumption made for the relevant halo models.

%%%%%%%%%%%%%%%
\subsection{Observational limits}
\label{sec:Constraints}
%%%%%%%%%%%%%%%

We are now prepared to develop constraints on kination scenarios. For each dark matter mass $m_\chi$ and temperature $\TKR$ of kinaton-radiation equality, we use the following procedure:
\begin{enumerate}[label={(\arabic*)}]
	\item We numerically integrate the background equations in Ref.~\cite{Redmond:2017} to determine the velocity-averaged dark matter annihilation cross section $\sigmav$ that generates the observed dark matter abundance ($\rhobar=33$~M$_\odot$kpc$^{-3}$, corresponding to $\Omega_\chi h^2=0.12$ \cite{Planck:2018vyg}) through thermal freeze out.
	\item We use Eq.~(\ref{decay}), with the given $m_\chi$ and established $\sigmav$, to convert the dark matter lifetime limits in Ref.~\cite{Blanco:2018esa} into a maximal allowed value of $\boost$.
	\item We use the P2H model, for the given $\TKR$, to determine the cutoff scale $\kcut$ that corresponds to this value of $\boost$.
    \item For dark matter that was once in kinetic equilibrium with the SM, we connect $\kcut$ to $\Tkd$ by evaluating the free-streaming scale, $\lfs=\sqrt{3/5}\kcut^{-1}$, as described in Section~\ref{sec:KDandFS}.
    \item We evaluate the corresponding value of $\TkdS$, the kinetic decoupling temperature for the same dark matter model in a standard thermal history (no kination). This quantity is useful because it is a property of the dark matter particle alone, and viable values for $\TkdS$ have been explored (e.g. Ref.~\cite{Cornell:2013rza}).
\end{enumerate}

\begin{figure}
	\centering
	\includegraphics[width=\columnwidth]{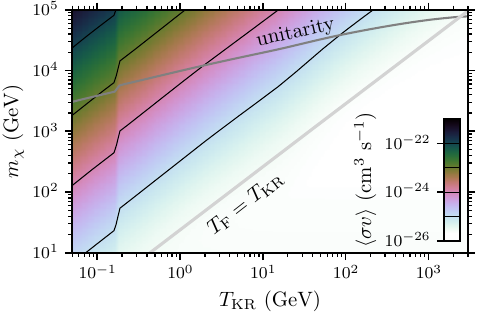}
	\caption{\label{fig:sigmav} The annihilation cross section $\sigmav$ for dark matter that freezes out with the observed abundance as a function of dark matter mass $m_\chi$ and temperature $\TKR$ of kinaton-radiation equality. The contours match the lines on the color bar and indicate successive powers of 10, starting at $10^{-25}$~cm$^3$~s$^{-1}$. Below the gray diagonal, the dark matter freezes out after the kination epoch ends, so $\sigmav\simeq 2\times 10^{-26}$~cm$^3$~s$^{-1}$ independently of $\TKR$. The upper gray line marks the unitarity limit \cite{Griest:1989wd}, which moves to smaller masses $m_\chi$ for dark matter that freezes out during kination because $\sigmav$ is higher.}
\end{figure}

Figure~\ref{fig:sigmav} shows how the value of $\sigmav$ that generates the observed dark matter density depends on $m_\chi$ and $\TKR$. 
As discussed in Ref.~\cite{Redmond:2017}, the required $\sigmav$ values exceed $ 2\times 10^{-26}\,\mathrm{cm^3\,s^{-1}}$ because the Hubble rate at a given temperature is higher during kination than it is during radiation domination, which causes dark matter to freeze out at a higher density. Since $H\sim T^3/(\TKR \mpl)$ during kination, the relic number density of dark matter particles that freeze out during kination would be independent of the freeze-out temperature $T_F$ if annihilation ceased when $H = n_\chi \sigmav$ and $T\propto a^{-1}$ thereafter.  However, annihilation continues to deplete the dark matter abundance throughout kination, which reduces the relic abundance by a factor of $\log(T_F/\TKR)$ \cite{Redmond:2017,DEramo:2017gpl}. Consequently, $\rhobar \propto m_\chi/[\sigmav\TKR\log(\Tf/\TKR)]$ when dark matter freezes out during kination.

\begin{figure}
	\centering
	\includegraphics[width=\columnwidth]{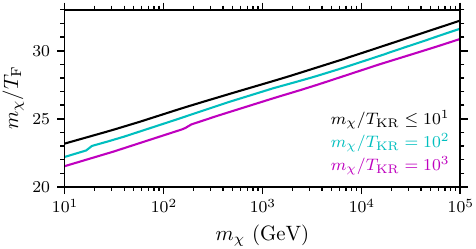}
	\caption{\label{fig:Tf} Ratio $m_\chi/\Tf$ between the dark matter mass and the freeze-out temperature for dark matter that has the observed density after freezing out from the SM radiation bath. 
 }
\end{figure}

The freeze-out temperature is defined by the relation $H(\Tf) = \sigmav n_{\chi,\mathrm{eq}}(\Tf)$, where $n_{\chi,\mathrm{eq}}(T)$ is the number density of dark matter particles in thermal equilibrium. Figure~\ref{fig:Tf} shows how $\Tf$ depends on $m_\chi$ and $\TKR$ when $\sigmav$ is chosen to give the correct freeze-out abundance.  Due to the exponential sensitivity of $n_{\chi,\mathrm{eq}}$ to $m_\chi/\Tf$, $m_\chi/\Tf \simeq 25$ for a wide range of $m_\chi$ values. Figure~\ref{fig:Tf} also demonstrates that $m_\chi/\Tf$ depends weakly on $m_\chi/\TKR$.
If the number of dark matter particles and the entropy of the radiation bath is conserved after freeze-out, then 
\begin{equation}
    n_{\chi,\mathrm{eq}}(\Tf) = \frac{\rhobar}{m_\chi}\left[\frac{g_{*S}(\Tf)\Tf^3}{g_{*S}(T_0)T_0^3}\right],
    \label{Tfdef}
\end{equation}
where $T_0$ is the present-day temperature and we continue to use $\rhobar$ to denote the present-day dark matter density.
In this case, it follows that $m_\chi/\Tf$ is independent of expansion history if $\sigmav$ is chosen to match the relic density to the observed density.  The persistent depletion of the dark matter during kination modifies Eq.~(\ref{Tfdef}): larger $m_\chi/\TKR$ values require smaller $m_\chi/\Tf$ values because a higher density at freeze-out is required to compensate for the loss of dark matter particles between freeze-out and the end of kination. 

\begin{figure}
	\centering
	\includegraphics[width=\columnwidth]{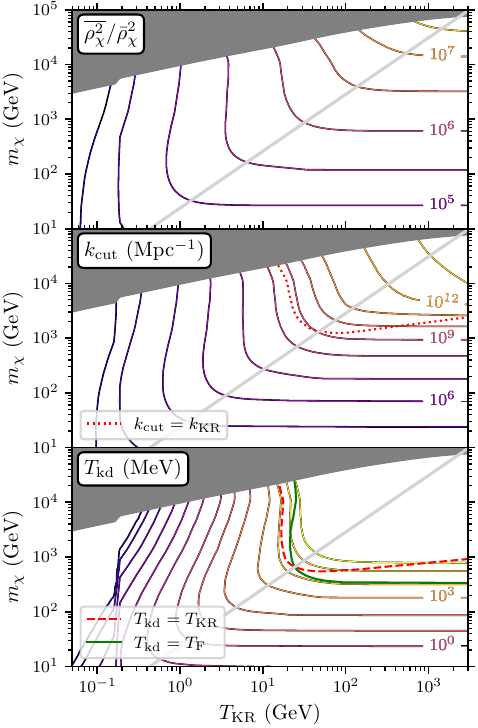}
	\caption{\label{fig:demo} Upper limits on the annihilation boost factor $\boost$ (upper panel), cutoff wavenumber $\kcut$ (middle panel), and kinetic decoupling temperature $\Tkd$ (lower panel) for dark matter that thermally decouples from the SM and annihilates into $b\bar b$, as a function of the dark matter mass $m_\chi$ and the kinaton-radiation equality temperature $\TKR$. These limits are derived from Ref.~\cite{Blanco:2018esa}'s limits on the dark matter lifetime, which employ the Fermi collaboration's measurement of the IGRB \cite{Fermi-LAT:2014ryh}. Contours mark successive powers of $\sqrt{10}$ (upper panel) or 10 (lower panels). The shaded region exceeds the unitarity limit \cite{Griest:1989wd}, while the gray diagonal marks where $\Tf=\TKR$; to its right, dark matter freezes out after kinaton-radiation equality, so limits on dark matter annihilation are no longer sensitive to $\TKR$. In the middle panel, we indicate where the upper limit on $\kcut$ is equal to $\kKR$, the wavenumber entering the horizon at kinaton-radiation equality. Below this dotted curve, kination's boost to the small-scale power spectrum cannot affect dark matter annihilation because it is erased by free streaming (see Fig.~\ref{fig:kinboost}). In the lower panel, we mark where the upper limit on $\Tkd$ is equal to $\TKR$; below this dashed curve, kination has no impact on the minimum halo mass (see Fig.~\ref{Fig:Mcut}) because the dark matter decouples after the kination epoch. Finally, the upper limit on $\Tkd$ exceeds $\Tf$ within the region to the upper right of the thick solid green curve, so all dark matter particles that kinetically decouple from the SM after they freeze out are allowed.}
\end{figure}

For the example case of dark matter annihilating into $b\bar b$, Fig.~\ref{fig:demo} shows the results of steps (3)--(5): the observational upper limits on $\boost$, $\kcut$, and $\Tkd$. These are evaluated under the conservative assumption that the prompt cusps become NFW cusps. We mark several noteworthy regimes here.
\begin{enumerate}[label={(\arabic*)}]
\item Below the gray diagonal line, the dark matter freezes out after kination ends ($\Tf<\TKR$), so the existence of a kination epoch does not affect this regime.
\item Above the gray diagonal line, the dark matter freezes out during kination. The upper limits on $\boost$ and $\kcut$ in this regime depend only weakly on $m_\chi$ because the annihilation cross section required to achieve the observed dark matter abundance is proportional to $(m_\chi/\TKR)/\log(\Tf/\TKR)$ \cite{Redmond:2017}.  It follows from Eq.~(\ref{decay}) that the effective dark matter lifetime is only logarithmically dependent on $\Tf \simeq m_\chi/25$. For $m_\chi\lesssim10^4$ GeV, the slight variation of the upper bounds on $\boost$ and $\kcut$ for different $m_\chi$ results from the $m_\chi$-dependence of the observational limits on the dark matter lifetime. The impact of the logarithmic $m_\chi$-dependence of the effective dark matter lifetime can be seen for larger masses.  
\item In the shaded region, the dark matter annihilation coupling strength exceeds the unitarity limit,
\begin{equation}
    \sigmav = 4\pi/m_\chi^2.
\end{equation}
\item Below the dotted curve in the middle panel, the cutoff wavenumber $\kcut$ is observationally constrained to be smaller than the wavenumber $\kKR$ that enters the horizon at kinaton-radiation equality. This implies that kination's boost to the small-scale power spectrum must be fully erased by free streaming.
\item Below the dashed curve in the lower panel, the kinetic decoupling temperature $\Tkd$ is constrained to be smaller than the temperature $\TKR$ of kination-radiation equality. Since the dark matter then decouples after kination ends, kination has no impact on structure (whether through its boost to the power spectrum or through its reduction of the minimum halo mass).
\item Finally, above the thick solid green curve in the lower panel, the observational upper limit on $\Tkd$ is higher than the freeze-out temperature $\Tf$. We do not expect dark matter to kinetically decouple while it is still in thermal equilibrium, and current observational limits on dark matter annihilation do not further restrict $\Tkd$ in this regime.
\end{enumerate}

\begin{figure*}
	%\centering
	\includegraphics[width=\linewidth]{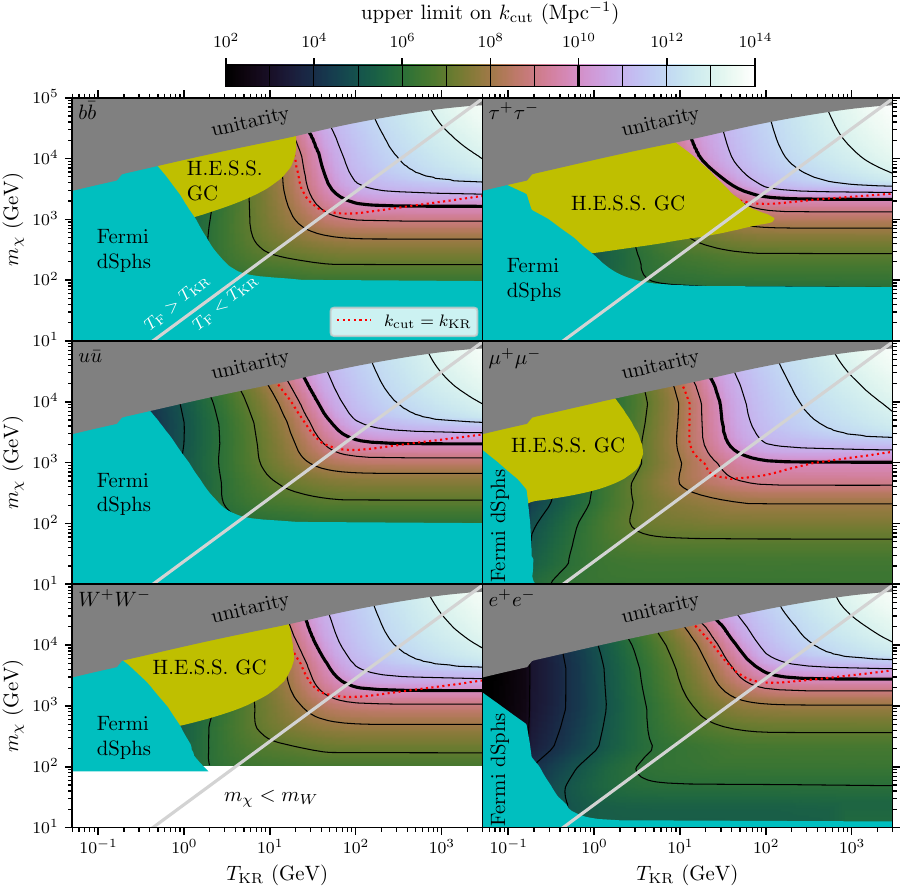}
	\caption{\label{fig:kcutlim} Allowed power spectrum cutoff wavenumber $\kcut$ for thermally produced dark matter in kination scenarios. 
    We plot the dark matter mass $m_\chi$ against the kinaton-radiation equality temperature $\TKR$ for a variety of annihilation channels (different panels). Shaded regions indicate where the unitarity bound is violated and where the Fermi and H.E.S.S. Collaborations' limits on dark matter annihilation in dwarf spheroidal galaxies and in the Galactic Center, respectively, are violated.
    In the space that remains, we plot in color the new upper limit on $\kcut$ based on the Fermi Collaboration's measurement of the IGRB, which we evaluate under the conservative assumption that microhalos develop NFW density profiles.
    The contours mark successive powers of 10 in $\kcut$, with the thick contour marking $\kcut=10^{10}$~Mpc$^{-1}$.
    The area above the red dotted curve is where $\kcut$ is allowed to exceed $\kKR$, i.e. kination could boost the power spectrum of density variations at small scales.
    }
\end{figure*}

\begin{figure*}
	%\centering
	\includegraphics[width=\linewidth]{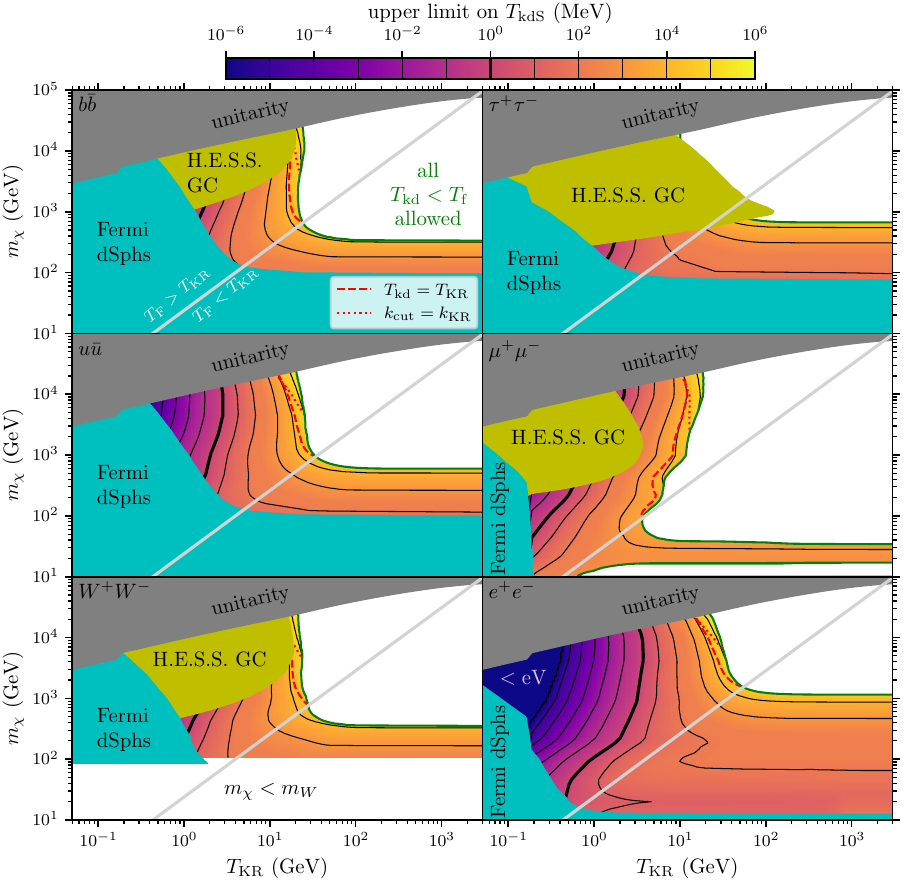}
	\caption{\label{fig:TkdSlim} Allowed dark matter coupling parameter $\TkdS$ for thermal relic dark matter in kination scenarios.
    As in Fig.~\ref{fig:kcutlim}, we shade regions ruled out by unitarity and by the Fermi and H.E.S.S. Collaborations' limits on dark matter annihilation that do not consider substructure.
    In the remaining space, we plot in color the upper limit on $\TkdS$ due to the Fermi Collaboration's IGRB measurement, again assuming conservatively that microhalos develop NFW density profiles.
    Since dark matter models are typically expected to have $\TkdS$ higher than around 1~MeV (thick contour) to 100~MeV \cite{Cornell:2013rza}, this new limit slightly improves upon the other limits.
    In the green-bounded white area, the upper limit on $\TkdS$ is high enough that kinetic decoupling would occur before freeze-out, which does not make physical sense, so this regime is observationally unconstrained.
    The area below the gray diagonal line is where $\Tf<\TKR$, i.e., we are effectively no longer studying a kination scenario.
    The small strip to the right of the red dashed curve is where $\TkdS$ is allowed to exceed $\TKR$, i.e. kination could possibly have an impact on dark matter structure.
    The even smaller strip to the right of the red dotted curve is where $\kcut$ is allowed to exceed $\kKR$, as in Fig.~\ref{fig:kcutlim}. 
    Evidently, the enhanced microhalo population that could result from a period of kination rarely affects limits on dark matter annihilation if dark matter decouples from the SM.
    }
\end{figure*}

\begin{figure}
	\centering
	\includegraphics[width=\columnwidth]{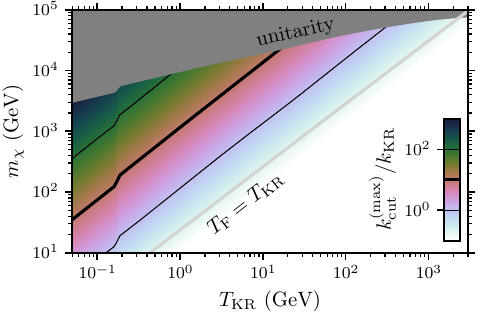}
	\caption{\label{fig:xfs_f} Maximum free-streaming wavenumber $\kcut$, obtained by assuming freeze-out and kinetic decoupling occur simultaneously, i.e. $\Tf=\Tkd$. We express $\kcut$ in units of the wavenumber $\kKR$ at kinaton-radiation equality. As Fig.~\ref{fig:kinboost} shows, $\kcut/\kKR>10$ is needed for kination's boost to the amplitudes of small-scale density variations to result in a nonnegligible increase in the annihilation rate. We mark $\kcut/\kKR=10$ here with a thick contour.}
\end{figure}

\begin{figure*}
	%\centering
	\includegraphics[width=\linewidth]{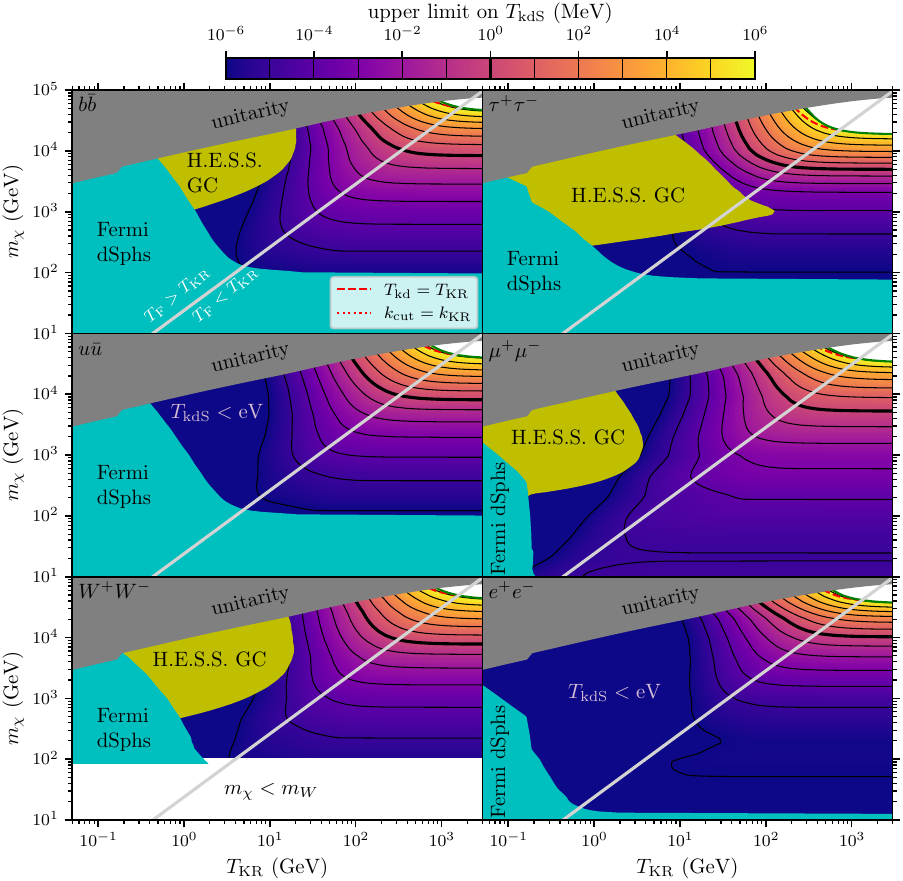}
	\caption{\label{fig:TkdSlimPC} Same as Fig.~\ref{fig:TkdSlim} but assuming that prompt cusps survive. If $\TkdS$ lies again around 1~MeV (thick contour) or higher (see Ref.~\cite{Cornell:2013rza}), most of the parameter space for dark matter that freezes out during a kination epoch is ruled out.
    The region between the red dashed and green curves again marks where kination's effect on small-scale structure is relevant. Its negligible size here justifies our use of Eq.~(\ref{promptcusps}) to evaluate the annihilation rate in prompt cusps, as this fitting form was derived for a standard thermal history.
    }
\end{figure*}

%\FloatBarrier

We now explore the parameter ranges allowed for thermally produced dark matter in kination scenarios. Figures \ref{fig:kcutlim} and~\ref{fig:TkdSlim} plot again the dark matter mass $m_\chi$ against the temperature $\TKR$ of kinaton-radiation equality. Different panels represent different annihilation channels. We mark regions ruled out by several different considerations: the unitarity bound (gray), the Fermi Collaboration's search for dark matter annihilation in dwarf spheroidal galaxies \cite{Fermi:Constraints} (cyan), and the H.E.S.S. Collaboration's search for dark matter annihilation in the Galactic Center \cite{HESS:2016mib} (yellow). The Fermi and H.E.S.S. upper limits on $\sigmav$ do not assume any boost to the annihilation rate due to subhalos. Nevertheless, they still severely restrict dark matter production during kination, %ruling out lower masses
owing to the large annihilation cross section $\sigmav$ required by such scenarios (see Fig.~\ref{fig:sigmav}).
Below the gray diagonal lines in Figs.~\ref{fig:kcutlim}~and~\ref{fig:TkdSlim}, dark matter freezes out after kinaton-radiation equality.
These considerations typically leave a fairly narrow viable region for dark matter that freezes out during a kination epoch, as was shown in Ref.~\cite{Redmond:2017}.

Within the allowed region, Fig.~\ref{fig:kcutlim} shows the upper limit on the power spectrum cutoff scale $\kcut$ that results from enforcing the IGRB bound on the dark matter effective decay rate \cite{Blanco:2018esa}, as described in this section.
We adopt the conservative assumption that the prompt cusps are softened by halo mergers or other interactions.
The region above the red dotted curve is where the upper limit on $\kcut$ is high enough that $\kcut>\kKR$ is possible, i.e., kination could boost the matter power spectrum and lead to early microhalo formation. Below this curve, the annihilation boost from standard substructure rules out the possibility of early microhalo formation due to a period of kination.

Figure~\ref{fig:TkdSlim} shows the upper limit on the dark matter kinetic coupling parameter $\TkdS$, as defined by Eq.~(\ref{eq:DefinitionTkds}), that results from these limits on $\kcut$ if the dark matter kinetically decoupled from the SM.
The absence of dark matter detections in terrestrial experiments typically requires $\TkdS$ to be larger than around 1~MeV (thick contour in Fig.~\ref{fig:TkdSlim}) to 100~MeV  \cite{Cornell:2013rza}, although some dark matter models can evade these limits. If we demand that $\TkdS > 1$ MeV, our new limits significantly reduce the viable parameter space for $\chi\chi\to e^+ e^-$ and $\chi\chi\to u \bar{u}$. 

For all annihilation channels, the upper right corner of the parameter space in Fig.~\ref{fig:TkdSlim} (white with green border; high $m_\chi$ and high $\TKR$) remains unconstrained because the limit that we derive on the kinetic decoupling temperature $\Tkd$ is higher than the freeze-out temperature. Meanwhile, the region to the right of the red dashed curve is where the upper limit on $\TkdS$ is high enough that kinetic decoupling could occur during the kination epoch. Everywhere else, kinetic decoupling during the kination epoch is forbidden.
Evidently, there is only a very narrow strip of the parameter space in which kination's effect on dark matter substructure affects our limits on dark matter annihilation.
The region to the right of the red dotted curve is where the upper limit on $\kcut$ is high enough that $\kcut>\kKR$ is possible, i.e., kination could boost the matter power spectrum and lead to early microhalo formation.
The regime where this outcome could affect our limits is even narrower. These considerations imply that our new limits on thermal relic dark matter in kination scenarios improve on previous limits almost entirely due to the same boost to dark matter annihilation as is present in standard (kination-free) thermal histories (e.g. that considered in Ref.~\cite{Fermi-LAT:2015qzw}).

Furthermore, Fig.~\ref{fig:kinboost} shows that $\kcut>10\kKR$ is needed for kination's boost to small-scale density variations to significantly increase the annihilation rate. 
If we set $\Tkd=\Tf$ so that the dark matter decouples as early as possible, the resulting maximum cutoff wavenumber $\kcut^\mathrm{(max)}$ only exceeds $10\kKR$ when $\TKR \lesssim 20$ GeV, as seen in Figure~\ref{fig:xfs_f}.  Figure~\ref{fig:TkdSlim} shows that gamma-ray observations demand that $\Tkd<\Tf$ when $\TKR \lesssim 20$, and in all such scenarios, $\kcut$ is restricted to values less than $10\kKR$.
Therefore, all scenarios in which a period of kination boosts the amplitudes of small-scale density variations are ruled out if dark matter is a thermal relic that respects the unitarity bound and kinetically decouples from the SM after it freezes out.

If dark matter is part of a colder hidden sector, then it is still possible for kination to enhance the microhalo population even if the particles that dark matter annihilates into promptly decay into SM particles. If the hidden-sector temperature is less than the SM temperature when dark matter freezes out from the hidden radiation bath, the SM temperature at freeze-out can exceed $m_\chi$.  Furthermore, the lower temperature of the dark matter particles increases $\kcut$ for a given value of the SM temperature when DM begins to free stream, as discussed in Sec.~\ref{sec:KDandFS}. Consequently, $\kcut/\kKR$ for dark matter in a hidden sector can greatly exceed the values shown in Fig.~\ref{fig:xfs_f} while still requiring that dark matter kinetically decouples after it thermally decouples.

The constraints on $\kcut$ shown in Fig.~\ref{fig:kcutlim} cannot be precisely applied to hidden-sector dark matter because the underlying calculation of $\sigmav$ assumes that dark matter decouples from the SM. If dark matter decouples within a colder hidden sector, a smaller value of $\sigmav$ is required to compensate for the fact that more annihilation occurs after freeze-out if freeze-out occurs at a higher value of $T/\TKR$.  Moreover, the dark matter abundance will be diluted when the particles within the hidden sector decay into SM particles.  However, $\rhobar$ is only logarithmically dependent on $\Tf/\TKR$, and subdominant hidden sector particles do not significantly increase the entropy of the visible sector when they decay.  Therefore, Fig.~\ref{fig:kcutlim} provides an accurate estimate of the maximum allowed value of $\kcut$ if dark matter freezes out within a forever-subdominant hidden sector whose lightest particles promptly decay into the relevant SM particles. 

The results so far have been derived under the assumption that prompt $\rho\propto r^{-3/2}$ cusps are softened during halo evolution. Constraints on dark matter freeze-out during kination are stronger if we assume that the prompt cusps persist, as suggested by recent simulations \cite{Delos:2022yhn}.  The resulting limits on $\TkdS$ are shown in Fig.~\ref{fig:TkdSlimPC}. 
For dark matter models with $\TkdS$ larger than around 1~MeV (thick contour), most of the parameter space for dark matter freeze-out during kination is ruled out. Also, the position of the dashed $\Tkd=\TKR$ curve here implies that there are essentially no allowed scenarios where
kination's effect on small-scale structure would impact our limits, which justifies our use of Eq.~(\ref{promptcusps}) to evaluate the annihilation rate in prompt cusps.

Finally, we remark on the importance of the foreground models of Ref.~\cite{Blanco:2018esa} to our conclusions. As we noted above, we derive limits on dark matter annihilation from Ref.~\cite{Blanco:2018esa}'s limits on dark matter decay via Eq.~(\ref{decay}). These decay limits were obtained from the Fermi collaboration's measurement of the IGRB \cite{Fermi-LAT:2014ryh} after subtraction of a model for unresolved astrophysical foregrounds, such as gamma rays from star-forming galaxies and active galactic nuclei. The model already accounts for most of the gamma-ray signal, leaving little room for a dark matter contribution. More conservative modeling choices can change our results significantly. For example, in Ref.~\cite{Fermi-LAT:2015qzw}, the Fermi collaboration published limits on dark matter annihilation based on their IGRB measurement but adopted much simpler and more conservative modeling assumptions. By scaling Ref.~\cite{Fermi-LAT:2015qzw}'s limit on $\sigmav$ by the ratio between the annihilation boost factors $\boost$ employed by Ref.~\cite{Fermi-LAT:2015qzw} and those that we derived above,\footnote{For the boost factor in Ref.~\cite{Fermi-LAT:2015qzw}, $\boostrho$ has a modest dependence on redshift, unlike our result in Fig.~\ref{fig:boostevo}. For simplicity, we use its value at $z=0$. Also, the Galactic contributions in our calculations and those of Ref.~\cite{Fermi-LAT:2015qzw} differ by a factor slightly different from the ratio of the $\boost$, due to different models of the Milky Way halo and different treatments of subhalos. We neglect this discrepancy in our estimate.} we computed alternative limits on $\TkdS$ as in Figs. \ref{fig:TkdSlim} and~\ref{fig:TkdSlimPC} (but only for the $b\bar b$, $\tau^+\tau^-$, $\mu^+\mu^-$, and $W^+W^-$ channels considered by Ref.~\cite{Fermi-LAT:2015qzw}). Under the conservative assumption that prompt cusps are softened, all scenarios with $\Tkd < \Tf$ that are not ruled out by Fermi observations of dwarf spheroidals and H.E.S.S. observations of the Galactic Center remain viable.  Even if all prompt cusps are assumed to survive, only a marginal improvement is achieved over the H.E.S.S. and Fermi limits that do not consider halo substructure.

%%%%%%%%%%%%%%%
\section{Conclusion}
\label{sec:Conclusion}
%%%%%%%%%%%%%%%

It is possible that the Universe went through a period of kination after inflation, during which a fast-rolling scalar field with $\rho \propto a^{-6}$ (the kinaton) dominated the energy density \cite{Spokoiny:1993, Joyce:1996, Ferreira:1997}.  If dark matter reaches thermal equilibrium and then freezes out during kination, its relic abundance is larger than it would be if freeze-out occurred during radiation domination.  Consequently, the dark matter particle's velocity-averaged annihilation cross section must exceed $2\times10^{-26} \, \mathrm{cm^3 s^{-1}}$ to avoid generating too much dark matter.  Such scenarios are tightly constrained by gamma-ray observations of dwarf spheroidal galaxies and the Galactic Center \cite{Redmond:2017}.  Kination also leaves an imprint on the small-scale matter power spectrum because subhorizon dark matter density perturbations grow linearly with the scale factor during kination \cite{Redmond:2018}.  We have determined how enhanced perturbation growth during kination affects the microhalo population and the dark matter annihilation rate.  

Although dark matter perturbations grow linearly with the scale factor during both kination and matter domination, the resulting rise of the dimensionless matter power spectrum ${\cal P}(k)$ due to kination is much shallower: ${\cal P}(k) \propto k^{n_s}$ for modes that enter the horizon during kination as opposed to ${\cal P}(k) \propto k^{n_s+3}$ for modes that enter the horizon during matter domination.  This slower increase in ${\cal P}(k)$ implies that kination only significantly increases density fluctuations on scales that enter the horizon well before the end of kination: if $M_\mathrm{KR}$ is the mass within the horizon at the time of kinaton-radiation equality, kination only increases the rms density contrast $\sigma(M)$ by more than 20\% on mass scales $M \lesssim 10^{-4}M_\mathrm{KR}$.  

If ${\cal P}(k) \propto \exp[-k^2/\kcut^2]$ due to dark matter free streaming, kination's imprint on the matter power spectrum is completely erased if $\kcut \lesssim 10 \kKR$, where $\kKR$ is the horizon wavenumber $aH$ evaluated at kinaton-radiation equality.  If dark matter is initially in equilibrium with the SM, obtaining \mbox{$\kcut \gtrsim 10 \kKR$} is difficult because the same increase in comoving drift velocity that is responsible for the rapid growth of dark matter perturbations during kination also increases the comoving dark matter free-streaming horizon for a given temperature at decoupling.  As a result, \mbox{$\kcut \gtrsim 10 \kKR$} is only possible for dark matter particles with $m_\chi/\TKR > 100$.  Nevertheless, decoupling during kination decreases $\kcut$ for a given dark matter particle because dark matter kinetically decouples earlier during kination than it does during radiation domination.  It follows that kination's impact on the microhalo population is two-fold. 
First, if dark matter kinetically decouples from the SM during kination, the minimum halo mass is reduced. Second, the boost to small-scale power causes halos with $M \lesssim 10^{-4}M_\mathrm{KR}$ to form earlier and hence be more internally dense, if the dark matter is cold enough to preserve such structures.
Both effects boost the dark matter annihilation rate, strengthening the power of gamma-ray observations to constrain dark matter freeze-out during kination.

If dark matter annihilation within unresolved microhalos dominates the total annihilation rate, the impact on the IGRB mimics the signal from decaying dark matter because the emission in both cases tracks the dark matter density (averaged over scales larger than the microhalos).  Therefore, microhalo-dominated annihilation can be characterized by an effective dark matter lifetime that is inversely proportional to the $\overline{\rhoc^2}/\rhobar^2$ annihilation boost factor.  We use lower limits on the dark matter lifetime \cite{Blanco:2018esa} derived from Fermi-LAT observations of the isotropic gamma-ray background (IGRB) \cite{Fermi-LAT:2014ryh} to constrain dark matter freeze-out during kination.  To compute the dark matter annihilation rate after a period of kination, we use the P2H method \cite{Delos:2019mxl,Delos:2022yhn} to characterize the microhalo population that is generated from a given matter power spectrum.  If the microhalos' initial $\rho \propto r^{-3/2}$ density cusps are subsequently softened to NFW profiles, the P2H method predicts a boost factor between $10^5$ and $10^6$ for a standard matter power spectrum with a wide range of $\kcut$ values, which is in agreement with simulation-calibrated halo-based computations.  If $\kcut/\kKR \gtrsim 10$, the growth of density fluctuations during kination enhances the boost factor, with $\rhoc^2/\rhobar^2 \gtrsim 10^7$ for $\kcut/\kKR \gtrsim 500$.  

Since the annihilation boost factor increases as $\kcut$ increases, we use IGRB observations to establish an upper bound on $\kcut$ for a given dark matter particle mass ($m_\chi$) and temperature at kinaton-radiation equality ($\TKR$).  For $\TKR \lesssim 10$ GeV, the maximum allowed value of $\kcut$ is less than $\kKR$, implying that all perturbation modes that enter the horizon during kination must be erased by dark matter free-streaming to avoid exceeding the allowed dark matter contribution to the IGRB.  If dark matter kinetically decouples from Standard Model particles, then the upper bound on $\kcut$ can be translated to an upper bound on the temperature when dark matter decoupled and began to free stream.  The temperature at dark matter freeze-out establishes an additional upper bound on $\kcut$ because dark matter does not decouple while it is still in thermal equilibrium.  We find that these two constraints imply that kination cannot enhance small-scale perturbations if dark matter is a thermal relic.  All scenarios in which dark matter freezes out early enough to allow $\kcut/\kKR \gtrsim 10$ require $\TKR \lesssim 20$ GeV, in which case $\sigmav$ is large enough that the IGRB rules out any enhancement to the small-scale power spectrum.  The constraints are even stronger if the initial $\rho \propto r^{-3/2}$ persists, as is indicated by recent high-resolution simulations \cite{Delos:2022yhn}.  In that case, nearly all scenarios in which dark matter kinetically decouples during kination are ruled out, and the kinetic decoupling temperature must be less than 1 MeV if $\TKR \lesssim 40$ GeV.

The growth of perturbations during kination can significantly enhance the dark matter annihilation rate if dark matter is part of a subdominant hidden sector that is colder than the Standard Model.  In this case, dark matter freezes out at a higher SM temperature and has a smaller free-streaming horizon for a given SM temperature at decoupling.  Both of these effects increase the maximum possible value of $\kcut/\kKR$ that is consistent with dark matter freeze-out preceding dark matter kinetic decoupling.  Meanwhile, the annihilation cross section required to generate the observed dark matter abundance is nearly the same because the relic density is only logarithmically dependent on the SM temperature at freeze-out when dark matter freezes out during kination.  If the dark matter annihilates to hidden-sector particles that promptly decay to Standard Model particles, then our upper bounds on $\kcut$ can be used to constrain hidden-sector dark matter that freezes out during kination.  

Finally, we note that the constraints presented here only apply to dark matter that freezes out from thermal equilibrium.  If dark matter never reaches thermal equilibrium (i.e. it freezes in \cite{Hall:2009bx, Redmond:2017}), or if it is produced gravitationally \cite{Chung:2001cb}, then its annihilation cross section can be much smaller than the values assumed in our analysis, and it can be cold enough for $\kcut/\kKR$ to significantly exceed 20.  In that case, the growth of dark matter perturbations during kination enhances the microhalo abundance for $M \lesssim 10^{-4}M_\mathrm{KR}$, which corresponds to $M \lesssim 10^{-3} M_\oplus$ for $\TKR \simeq 10$ MeV.  Microhalos this small are difficult to detect gravitationally, but the microhalos generated from enhanced small-scale fluctuations form earlier and are therefore denser than standard microhalos.  As a result, there are a few promising detection prospects.
Dense sub-earth-mass microhalos leave a potentially detectable imprint on the light curves of high-redshift stars that are microlensed by intervening stars while near a galaxy cluster lens caustic \cite{Dai:2019lud, Blinov:2021axd}.
Pulsar timing observations can detect the motion of pulsars due to passing sub-earth-mass halos \cite{Dror:2019twh, Ramani:2020hdo, Lee:2020wfn}: a pulsar timing array array consisting of 100 pulsars with 10-ns residuals are capable of detecting dense microhalos with masses as small as $10^{-6} M_\odot$ after 40 years of observations \cite{Delos:2021rqs}.  
If fluctuations are enhanced to a sufficient degree to form microhalos before the matter epoch (e.g. \cite{2010PhRvD..81j3529B,Berezinsky:2013fxa,Blanco:2019eij,StenDelos:2022jld}), they could even be compact enough to microlens stars of our Local Group at detectable levels \cite{Delos:2023fpm}.
These probes are capable of detecting the microhalos that form after an early matter-dominated era \cite{Blinov:2021axd, Lee:2020wfn, Delos:2021rqs}, but it remains to be seen if the shallower rise of ${\cal P}(k)$ generated by kination can generate halos that are large and dense enough to be detected gravitationally.  

\section*{Acknowledgements}
All authors received support from NSF CAREER grant PHY-1752752 (P.I. Erickcek) while working on this project.

\appendix

%%%%%%%%%%%%%%%
\section{Tidal suppression of the annihilation rate}
\label{sec:tidal}
%%%%%%%%%%%%%%%

In this appendix, we estimate how the tidal influence of larger structures suppresses the annihilation rate within their subhalos. We employ the fitting function presented by Ref.~\cite{StenDelos:2019xdk}, which is based on the simulation-tuned tidal evolution model developed in Ref.~\cite{Delos:2019lik} and describes the orbit-averaged scaling factor $S(\rho_s/P_s,t\sqrt{GP_s},c)$ for all subhalos of scale density $\rho_s$ orbiting a host with scale density $P_s$ and concentration $c=R_s/R_\mathrm{vir}$ for the duration $t$.

Our strategy will be to use a conventional halo model (applicable to a scenario without kination) to quantify the population of potential host halos for our kination-boosted microhalos. Denoting by $\bar S(\rho_s,t)$ the global factor by which the annihilation rate within microhalos of scale density $\rho_s$ is scaled due to tidal evolution for the duration $t$, we may write
\begin{equation}\label{Sglob}
\bar S(\rho_s,t)=1-\frac{1}{\rhobar}\int_{\Mmin}^\infty\!\!\!\!\!\!\!\!\!\!\diff M \!\left[1\!-\!S\!\left(\frac{\rho_s}{P_s},t\sqrt{GP_s},c\right)\right]\!M f_s\frac{\diff n}{\diff M},
\end{equation}
where $P_s$ and $c$ are assumed to be functions of $M$. Note that $\diff n/\diff M$ here includes subhalos as well as host halos, and we include the factor $f_s$ to denote the fraction of material within a halo of mass $M$ that is not in subhalos. There are now three components to completing this calculation---the microhalo scale density $\rho_s$, the halo mass function $\diff n/\diff M$ (including subhalos), and the concentration-mass relation $c(M)$ (which also sets $P_s$)---and we describe how we handle each in turn.

\begin{figure}
	\centering
	\includegraphics[width=\columnwidth]{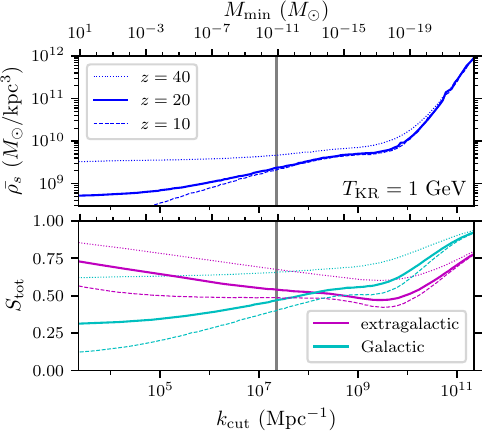}
	\caption{\label{fig:supp} Top: The $J$-weighted average of microhalos' scale density $\rho_s$, as a function of the cutoff scale $\kcut$, for a kination scenario with $\TKR=1$ GeV. We evaluate $\bar{\rho}_s$ at three different redshifts. Kination's boost to the small-scale power spectrum significantly raises $\bar\rho_s$ as long as $\kcut\gtrsim 10^2\kKR$; $\kKR$ is marked by the vertical line. Bottom: The tidal suppression factor $S_\mathrm{tot}$ computed for both Galactic and extragalactic microhalos in the same scenarios, assuming that their initial density profiles for tidal evolution are fixed at redshift $z$.}
\end{figure}

To estimate the scale density $\rho_s$ of microhalos, we use Ref.~\cite{Delos:2019mxl}'s prescription for quantifying the broader halo that forms around a density peak. Specifically, we predict the radius $\rmax$ of maximum circular velocity and its associated enclosed mass $\mmax$ using the ``$s=0$ adiabatic contraction'' model in Ref.~\cite{Delos:2019mxl}, and we assume an NFW density profile to then obtain $\rho_s$. Reference~\cite{Delos:2019mxl} found that the accuracy of these predictions is only mildly sensitive to frequency of halo mergers.
We pick the redshift $z=20$ at which to compute $\rho_s$, guessing at a typical redshift at which a microhalo might be expected to accrete onto a larger halo, but we also show the impact on $\bar S$ if we instead pick $z=10$ or $z=40$. The $J$-weighted average $\rho_s$ for these three redshift choices is shown in Fig.~\ref{fig:supp} for an example kination scenario.

We use the spherical-overdensity mass function of Ref.~\cite{Watson:2012mt} to describe the mass function of field (not sub-) halos, using the matter power spectrum (with no kination and no cutoff) of Ref.~\cite{Eisenstein:1997}. We normalize this power spectrum so that the rms fractional variance in the linear density field, smoothed on the scale $8h^{-1}$ Mpc (where $h$ is the Hubble parameter), is $\sigma_8=0.81$ \cite{Planck:2018vyg}. For the subhalo mass function, we assume $\diff N/\diff \ln M=A (M/M_\mathrm{host})^{1-\alpha}$ with $A=0.012$ and $\alpha=2$ as in Ref.~\cite{Sanchez-Conde:2013yxa}. $\alpha$ is typically measured to lie between $1.9$ and $2$ in simulations (e.g., Refs.~\cite{Madau:2008fr,Springel:2008cc}); $\alpha=2$ produces more subhalos, so it is the conservative choice in that it yields more tidal suppression of the smallest halos. The mass function including both field halos and subhalos is then
\begin{equation}\label{dndlnM_tot}
\left(\frac{\diff n}{\diff\ln M}\right)_\mathrm{tot} = \sum_{i=0}^\infty \left(\frac{\diff n}{\diff\ln M}\right)_i,
\end{equation}
where $(\diff n/\diff\ln M)_0$ is the field halo mass function and
\begin{equation}\label{dndlnM_sub}
\left(\frac{\diff n}{\diff\ln M}\right)_i = \int_M^\infty\!\!\!\!\! \diff\ln M^\prime \left(\frac{\diff n}{\diff\ln M^\prime}\right)_{i-1}\! \left.\frac{\diff N}{\diff\ln M}\right|_{M_\mathrm{host}=M^\prime}
\end{equation}
for $i>0$. We evaluate Eq.~(\ref{dndlnM_tot}) up to $i=8$, although convergence at the 1\% level is achieved at $i=5$. The fraction $f_s$ of halo mass not in subhalos is
\begin{equation}\label{subfrac}
f_s(M) = 1-\frac{1}{M}\int_{\Mmin}^M\diff \ln M^\prime \left.\frac{\diff N}{\diff\ln M^\prime}\right|_{M_\mathrm{host}=M}M^\prime
\end{equation}
for a halo of mass $M$.

Finally, we use the concentration-mass relation presented in Ref.~\cite{Diemer:2018vmz} to evaluate the concentration $c$ and scale density $P_s$ for each halo mass $M$, again using the matter power spectrum of Ref.~\cite{Eisenstein:1997}. With all of these ingredients, we are prepared to evaluate $\bar S(\rho_s,t)$. We conservatively set $t=t_\mathrm{age}$, the age of the Universe, and we evaluate $\bar S$ using Eq.~(\ref{Sglob}) for the scale density $\rho_s$ of each sampled microhalo. We set $\Mmin=(4/3)\pi \bar{\rho}_m \kcut^{-3}$. The overall tidal scaling factor for a given kination scenario is then the $J$-weighted average of the $\bar S$ for each halo, or
\begin{equation}\label{Stot}
S_\mathrm{tot} = \left.\left[\sum_{i=1}^N J_i \bar S(\rho_{s,i},t_\mathrm{age})\right]\right/\left(\sum_{i=1}^N J_i \right).
\end{equation}
We plot $S_\mathrm{tot}$ for an example kination scenario in the lower panel of Fig.~\ref{fig:supp} (magenta curves). The general behavior is that increasing $\kcut$ raises the scale density of the microhalos, making them less susceptible to tidal suppression, but also boosts the amount of host structure causing this suppression. If $\kcut \lesssim 10^2 \kKR$, the latter effect dominates and increasing $\kcut$ causes more suppression of the annihilation rate ($S_\mathrm{tot}$ decreases). If $\kcut \gtrsim 10^2 \kKR$, the $\rho_s$ are sufficiently sensitive to $\kcut$ that increasing $\kcut$ reduces the level of tidal suppression.

The above procedure accounts for the tidal suppression of the extragalactic annihilation signal. However, the majority of dark matter's contribution to the IGRB comes from the Galactic halo \cite{Blanco:2018esa}. To estimate the tidal suppression factor for this contribution, we apply the tidal evolution model of Ref.~\cite{Delos:2019lik} using the Galactic halo as the host. We assume the Galactic halo has an NFW density profile with scale radius 20 kpc and scale density set so that the local dark matter density at radius 8.25 kpc is 0.4 GeV/cm$^3$. At each radius $R$, we average the model prediction $J/J_\mathrm{init}$ over subhalo orbits, assuming the isotropic distribution function of Ref.~\cite{2000ApJS..131...39W} (see Ref.~\cite{StenDelos:2019xdk} for detail). We thereby obtain the tidal scaling factor $s(R,\rho_s,t)$ as a function of Galactocentric radius $R$, which we in turn average over the Galactic halo's density profile along the line of sight perpendicular to the Galactic plane. By averaging this factor over the microhalo population as before and setting $t=t_\mathrm{age}$, we obtain the Galactic tidal scaling factor, an example of which is shown in the lower panel of Fig.~\ref{fig:supp} (cyan curves). In this calculation, $\kcut$'s only relevant influence is on the scale density $\rho_s$ of the microhalos, so increasing $\kcut$ reduces the amount of tidal suppression.

Galactic and extragalactic dark matter contribute different gamma-ray spectra to the IGRB (see Ref.~\cite{Blanco:2018esa}). Consequently, if we change their relative contributions by applying different tidal scaling factors, then direct conversion from decay to annihilation bounds (Eq.~\ref{decay}) is no longer possible. To maintain this conversion, we instead make the conservative choice to apply a universal tidal scaling factor equal to the smaller of the Galactic and extragalactic factors. This choice is also motivated by the likelihood that an unknown fraction of Galactic microhalos should experience suppression closer to the extragalactic factor due to the presence of larger Galactic substructure. In Fig.~\ref{fig:kinboost}, we plot the global boost factor $\boost$ tidally suppressed in this way. For scenarios without a significant kination boost (black curves), tidal effects suppress the annihilation rate by a factor of about 3.

%%%%%%%%%%%%%%%
\section{Comparing peak and halo model predictions for the annihilation boost}
\label{sec:halomodel}
%%%%%%%%%%%%%%%

\begin{figure}
	\centering
	\includegraphics[width=\columnwidth]{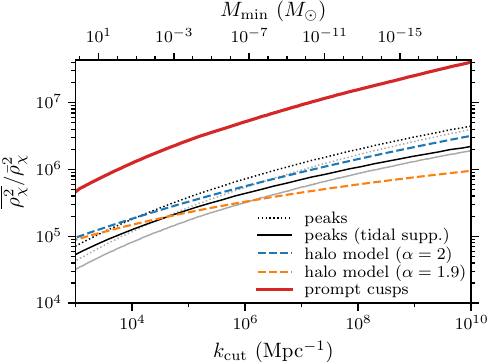}
	\caption{\label{fig:stdboost} The global boost factor $\boost$ for a conventional (kination-free) cosmological scenario, as a function of cutoff scale $\kcut$ (or equivalently minimum halo mass $\Mmin$). Halo model predictions with the $\alpha=2$ and $\alpha=1.9$ subhalo mass functions (dashed curves) neatly bracket the prediction of our peak model in the same scenario (solid black curve) for relevant values of $\kcut$. The peak model prediction includes the extragalactic tidal scaling factor; we also show as the black dotted curve the peak model prediction without this scaling. Peak and halo model predictions here both neglect small-scale baryonic suppression (see the text).
    The faint gray curves show the respective peak model predictions when baryonic suppression is instead taken into account.
    Finally, we also show the boost factor if prompt cusps are assumed to survive \cite{Delos:2022bhp} (thick red curve), also with baryonic suppression accounted for.}
\end{figure}

In this appendix, we compare the results of our peak-based calculation of the dark matter annihilation boost to that of halo models. To make this comparison, we again use the spherical overdensity mass function (at $z=0$) of Ref.~\cite{Watson:2012mt} and the concentration-mass relation of Ref.~\cite{Diemer:2018vmz}, both evaluated using the power spectrum of Ref.~\cite{Eisenstein:1997} normalized to $\sigma_8=0.8102$. For the subhalo population, we assume $\diff N/\diff \ln M=A (M/M_\mathrm{host})^{1-\alpha}$ as before but consider both $(A,\alpha)=(0.012,2)$ and $(A,\alpha)=(0.030,1.9)$ \cite{Sanchez-Conde:2013yxa}. For a given concentration $c$ and mass $M$, a halo's volume-integrated squared density $J=\int\rho^2 \diff V$, which is proportional to the annihilation rate, is
\begin{equation}\label{Jhalo}
J=\frac{\Delta}{9} \frac{c^3}{[g(c)]^2}\left[1-(1+c)^3\right] M \rho_\mathrm{crit},
\end{equation}
where $g(c)=\ln(1+c)-c/(1+c)$, $\rho_\mathrm{crit}$ is the critical density, and we take $\Delta=200$ as the virial overdensity. Each field halo's annihilation rate is scaled by the factor $1+B$ due to the presence of substructure, where
\begin{equation}\label{subhaloboost}
B(M) = \frac{1}{J(M)}\int_{\Mmin}^M
\!\!\!\!\!\!\!\!\! \diff \ln m \left.\frac{\diff N}{\diff\ln m}\right|_{M_\mathrm{host}=M}
\!\!\!\!\!\!\!\!\!\!\!\!\!\!\!\!\!\! [1+B(m)] J(m),
\end{equation}
an equation that we evaluate iteratively (beginning with $B=0$) up to eight iterations. The mean squared density (from which $\boost$ follows) is then
\begin{equation}\label{haloboost}
\overline{\rhoc^2} = \int_{\Mmin}^\infty \diff\ln M \frac{\diff n}{\diff\ln M} [1+B(M)] J(M),
\end{equation}
where $\diff n/\diff\ln M$ is the mass function of field halos. Figure~\ref{fig:stdboost} shows how $\boost$ depends on $\Mmin$ for both the $\alpha=2$ and $\alpha=1.9$ subhalo mass functions.

We first seek to compare the halo model prediction in Fig.~\ref{fig:stdboost} with our conservative peak model predictions as in Fig.~\ref{fig:kinboost}. The latter are plotted in Fig.~\ref{fig:stdboost} as the faint gray solid and dotted lines (with and without extragalactic tidal suppression, respectively). However, there is one further consideration. As we noted in Sec.~\ref{sec:boost}, baryonic matter does not cluster at mass scales below about $10^5 M_\odot$ \cite{Bertschinger:2006}, so the growth rate of dark matter structures below this scale is suppressed. Our peak model accounts for this effect, but the field halo mass function, subhalo mass functions, and concentration-mass relation that the halo model prediction employed do not. Consequently, to make a fair comparison, we repeat our boost computation using the peak model, but we leave out the baryonic correction and employ the same power spectrum that we used for the halo model computations, which also does not account for baryons' nonclustering at small scales. We scale this power spectrum by the exponential cutoff $\exp[-(k/\kcut)^2)]$ with $\Mmin=(4/3)\pi \rhom \kcut^{-3}$, where $\rhom$ is the mean matter density. The boost predicted by the peak model with small-scale baryonic suppression neglected are shown in Fig.~\ref{fig:stdboost} as black curves. With extragalactic tidal suppression (black solid curve), the peak model predictions are bracketed by the halo model predictions with $\alpha=2$ and $\alpha=1.9$ for most relevant values of $\kcut$ (or $\Mmin$). That is, the peak model yields predictions comparable to those of the established halo model.
Note that since the halo model predictions represent the extragalactic case, comparison to the peak model with extragalactic suppression only is appropriate.

We also show in Fig.~\ref{fig:stdboost} the boost factor if prompt $\rho\propto r^{-3/2}$ cusps are assumed to survive \cite{Delos:2022bhp} (thick red curve). Nonclustering of the baryons is accounted for in this case. This assumption raises the annihilation rate by a factor of about 10.

%\clearpage
\bibliography{kination}

\begin{thebibliography}{109}%
\makeatletter
\providecommand \@ifxundefined [1]{%
 \@ifx{#1\undefined}
}%
\providecommand \@ifnum [1]{%
 \ifnum #1\expandafter \@firstoftwo
 \else \expandafter \@secondoftwo
 \fi
}%
\providecommand \@ifx [1]{%
 \ifx #1\expandafter \@firstoftwo
 \else \expandafter \@secondoftwo
 \fi
}%
\providecommand \natexlab [1]{#1}%
\providecommand \enquote  [1]{``#1''}%
\providecommand \bibnamefont  [1]{#1}%
\providecommand \bibfnamefont [1]{#1}%
\providecommand \citenamefont [1]{#1}%
\providecommand \href@noop [0]{\@secondoftwo}%
\providecommand \href [0]{\begingroup \@sanitize@url \@href}%
\providecommand \@href[1]{\@@startlink{#1}\@@href}%
\providecommand \@@href[1]{\endgroup#1\@@endlink}%
\providecommand \@sanitize@url [0]{\catcode `\\12\catcode `\$12\catcode
  `\&12\catcode `\#12\catcode `\^12\catcode `\_12\catcode `\%12\relax}%
\providecommand \@@startlink[1]{}%
\providecommand \@@endlink[0]{}%
\providecommand \url  [0]{\begingroup\@sanitize@url \@url }%
\providecommand \@url [1]{\endgroup\@href {#1}{\urlprefix }}%
\providecommand \urlprefix  [0]{URL }%
\providecommand \Eprint [0]{\href }%
\providecommand \doibase [0]{http://dx.doi.org/}%
\providecommand \selectlanguage [0]{\@gobble}%
\providecommand \bibinfo  [0]{\@secondoftwo}%
\providecommand \bibfield  [0]{\@secondoftwo}%
\providecommand \translation [1]{[#1]}%
\providecommand \BibitemOpen [0]{}%
\providecommand \bibitemStop [0]{}%
\providecommand \bibitemNoStop [0]{.\EOS\space}%
\providecommand \EOS [0]{\spacefactor3000\relax}%
\providecommand \BibitemShut  [1]{\csname bibitem#1\endcsname}%
\let\auto@bib@innerbib\@empty
%</preamble>
\bibitem [{\citenamefont {Allahverdi}\ \emph {et~al.}(2021)\citenamefont
  {Allahverdi} \emph {et~al.}}]{Allahverdi:2020bys}%
  \BibitemOpen
  \bibfield  {author} {\bibinfo {author} {\bibfnamefont {R.}~\bibnamefont
  {Allahverdi}} \emph {et~al.},\ }\href {\doibase 10.21105/astro.2006.16182}
  {\bibfield  {journal} {\bibinfo  {journal} {Open J. Astrophys.}\ }\textbf
  {\bibinfo {volume} {4}},\ \bibinfo {eid} {1} (\bibinfo {year} {2021})},\
  \Eprint {http://arxiv.org/abs/2006.16182} {arXiv:2006.16182 [astro-ph.CO]}
  \BibitemShut {NoStop}%
\bibitem [{\citenamefont {Kawasaki}\ \emph {et~al.}(1999)\citenamefont
  {Kawasaki}, \citenamefont {Kohri},\ and\ \citenamefont
  {Sugiyama}}]{Kawasaki:1999}%
  \BibitemOpen
  \bibfield  {author} {\bibinfo {author} {\bibfnamefont {M.}~\bibnamefont
  {Kawasaki}}, \bibinfo {author} {\bibfnamefont {K.}~\bibnamefont {Kohri}}, \
  and\ \bibinfo {author} {\bibfnamefont {N.}~\bibnamefont {Sugiyama}},\ }\href
  {\doibase 10.1103/PhysRevLett.82.4168} {\bibfield  {journal} {\bibinfo
  {journal} {Phys. Rev. Lett.}\ }\textbf {\bibinfo {volume} {82}},\ \bibinfo
  {pages} {4168} (\bibinfo {year} {1999})},\ \Eprint
  {http://arxiv.org/abs/astro-ph/9811437} {arXiv:astro-ph/9811437 [astro-ph]}
  \BibitemShut {NoStop}%
%%CITATION = ASTRO-PH/9811437;%%
\bibitem [{\citenamefont {Kawasaki}\ \emph {et~al.}(2000)\citenamefont
  {Kawasaki}, \citenamefont {Kohri},\ and\ \citenamefont
  {Sugiyama}}]{Kawasaki:2000}%
  \BibitemOpen
  \bibfield  {author} {\bibinfo {author} {\bibfnamefont {M.}~\bibnamefont
  {Kawasaki}}, \bibinfo {author} {\bibfnamefont {K.}~\bibnamefont {Kohri}}, \
  and\ \bibinfo {author} {\bibfnamefont {N.}~\bibnamefont {Sugiyama}},\ }\href
  {\doibase 10.1103/PhysRevD.62.023506} {\bibfield  {journal} {\bibinfo
  {journal} {Phys. Rev.}\ }\textbf {\bibinfo {volume} {D62}},\ \bibinfo {pages}
  {023506} (\bibinfo {year} {2000})},\ \Eprint
  {http://arxiv.org/abs/astro-ph/0002127} {arXiv:astro-ph/0002127 [astro-ph]}
  \BibitemShut {NoStop}%
%%CITATION = ASTRO-PH/0002127;%%
\bibitem [{\citenamefont {Hannestad}(2004)}]{Hannestad:2004}%
  \BibitemOpen
  \bibfield  {author} {\bibinfo {author} {\bibfnamefont {S.}~\bibnamefont
  {Hannestad}},\ }\href {\doibase 10.1103/PhysRevD.70.043506} {\bibfield
  {journal} {\bibinfo  {journal} {Phys. Rev.}\ }\textbf {\bibinfo {volume}
  {D70}},\ \bibinfo {pages} {043506} (\bibinfo {year} {2004})},\ \Eprint
  {http://arxiv.org/abs/astro-ph/0403291} {arXiv:astro-ph/0403291 [astro-ph]}
  \BibitemShut {NoStop}%
%%CITATION = ASTRO-PH/0403291;%%
\bibitem [{\citenamefont {Ichikawa}\ \emph {et~al.}(2005)\citenamefont
  {Ichikawa}, \citenamefont {Kawasaki},\ and\ \citenamefont
  {Takahashi}}]{Ichikawa:2005}%
  \BibitemOpen
  \bibfield  {author} {\bibinfo {author} {\bibfnamefont {K.}~\bibnamefont
  {Ichikawa}}, \bibinfo {author} {\bibfnamefont {M.}~\bibnamefont {Kawasaki}},
  \ and\ \bibinfo {author} {\bibfnamefont {F.}~\bibnamefont {Takahashi}},\
  }\href {\doibase 10.1103/PhysRevD.72.043522} {\bibfield  {journal} {\bibinfo
  {journal} {Phys. Rev.}\ }\textbf {\bibinfo {volume} {D72}},\ \bibinfo {pages}
  {043522} (\bibinfo {year} {2005})},\ \Eprint
  {http://arxiv.org/abs/astro-ph/0505395} {arXiv:astro-ph/0505395 [astro-ph]}
  \BibitemShut {NoStop}%
%%CITATION = ASTRO-PH/0505395;%%
\bibitem [{\citenamefont {Ichikawa}\ \emph {et~al.}(2007)\citenamefont
  {Ichikawa}, \citenamefont {Kawasaki},\ and\ \citenamefont
  {Takahashi}}]{Ichikawa:2006}%
  \BibitemOpen
  \bibfield  {author} {\bibinfo {author} {\bibfnamefont {K.}~\bibnamefont
  {Ichikawa}}, \bibinfo {author} {\bibfnamefont {M.}~\bibnamefont {Kawasaki}},
  \ and\ \bibinfo {author} {\bibfnamefont {F.}~\bibnamefont {Takahashi}},\
  }\href {\doibase 10.1088/1475-7516/2007/05/007} {\bibfield  {journal}
  {\bibinfo  {journal} {J. Cosmol. Astropart. Phys.}\ }\textbf {\bibinfo
  {volume} {0705}},\ \bibinfo {pages} {007} (\bibinfo {year} {2007})},\ \Eprint
  {http://arxiv.org/abs/astro-ph/0611784} {arXiv:astro-ph/0611784 [astro-ph]}
  \BibitemShut {NoStop}%
%%CITATION = ASTRO-PH/0611784;%%
\bibitem [{\citenamefont {Hasegawa}\ \emph {et~al.}(2019)\citenamefont
  {Hasegawa}, \citenamefont {Hiroshima}, \citenamefont {Kohri}, \citenamefont
  {Hansen}, \citenamefont {Tram},\ and\ \citenamefont
  {Hannestad}}]{Hasegawa:2019jsa}%
  \BibitemOpen
  \bibfield  {author} {\bibinfo {author} {\bibfnamefont {T.}~\bibnamefont
  {Hasegawa}}, \bibinfo {author} {\bibfnamefont {N.}~\bibnamefont {Hiroshima}},
  \bibinfo {author} {\bibfnamefont {K.}~\bibnamefont {Kohri}}, \bibinfo
  {author} {\bibfnamefont {R.~S.}\ \bibnamefont {Hansen}}, \bibinfo {author}
  {\bibfnamefont {T.}~\bibnamefont {Tram}}, \ and\ \bibinfo {author}
  {\bibfnamefont {S.}~\bibnamefont {Hannestad}},\ }\href {\doibase
  10.1088/1475-7516/2019/12/012} {\bibfield  {journal} {\bibinfo  {journal} {J.
  Cosmol. Astropart. Phys.}\ }\textbf {\bibinfo {volume} {12}},\ \bibinfo
  {pages} {012} (\bibinfo {year} {2019})},\ \Eprint
  {http://arxiv.org/abs/1908.10189} {arXiv:1908.10189 [hep-ph]} \BibitemShut
  {NoStop}%
\bibitem [{\citenamefont {{de Bernardis}}\ \emph {et~al.}(2008)\citenamefont
  {{de Bernardis}}, \citenamefont {{Pagano}},\ and\ \citenamefont
  {{Melchiorri}}}]{dBPM08}%
  \BibitemOpen
  \bibfield  {author} {\bibinfo {author} {\bibfnamefont {F.}~\bibnamefont {{de
  Bernardis}}}, \bibinfo {author} {\bibfnamefont {L.}~\bibnamefont {{Pagano}}},
  \ and\ \bibinfo {author} {\bibfnamefont {A.}~\bibnamefont {{Melchiorri}}},\
  }\href {\doibase 10.1016/j.astropartphys.2008.09.005} {\bibfield  {journal}
  {\bibinfo  {journal} {Astroparticle Physics}\ }\textbf {\bibinfo {volume}
  {30}},\ \bibinfo {pages} {192} (\bibinfo {year} {2008})}\BibitemShut
  {NoStop}%
\bibitem [{\citenamefont {de~Salas}\ \emph {et~al.}(2015)\citenamefont
  {de~Salas}, \citenamefont {Lattanzi}, \citenamefont {Mangano}, \citenamefont
  {Miele}, \citenamefont {Pastor},\ and\ \citenamefont
  {Pisanti}}]{deSalas:2015glj}%
  \BibitemOpen
  \bibfield  {author} {\bibinfo {author} {\bibfnamefont {P.}~\bibnamefont
  {de~Salas}}, \bibinfo {author} {\bibfnamefont {M.}~\bibnamefont {Lattanzi}},
  \bibinfo {author} {\bibfnamefont {G.}~\bibnamefont {Mangano}}, \bibinfo
  {author} {\bibfnamefont {G.}~\bibnamefont {Miele}}, \bibinfo {author}
  {\bibfnamefont {S.}~\bibnamefont {Pastor}}, \ and\ \bibinfo {author}
  {\bibfnamefont {O.}~\bibnamefont {Pisanti}},\ }\href {\doibase
  10.1103/PhysRevD.92.123534} {\bibfield  {journal} {\bibinfo  {journal} {Phys.
  Rev. D}\ }\textbf {\bibinfo {volume} {92}},\ \bibinfo {pages} {123534}
  (\bibinfo {year} {2015})},\ \Eprint {http://arxiv.org/abs/1511.00672}
  {arXiv:1511.00672 [astro-ph.CO]} \BibitemShut {NoStop}%
\bibitem [{\citenamefont {Boyle}\ and\ \citenamefont
  {Steinhardt}(2008)}]{Boyle:2005GW}%
  \BibitemOpen
  \bibfield  {author} {\bibinfo {author} {\bibfnamefont {L.~A.}\ \bibnamefont
  {Boyle}}\ and\ \bibinfo {author} {\bibfnamefont {P.~J.}\ \bibnamefont
  {Steinhardt}},\ }\href {\doibase 10.1103/PhysRevD.77.063504} {\bibfield
  {journal} {\bibinfo  {journal} {Phys. Rev.}\ }\textbf {\bibinfo {volume}
  {D77}},\ \bibinfo {pages} {063504} (\bibinfo {year} {2008})},\ \Eprint
  {http://arxiv.org/abs/astro-ph/0512014} {arXiv:astro-ph/0512014 [astro-ph]}
  \BibitemShut {NoStop}%
%%CITATION = ASTRO-PH/0512014;%%
\bibitem [{\citenamefont {Easther}\ and\ \citenamefont
  {Lim}(2006)}]{Easther:2006GW}%
  \BibitemOpen
  \bibfield  {author} {\bibinfo {author} {\bibfnamefont {R.}~\bibnamefont
  {Easther}}\ and\ \bibinfo {author} {\bibfnamefont {E.~A.}\ \bibnamefont
  {Lim}},\ }\href {\doibase 10.1088/1475-7516/2006/04/010} {\bibfield
  {journal} {\bibinfo  {journal} {J. Cosmol. Astropart. Phys.}\ }\textbf
  {\bibinfo {volume} {0604}},\ \bibinfo {pages} {010} (\bibinfo {year}
  {2006})},\ \Eprint {http://arxiv.org/abs/astro-ph/0601617}
  {arXiv:astro-ph/0601617 [astro-ph]} \BibitemShut {NoStop}%
%%CITATION = ASTRO-PH/0601617;%%
\bibitem [{\citenamefont {Easther}\ \emph {et~al.}(2008)\citenamefont
  {Easther}, \citenamefont {Giblin}, \citenamefont {Lim}, \citenamefont
  {Park},\ and\ \citenamefont {Stewart}}]{Easther:2008GW}%
  \BibitemOpen
  \bibfield  {author} {\bibinfo {author} {\bibfnamefont {R.}~\bibnamefont
  {Easther}}, \bibinfo {author} {\bibfnamefont {J.~T.}\ \bibnamefont {Giblin},
  \bibfnamefont {Jr.}}, \bibinfo {author} {\bibfnamefont {E.~A.}\ \bibnamefont
  {Lim}}, \bibinfo {author} {\bibfnamefont {W.-I.}\ \bibnamefont {Park}}, \
  and\ \bibinfo {author} {\bibfnamefont {E.~D.}\ \bibnamefont {Stewart}},\
  }\href {\doibase 10.1088/1475-7516/2008/05/013} {\bibfield  {journal}
  {\bibinfo  {journal} {J. Cosmol. Astropart. Phys.}\ }\textbf {\bibinfo
  {volume} {0805}},\ \bibinfo {pages} {013} (\bibinfo {year} {2008})},\ \Eprint
  {http://arxiv.org/abs/0801.4197} {arXiv:0801.4197 [astro-ph]} \BibitemShut
  {NoStop}%
%%CITATION = ARXIV:0801.4197;%%
\bibitem [{\citenamefont {Amin}\ \emph {et~al.}(2014)\citenamefont {Amin},
  \citenamefont {Hertzberg}, \citenamefont {Kaiser},\ and\ \citenamefont
  {Karouby}}]{Amin:2014}%
  \BibitemOpen
  \bibfield  {author} {\bibinfo {author} {\bibfnamefont {M.~A.}\ \bibnamefont
  {Amin}}, \bibinfo {author} {\bibfnamefont {M.~P.}\ \bibnamefont {Hertzberg}},
  \bibinfo {author} {\bibfnamefont {D.~I.}\ \bibnamefont {Kaiser}}, \ and\
  \bibinfo {author} {\bibfnamefont {J.}~\bibnamefont {Karouby}},\ }\href
  {\doibase 10.1142/S0218271815300037} {\bibfield  {journal} {\bibinfo
  {journal} {Int. J. Mod. Phys.}\ }\textbf {\bibinfo {volume} {D24}},\ \bibinfo
  {pages} {1530003} (\bibinfo {year} {2014})},\ \Eprint
  {http://arxiv.org/abs/1410.3808} {arXiv:1410.3808 [hep-ph]} \BibitemShut
  {NoStop}%
%%CITATION = ARXIV:1410.3808;%%
\bibitem [{\citenamefont {Giblin}\ and\ \citenamefont
  {Thrane}(2014)}]{Giblin:2014GW}%
  \BibitemOpen
  \bibfield  {author} {\bibinfo {author} {\bibfnamefont {J.~T.}\ \bibnamefont
  {Giblin}}\ and\ \bibinfo {author} {\bibfnamefont {E.}~\bibnamefont
  {Thrane}},\ }\href {\doibase 10.1103/PhysRevD.90.107502} {\bibfield
  {journal} {\bibinfo  {journal} {Phys. Rev.}\ }\textbf {\bibinfo {volume}
  {D90}},\ \bibinfo {pages} {107502} (\bibinfo {year} {2014})},\ \Eprint
  {http://arxiv.org/abs/1410.4779} {arXiv:1410.4779 [gr-qc]} \BibitemShut
  {NoStop}%
%%CITATION = ARXIV:1410.4779;%%
\bibitem [{\citenamefont {Figueroa}\ and\ \citenamefont
  {Tanin}(2019{\natexlab{a}})}]{Figueroa:2019paj}%
  \BibitemOpen
  \bibfield  {author} {\bibinfo {author} {\bibfnamefont {D.~G.}\ \bibnamefont
  {Figueroa}}\ and\ \bibinfo {author} {\bibfnamefont {E.~H.}\ \bibnamefont
  {Tanin}},\ }\href {\doibase 10.1088/1475-7516/2019/08/011} {\bibfield
  {journal} {\bibinfo  {journal} {J. Cosmol. Astropart. Phys.}\ }\textbf
  {\bibinfo {volume} {08}},\ \bibinfo {pages} {011} (\bibinfo {year}
  {2019}{\natexlab{a}})},\ \Eprint {http://arxiv.org/abs/1905.11960}
  {arXiv:1905.11960 [astro-ph.CO]} \BibitemShut {NoStop}%
\bibitem [{\citenamefont {Cui}\ \emph {et~al.}(2018)\citenamefont {Cui},
  \citenamefont {Lewicki}, \citenamefont {Morrissey},\ and\ \citenamefont
  {Wells}}]{Cui:2017GW}%
  \BibitemOpen
  \bibfield  {author} {\bibinfo {author} {\bibfnamefont {Y.}~\bibnamefont
  {Cui}}, \bibinfo {author} {\bibfnamefont {M.}~\bibnamefont {Lewicki}},
  \bibinfo {author} {\bibfnamefont {D.~E.}\ \bibnamefont {Morrissey}}, \ and\
  \bibinfo {author} {\bibfnamefont {J.~D.}\ \bibnamefont {Wells}},\ }\href
  {\doibase 10.1103/PhysRevD.97.123505} {\bibfield  {journal} {\bibinfo
  {journal} {Phys. Rev.}\ }\textbf {\bibinfo {volume} {D97}},\ \bibinfo {pages}
  {123505} (\bibinfo {year} {2018})},\ \Eprint
  {http://arxiv.org/abs/1711.03104} {arXiv:1711.03104 [hep-ph]} \BibitemShut
  {NoStop}%
%%CITATION = ARXIV:1711.03104;%%
\bibitem [{\citenamefont {Erickcek}\ and\ \citenamefont
  {Sigurdson}(2011)}]{Erickcek:2011}%
  \BibitemOpen
  \bibfield  {author} {\bibinfo {author} {\bibfnamefont {A.~L.}\ \bibnamefont
  {Erickcek}}\ and\ \bibinfo {author} {\bibfnamefont {K.}~\bibnamefont
  {Sigurdson}},\ }\href {\doibase 10.1103/PhysRevD.84.083503} {\bibfield
  {journal} {\bibinfo  {journal} {Phys. Rev.}\ }\textbf {\bibinfo {volume}
  {D84}},\ \bibinfo {pages} {083503} (\bibinfo {year} {2011})},\ \Eprint
  {http://arxiv.org/abs/1106.0536} {arXiv:1106.0536 [astro-ph.CO]} \BibitemShut
  {NoStop}%
%%CITATION = ARXIV:1106.0536;%%
\bibitem [{\citenamefont {Barenboim}\ and\ \citenamefont
  {Rasero}(2014)}]{Barenboim:2013}%
  \BibitemOpen
  \bibfield  {author} {\bibinfo {author} {\bibfnamefont {G.}~\bibnamefont
  {Barenboim}}\ and\ \bibinfo {author} {\bibfnamefont {J.}~\bibnamefont
  {Rasero}},\ }\href {\doibase 10.1007/JHEP04(2014)138} {\bibfield  {journal}
  {\bibinfo  {journal} {JHEP}\ }\textbf {\bibinfo {volume} {04}},\ \bibinfo
  {pages} {138} (\bibinfo {year} {2014})},\ \Eprint
  {http://arxiv.org/abs/1311.4034} {arXiv:1311.4034 [hep-ph]} \BibitemShut
  {NoStop}%
%%CITATION = ARXIV:1311.4034;%%
\bibitem [{\citenamefont {Fan}\ \emph {et~al.}(2014)\citenamefont {Fan},
  \citenamefont {Özsoy},\ and\ \citenamefont {Watson}}]{Fan:2014}%
  \BibitemOpen
  \bibfield  {author} {\bibinfo {author} {\bibfnamefont {J.}~\bibnamefont
  {Fan}}, \bibinfo {author} {\bibfnamefont {O.}~\bibnamefont {Özsoy}}, \ and\
  \bibinfo {author} {\bibfnamefont {S.}~\bibnamefont {Watson}},\ }\href
  {\doibase 10.1103/PhysRevD.90.043536} {\bibfield  {journal} {\bibinfo
  {journal} {Phys. Rev.}\ }\textbf {\bibinfo {volume} {D90}},\ \bibinfo {pages}
  {043536} (\bibinfo {year} {2014})},\ \Eprint {http://arxiv.org/abs/1405.7373}
  {arXiv:1405.7373 [hep-ph]} \BibitemShut {NoStop}%
%%CITATION = ARXIV:1405.7373;%%
\bibitem [{\citenamefont {Erickcek}(2015)}]{Erickcek:2015}%
  \BibitemOpen
  \bibfield  {author} {\bibinfo {author} {\bibfnamefont {A.~L.}\ \bibnamefont
  {Erickcek}},\ }\href {\doibase 10.1103/PhysRevD.92.103505} {\bibfield
  {journal} {\bibinfo  {journal} {Phys. Rev.}\ }\textbf {\bibinfo {volume}
  {D92}},\ \bibinfo {pages} {103505} (\bibinfo {year} {2015})},\ \Eprint
  {http://arxiv.org/abs/1504.03335} {arXiv:1504.03335 [astro-ph.CO]}
  \BibitemShut {NoStop}%
%%CITATION = ARXIV:1504.03335;%%
\bibitem [{\citenamefont {Erickcek}\ \emph {et~al.}(2016)\citenamefont
  {Erickcek}, \citenamefont {Sinha},\ and\ \citenamefont
  {Watson}}]{Erickcek:Sinha:Watson:2015}%
  \BibitemOpen
  \bibfield  {author} {\bibinfo {author} {\bibfnamefont {A.~L.}\ \bibnamefont
  {Erickcek}}, \bibinfo {author} {\bibfnamefont {K.}~\bibnamefont {Sinha}}, \
  and\ \bibinfo {author} {\bibfnamefont {S.}~\bibnamefont {Watson}},\ }\href
  {\doibase 10.1103/PhysRevD.94.063502} {\bibfield  {journal} {\bibinfo
  {journal} {Phys. Rev.}\ }\textbf {\bibinfo {volume} {D94}},\ \bibinfo {pages}
  {063502} (\bibinfo {year} {2016})},\ \Eprint
  {http://arxiv.org/abs/1510.04291} {arXiv:1510.04291 [hep-ph]} \BibitemShut
  {NoStop}%
%%CITATION = ARXIV:1510.04291;%%
\bibitem [{\citenamefont {Blanco}\ \emph {et~al.}(2019)\citenamefont {Blanco},
  \citenamefont {Delos}, \citenamefont {Erickcek},\ and\ \citenamefont
  {Hooper}}]{Blanco:2019eij}%
  \BibitemOpen
  \bibfield  {author} {\bibinfo {author} {\bibfnamefont {C.}~\bibnamefont
  {Blanco}}, \bibinfo {author} {\bibfnamefont {M.~S.}\ \bibnamefont {Delos}},
  \bibinfo {author} {\bibfnamefont {A.~L.}\ \bibnamefont {Erickcek}}, \ and\
  \bibinfo {author} {\bibfnamefont {D.}~\bibnamefont {Hooper}},\ }\href
  {\doibase 10.1103/PhysRevD.100.103010} {\bibfield  {journal} {\bibinfo
  {journal} {Phys. Rev. D}\ }\textbf {\bibinfo {volume} {100}},\ \bibinfo
  {pages} {103010} (\bibinfo {year} {2019})},\ \Eprint
  {http://arxiv.org/abs/1906.00010} {arXiv:1906.00010 [astro-ph.CO]}
  \BibitemShut {NoStop}%
\bibitem [{\citenamefont {Delos}\ \emph
  {et~al.}(2019{\natexlab{a}})\citenamefont {Delos}, \citenamefont {Linden},\
  and\ \citenamefont {Erickcek}}]{StenDelos:2019xdk}%
  \BibitemOpen
  \bibfield  {author} {\bibinfo {author} {\bibfnamefont {M.~S.}\ \bibnamefont
  {Delos}}, \bibinfo {author} {\bibfnamefont {T.}~\bibnamefont {Linden}}, \
  and\ \bibinfo {author} {\bibfnamefont {A.~L.}\ \bibnamefont {Erickcek}},\
  }\href {\doibase 10.1103/PhysRevD.100.123546} {\bibfield  {journal} {\bibinfo
   {journal} {Phys. Rev. D}\ }\textbf {\bibinfo {volume} {100}},\ \bibinfo
  {pages} {123546} (\bibinfo {year} {2019}{\natexlab{a}})},\ \Eprint
  {http://arxiv.org/abs/1910.08553} {arXiv:1910.08553 [astro-ph.CO]}
  \BibitemShut {NoStop}%
\bibitem [{\citenamefont {Erickcek}\ \emph {et~al.}(2021)\citenamefont
  {Erickcek}, \citenamefont {Ralegankar},\ and\ \citenamefont
  {Shelton}}]{Erickcek:2020wzd}%
  \BibitemOpen
  \bibfield  {author} {\bibinfo {author} {\bibfnamefont {A.~L.}\ \bibnamefont
  {Erickcek}}, \bibinfo {author} {\bibfnamefont {P.}~\bibnamefont
  {Ralegankar}}, \ and\ \bibinfo {author} {\bibfnamefont {J.}~\bibnamefont
  {Shelton}},\ }\href {\doibase 10.1103/PhysRevD.103.103508} {\bibfield
  {journal} {\bibinfo  {journal} {Phys. Rev. D}\ }\textbf {\bibinfo {volume}
  {103}},\ \bibinfo {pages} {103508} (\bibinfo {year} {2021})},\ \Eprint
  {http://arxiv.org/abs/2008.04311} {arXiv:2008.04311 [astro-ph.CO]}
  \BibitemShut {NoStop}%
\bibitem [{\citenamefont {Erickcek}\ \emph {et~al.}(2022)\citenamefont
  {Erickcek}, \citenamefont {Ralegankar},\ and\ \citenamefont
  {Shelton}}]{Erickcek:2021fsu}%
  \BibitemOpen
  \bibfield  {author} {\bibinfo {author} {\bibfnamefont {A.~L.}\ \bibnamefont
  {Erickcek}}, \bibinfo {author} {\bibfnamefont {P.}~\bibnamefont
  {Ralegankar}}, \ and\ \bibinfo {author} {\bibfnamefont {J.}~\bibnamefont
  {Shelton}},\ }\href {\doibase 10.1088/1475-7516/2022/01/017} {\bibfield
  {journal} {\bibinfo  {journal} {JCAP}\ }\textbf {\bibinfo {volume} {01}},\
  \bibinfo {pages} {017} (\bibinfo {year} {2022})},\ \Eprint
  {http://arxiv.org/abs/2106.09041} {arXiv:2106.09041 [hep-ph]} \BibitemShut
  {NoStop}%
\bibitem [{\citenamefont {Spokoiny}(1993)}]{Spokoiny:1993}%
  \BibitemOpen
  \bibfield  {author} {\bibinfo {author} {\bibfnamefont {B.}~\bibnamefont
  {Spokoiny}},\ }\href {\doibase 10.1016/0370-2693(93)90155-B} {\bibfield
  {journal} {\bibinfo  {journal} {Phys. Lett.}\ }\textbf {\bibinfo {volume}
  {B315}},\ \bibinfo {pages} {40} (\bibinfo {year} {1993})},\ \Eprint
  {http://arxiv.org/abs/gr-qc/9306008} {arXiv:gr-qc/9306008 [gr-qc]}
  \BibitemShut {NoStop}%
%%CITATION = GR-QC/9306008;%%
\bibitem [{\citenamefont {Joyce}(1997)}]{Joyce:1996}%
  \BibitemOpen
  \bibfield  {author} {\bibinfo {author} {\bibfnamefont {M.}~\bibnamefont
  {Joyce}},\ }\href {\doibase 10.1103/PhysRevD.55.1875} {\bibfield  {journal}
  {\bibinfo  {journal} {Phys. Rev.}\ }\textbf {\bibinfo {volume} {D55}},\
  \bibinfo {pages} {1875} (\bibinfo {year} {1997})},\ \Eprint
  {http://arxiv.org/abs/hep-ph/9606223} {arXiv:hep-ph/9606223 [hep-ph]}
  \BibitemShut {NoStop}%
%%CITATION = HEP-PH/9606223;%%
\bibitem [{\citenamefont {Ferreira}\ and\ \citenamefont
  {Joyce}(1998)}]{Ferreira:1997}%
  \BibitemOpen
  \bibfield  {author} {\bibinfo {author} {\bibfnamefont {P.~G.}\ \bibnamefont
  {Ferreira}}\ and\ \bibinfo {author} {\bibfnamefont {M.}~\bibnamefont
  {Joyce}},\ }\href {\doibase 10.1103/PhysRevD.58.023503} {\bibfield  {journal}
  {\bibinfo  {journal} {Phys. Rev.}\ }\textbf {\bibinfo {volume} {D58}},\
  \bibinfo {pages} {023503} (\bibinfo {year} {1998})},\ \Eprint
  {http://arxiv.org/abs/astro-ph/9711102} {arXiv:astro-ph/9711102 [astro-ph]}
  \BibitemShut {NoStop}%
%%CITATION = ASTRO-PH/9711102;%%
\bibitem [{\citenamefont {Conlon}\ and\ \citenamefont
  {Revello}(2022)}]{Conlon:2022pnx}%
  \BibitemOpen
  \bibfield  {author} {\bibinfo {author} {\bibfnamefont {J.~P.}\ \bibnamefont
  {Conlon}}\ and\ \bibinfo {author} {\bibfnamefont {F.}~\bibnamefont
  {Revello}},\ }\href {\doibase 10.1007/JHEP11(2022)155} {\bibfield  {journal}
  {\bibinfo  {journal} {JHEP}\ }\textbf {\bibinfo {volume} {11}},\ \bibinfo
  {pages} {155} (\bibinfo {year} {2022})},\ \Eprint
  {http://arxiv.org/abs/2207.00567} {arXiv:2207.00567 [hep-th]} \BibitemShut
  {NoStop}%
\bibitem [{\citenamefont {Apers}\ \emph {et~al.}(2022)\citenamefont {Apers},
  \citenamefont {Conlon}, \citenamefont {Mosny},\ and\ \citenamefont
  {Revello}}]{Apers:2022cyl}%
  \BibitemOpen
  \bibfield  {author} {\bibinfo {author} {\bibfnamefont {F.}~\bibnamefont
  {Apers}}, \bibinfo {author} {\bibfnamefont {J.~P.}\ \bibnamefont {Conlon}},
  \bibinfo {author} {\bibfnamefont {M.}~\bibnamefont {Mosny}}, \ and\ \bibinfo
  {author} {\bibfnamefont {F.}~\bibnamefont {Revello}},\ }\href@noop {} {\
  (\bibinfo {year} {2022})},\ \Eprint {http://arxiv.org/abs/2212.10293}
  {arXiv:2212.10293 [hep-th]} \BibitemShut {NoStop}%
\bibitem [{\citenamefont {Figueroa}\ and\ \citenamefont
  {Tanin}(2019{\natexlab{b}})}]{Figueroa:2018twl}%
  \BibitemOpen
  \bibfield  {author} {\bibinfo {author} {\bibfnamefont {D.~G.}\ \bibnamefont
  {Figueroa}}\ and\ \bibinfo {author} {\bibfnamefont {E.~H.}\ \bibnamefont
  {Tanin}},\ }\href {\doibase 10.1088/1475-7516/2019/10/050} {\bibfield
  {journal} {\bibinfo  {journal} {J. Cosmol. Astropart. Phys.}\ }\textbf
  {\bibinfo {volume} {10}},\ \bibinfo {pages} {050} (\bibinfo {year}
  {2019}{\natexlab{b}})},\ \Eprint {http://arxiv.org/abs/1811.04093}
  {arXiv:1811.04093 [astro-ph.CO]} \BibitemShut {NoStop}%
\bibitem [{\citenamefont {Peebles}\ and\ \citenamefont
  {Vilenkin}(1999)}]{Peebles:1998}%
  \BibitemOpen
  \bibfield  {author} {\bibinfo {author} {\bibfnamefont {P.~J.~E.}\
  \bibnamefont {Peebles}}\ and\ \bibinfo {author} {\bibfnamefont
  {A.}~\bibnamefont {Vilenkin}},\ }\href {\doibase 10.1103/PhysRevD.59.063505}
  {\bibfield  {journal} {\bibinfo  {journal} {Phys. Rev.}\ }\textbf {\bibinfo
  {volume} {D59}},\ \bibinfo {pages} {063505} (\bibinfo {year} {1999})},\
  \Eprint {http://arxiv.org/abs/astro-ph/9810509} {arXiv:astro-ph/9810509
  [astro-ph]} \BibitemShut {NoStop}%
%%CITATION = ASTRO-PH/9810509;%%
\bibitem [{\citenamefont {Dimopoulos}\ and\ \citenamefont
  {Valle}(2002)}]{Dimopoulos:2001}%
  \BibitemOpen
  \bibfield  {author} {\bibinfo {author} {\bibfnamefont {K.}~\bibnamefont
  {Dimopoulos}}\ and\ \bibinfo {author} {\bibfnamefont {J.~W.~F.}\ \bibnamefont
  {Valle}},\ }\href {\doibase 10.1016/S0927-6505(02)00115-9} {\bibfield
  {journal} {\bibinfo  {journal} {Astropart. Phys.}\ }\textbf {\bibinfo
  {volume} {18}},\ \bibinfo {pages} {287} (\bibinfo {year} {2002})},\ \Eprint
  {http://arxiv.org/abs/astro-ph/0111417} {arXiv:astro-ph/0111417 [astro-ph]}
  \BibitemShut {NoStop}%
%%CITATION = ASTRO-PH/0111417;%%
\bibitem [{\citenamefont {Dimopoulos}(2003)}]{Dimopoulos:2002Curvaton}%
  \BibitemOpen
  \bibfield  {author} {\bibinfo {author} {\bibfnamefont {K.}~\bibnamefont
  {Dimopoulos}},\ }\href {\doibase 10.1103/PhysRevD.68.123506} {\bibfield
  {journal} {\bibinfo  {journal} {Phys. Rev.}\ }\textbf {\bibinfo {volume}
  {D68}},\ \bibinfo {pages} {123506} (\bibinfo {year} {2003})},\ \Eprint
  {http://arxiv.org/abs/astro-ph/0212264} {arXiv:astro-ph/0212264 [astro-ph]}
  \BibitemShut {NoStop}%
%%CITATION = ASTRO-PH/0212264;%%
\bibitem [{\citenamefont {Chung}\ \emph {et~al.}(2007)\citenamefont {Chung},
  \citenamefont {Everett}, \citenamefont {Kong},\ and\ \citenamefont
  {Matchev}}]{Chung:2007}%
  \BibitemOpen
  \bibfield  {author} {\bibinfo {author} {\bibfnamefont {D.~J.~H.}\
  \bibnamefont {Chung}}, \bibinfo {author} {\bibfnamefont {L.~L.}\ \bibnamefont
  {Everett}}, \bibinfo {author} {\bibfnamefont {K.}~\bibnamefont {Kong}}, \
  and\ \bibinfo {author} {\bibfnamefont {K.~T.}\ \bibnamefont {Matchev}},\
  }\href {\doibase 10.1088/1126-6708/2007/10/016} {\bibfield  {journal}
  {\bibinfo  {journal} {JHEP}\ }\textbf {\bibinfo {volume} {10}},\ \bibinfo
  {pages} {016} (\bibinfo {year} {2007})},\ \Eprint
  {http://arxiv.org/abs/0706.2375} {arXiv:0706.2375 [hep-ph]} \BibitemShut
  {NoStop}%
%%CITATION = ARXIV:0706.2375;%%
\bibitem [{\citenamefont {Li}\ \emph {et~al.}(2014)\citenamefont {Li},
  \citenamefont {Rindler-Daller},\ and\ \citenamefont {Shapiro}}]{Li:2013nal}%
  \BibitemOpen
  \bibfield  {author} {\bibinfo {author} {\bibfnamefont {B.}~\bibnamefont
  {Li}}, \bibinfo {author} {\bibfnamefont {T.}~\bibnamefont {Rindler-Daller}},
  \ and\ \bibinfo {author} {\bibfnamefont {P.~R.}\ \bibnamefont {Shapiro}},\
  }\href {\doibase 10.1103/PhysRevD.89.083536} {\bibfield  {journal} {\bibinfo
  {journal} {Phys. Rev. D}\ }\textbf {\bibinfo {volume} {89}},\ \bibinfo
  {pages} {083536} (\bibinfo {year} {2014})},\ \Eprint
  {http://arxiv.org/abs/1310.6061} {arXiv:1310.6061 [astro-ph.CO]} \BibitemShut
  {NoStop}%
\bibitem [{\citenamefont {Li}\ \emph {et~al.}(2017)\citenamefont {Li},
  \citenamefont {Shapiro},\ and\ \citenamefont {Rindler-Daller}}]{Li:2016mmc}%
  \BibitemOpen
  \bibfield  {author} {\bibinfo {author} {\bibfnamefont {B.}~\bibnamefont
  {Li}}, \bibinfo {author} {\bibfnamefont {P.~R.}\ \bibnamefont {Shapiro}}, \
  and\ \bibinfo {author} {\bibfnamefont {T.}~\bibnamefont {Rindler-Daller}},\
  }\href {\doibase 10.1103/PhysRevD.96.063505} {\bibfield  {journal} {\bibinfo
  {journal} {Phys. Rev. D}\ }\textbf {\bibinfo {volume} {96}},\ \bibinfo
  {pages} {063505} (\bibinfo {year} {2017})},\ \Eprint
  {http://arxiv.org/abs/1611.07961} {arXiv:1611.07961 [astro-ph.CO]}
  \BibitemShut {NoStop}%
\bibitem [{\citenamefont {Li}\ and\ \citenamefont
  {Shapiro}(2021)}]{Li:2021htg}%
  \BibitemOpen
  \bibfield  {author} {\bibinfo {author} {\bibfnamefont {B.}~\bibnamefont
  {Li}}\ and\ \bibinfo {author} {\bibfnamefont {P.~R.}\ \bibnamefont
  {Shapiro}},\ }\href {\doibase 10.1088/1475-7516/2021/10/024} {\bibfield
  {journal} {\bibinfo  {journal} {JCAP}\ }\textbf {\bibinfo {volume} {10}},\
  \bibinfo {pages} {024} (\bibinfo {year} {2021})},\ \Eprint
  {http://arxiv.org/abs/2107.12229} {arXiv:2107.12229 [astro-ph.CO]}
  \BibitemShut {NoStop}%
\bibitem [{\citenamefont {Redmond}\ \emph {et~al.}(2018)\citenamefont
  {Redmond}, \citenamefont {Trezza},\ and\ \citenamefont
  {Erickcek}}]{Redmond:2018}%
  \BibitemOpen
  \bibfield  {author} {\bibinfo {author} {\bibfnamefont {K.}~\bibnamefont
  {Redmond}}, \bibinfo {author} {\bibfnamefont {A.}~\bibnamefont {Trezza}}, \
  and\ \bibinfo {author} {\bibfnamefont {A.~L.}\ \bibnamefont {Erickcek}},\
  }\href {\doibase 10.1103/PhysRevD.98.063504} {\bibfield  {journal} {\bibinfo
  {journal} {Phys. Rev.}\ }\textbf {\bibinfo {volume} {D98}},\ \bibinfo {pages}
  {063504} (\bibinfo {year} {2018})},\ \Eprint
  {http://arxiv.org/abs/1807.01327} {arXiv:1807.01327 [astro-ph.CO]}
  \BibitemShut {NoStop}%
%%CITATION = ARXIV:1807.01327;%%
\bibitem [{\citenamefont {Profumo}\ and\ \citenamefont
  {Ullio}(2003)}]{Profumo:2003}%
  \BibitemOpen
  \bibfield  {author} {\bibinfo {author} {\bibfnamefont {S.}~\bibnamefont
  {Profumo}}\ and\ \bibinfo {author} {\bibfnamefont {P.}~\bibnamefont
  {Ullio}},\ }\href {\doibase 10.1088/1475-7516/2003/11/006} {\bibfield
  {journal} {\bibinfo  {journal} {J. Cosmol. Astropart. Phys.}\ }\textbf
  {\bibinfo {volume} {0311}},\ \bibinfo {pages} {006} (\bibinfo {year}
  {2003})},\ \Eprint {http://arxiv.org/abs/hep-ph/0309220}
  {arXiv:hep-ph/0309220 [hep-ph]} \BibitemShut {NoStop}%
%%CITATION = HEP-PH/0309220;%%
\bibitem [{\citenamefont {Pallis}(2005)}]{Pallis:2005}%
  \BibitemOpen
  \bibfield  {author} {\bibinfo {author} {\bibfnamefont {C.}~\bibnamefont
  {Pallis}},\ }\href {\doibase 10.1088/1475-7516/2005/10/015} {\bibfield
  {journal} {\bibinfo  {journal} {J. Cosmol. Astropart. Phys.}\ }\textbf
  {\bibinfo {volume} {0510}},\ \bibinfo {pages} {015} (\bibinfo {year}
  {2005})},\ \Eprint {http://arxiv.org/abs/hep-ph/0503080}
  {arXiv:hep-ph/0503080 [hep-ph]} \BibitemShut {NoStop}%
%%CITATION = HEP-PH/0503080;%%
\bibitem [{\citenamefont {Pallis}(2006)}]{Pallis:2006}%
  \BibitemOpen
  \bibfield  {author} {\bibinfo {author} {\bibfnamefont {C.}~\bibnamefont
  {Pallis}},\ }in\ \href@noop {} {\emph {\bibinfo {booktitle} {{Proceedings,
  6th International Workshop on The identification of dark matter (IDM 2006):
  Rhodes, Greece, September 11-16, 2006}}}}\ (\bibinfo {year} {2006})\ pp.\
  \bibinfo {pages} {602--608},\ \Eprint {http://arxiv.org/abs/hep-ph/0610433}
  {arXiv:hep-ph/0610433 [hep-ph]} \BibitemShut {NoStop}%
%%CITATION = HEP-PH/0610433;%%
\bibitem [{\citenamefont {Gomez}\ \emph {et~al.}(2009)\citenamefont {Gomez},
  \citenamefont {Lola}, \citenamefont {Pallis},\ and\ \citenamefont
  {Rodriguez-Quintero}}]{Gomez:2008}%
  \BibitemOpen
  \bibfield  {author} {\bibinfo {author} {\bibfnamefont {M.~E.}\ \bibnamefont
  {Gomez}}, \bibinfo {author} {\bibfnamefont {S.}~\bibnamefont {Lola}},
  \bibinfo {author} {\bibfnamefont {C.}~\bibnamefont {Pallis}}, \ and\ \bibinfo
  {author} {\bibfnamefont {J.}~\bibnamefont {Rodriguez-Quintero}},\ }\href
  {\doibase 10.1088/1475-7516/2009/01/027} {\bibfield  {journal} {\bibinfo
  {journal} {J. Cosmol. Astropart. Phys.}\ }\textbf {\bibinfo {volume}
  {0901}},\ \bibinfo {pages} {027} (\bibinfo {year} {2009})},\ \Eprint
  {http://arxiv.org/abs/0809.1859} {arXiv:0809.1859 [hep-ph]} \BibitemShut
  {NoStop}%
%%CITATION = ARXIV:0809.1859;%%
\bibitem [{\citenamefont {Lola}\ \emph {et~al.}(2009)\citenamefont {Lola},
  \citenamefont {Pallis},\ and\ \citenamefont {Tzelati}}]{Lola:2009}%
  \BibitemOpen
  \bibfield  {author} {\bibinfo {author} {\bibfnamefont {S.}~\bibnamefont
  {Lola}}, \bibinfo {author} {\bibfnamefont {C.}~\bibnamefont {Pallis}}, \ and\
  \bibinfo {author} {\bibfnamefont {E.}~\bibnamefont {Tzelati}},\ }\href
  {\doibase 10.1088/1475-7516/2009/11/017} {\bibfield  {journal} {\bibinfo
  {journal} {J. Cosmol. Astropart. Phys.}\ }\textbf {\bibinfo {volume}
  {0911}},\ \bibinfo {pages} {017} (\bibinfo {year} {2009})},\ \Eprint
  {http://arxiv.org/abs/0907.2941} {arXiv:0907.2941 [hep-ph]} \BibitemShut
  {NoStop}%
%%CITATION = ARXIV:0907.2941;%%
\bibitem [{\citenamefont {Pallis}(2010)}]{Pallis:2009}%
  \BibitemOpen
  \bibfield  {author} {\bibinfo {author} {\bibfnamefont {C.}~\bibnamefont
  {Pallis}},\ }\href {\doibase 10.1016/j.nuclphysb.2010.01.015} {\bibfield
  {journal} {\bibinfo  {journal} {Nucl. Phys.}\ }\textbf {\bibinfo {volume}
  {B831}},\ \bibinfo {pages} {217} (\bibinfo {year} {2010})},\ \Eprint
  {http://arxiv.org/abs/0909.3026} {arXiv:0909.3026 [hep-ph]} \BibitemShut
  {NoStop}%
%%CITATION = ARXIV:0909.3026;%%
\bibitem [{\citenamefont {D'Eramo}\ \emph {et~al.}(2017)\citenamefont
  {D'Eramo}, \citenamefont {Fernandez},\ and\ \citenamefont
  {Profumo}}]{DEramo:2017gpl}%
  \BibitemOpen
  \bibfield  {author} {\bibinfo {author} {\bibfnamefont {F.}~\bibnamefont
  {D'Eramo}}, \bibinfo {author} {\bibfnamefont {N.}~\bibnamefont {Fernandez}},
  \ and\ \bibinfo {author} {\bibfnamefont {S.}~\bibnamefont {Profumo}},\ }\href
  {\doibase 10.1088/1475-7516/2017/05/012} {\bibfield  {journal} {\bibinfo
  {journal} {J. Cosmol. Astropart. Phys.}\ }\textbf {\bibinfo {volume} {05}},\
  \bibinfo {pages} {012} (\bibinfo {year} {2017})},\ \Eprint
  {http://arxiv.org/abs/1703.04793} {arXiv:1703.04793 [hep-ph]} \BibitemShut
  {NoStop}%
\bibitem [{\citenamefont {Redmond}\ and\ \citenamefont
  {Erickcek}(2017)}]{Redmond:2017}%
  \BibitemOpen
  \bibfield  {author} {\bibinfo {author} {\bibfnamefont {K.}~\bibnamefont
  {Redmond}}\ and\ \bibinfo {author} {\bibfnamefont {A.~L.}\ \bibnamefont
  {Erickcek}},\ }\href {\doibase 10.1103/PhysRevD.96.043511} {\bibfield
  {journal} {\bibinfo  {journal} {Phys. Rev.}\ }\textbf {\bibinfo {volume}
  {D96}},\ \bibinfo {pages} {043511} (\bibinfo {year} {2017})},\ \Eprint
  {http://arxiv.org/abs/1704.01056} {arXiv:1704.01056 [hep-ph]} \BibitemShut
  {NoStop}%
%%CITATION = ARXIV:1704.01056;%%
\bibitem [{\citenamefont {D'Eramo}\ \emph {et~al.}(2018)\citenamefont
  {D'Eramo}, \citenamefont {Fernandez},\ and\ \citenamefont
  {Profumo}}]{DEramo:2017ecx}%
  \BibitemOpen
  \bibfield  {author} {\bibinfo {author} {\bibfnamefont {F.}~\bibnamefont
  {D'Eramo}}, \bibinfo {author} {\bibfnamefont {N.}~\bibnamefont {Fernandez}},
  \ and\ \bibinfo {author} {\bibfnamefont {S.}~\bibnamefont {Profumo}},\ }\href
  {\doibase 10.1088/1475-7516/2018/02/046} {\bibfield  {journal} {\bibinfo
  {journal} {J. Cosmol. Astropart. Phys.}\ }\textbf {\bibinfo {volume}
  {1802}},\ \bibinfo {pages} {046} (\bibinfo {year} {2018})},\ \Eprint
  {http://arxiv.org/abs/1712.07453} {arXiv:1712.07453 [hep-ph]} \BibitemShut
  {NoStop}%
%%CITATION = ARXIV:1712.07453;%%
\bibitem [{\citenamefont {Visinelli}(2018)}]{Visinelli:2017qga}%
  \BibitemOpen
  \bibfield  {author} {\bibinfo {author} {\bibfnamefont {L.}~\bibnamefont
  {Visinelli}},\ }\href {\doibase 10.3390/sym10110546} {\bibfield  {journal}
  {\bibinfo  {journal} {Symmetry}\ }\textbf {\bibinfo {volume} {10}},\ \bibinfo
  {pages} {546} (\bibinfo {year} {2018})},\ \Eprint
  {http://arxiv.org/abs/1710.11006} {arXiv:1710.11006 [astro-ph.CO]}
  \BibitemShut {NoStop}%
\bibitem [{\citenamefont {Ackermann}\ \emph
  {et~al.}(2015{\natexlab{a}})\citenamefont {Ackermann} \emph
  {et~al.}}]{Fermi:Constraints}%
  \BibitemOpen
  \bibfield  {author} {\bibinfo {author} {\bibfnamefont {M.}~\bibnamefont
  {Ackermann}} \emph {et~al.} (\bibinfo {collaboration} {Fermi-LAT}),\ }\href
  {\doibase 10.1103/PhysRevLett.115.231301} {\bibfield  {journal} {\bibinfo
  {journal} {Phys. Rev. Lett.}\ }\textbf {\bibinfo {volume} {115}},\ \bibinfo
  {pages} {231301} (\bibinfo {year} {2015}{\natexlab{a}})},\ \Eprint
  {http://arxiv.org/abs/1503.02641} {arXiv:1503.02641 [astro-ph.HE]}
  \BibitemShut {NoStop}%
%%CITATION = ARXIV:1503.02641;%%
\bibitem [{\citenamefont {Lefranc}\ and\ \citenamefont
  {Moulin}(2016)}]{HESS:Constraints}%
  \BibitemOpen
  \bibfield  {author} {\bibinfo {author} {\bibfnamefont {V.}~\bibnamefont
  {Lefranc}}\ and\ \bibinfo {author} {\bibfnamefont {E.}~\bibnamefont {Moulin}}
  (\bibinfo {collaboration} {H.E.S.S.}),\ }\bibfield  {booktitle} {\emph
  {\bibinfo {booktitle} {{Proceedings, 34th International Cosmic Ray Conference
  (ICRC 2015): The Hague, The Netherlands, July 30-August 6, 2015}}},\
  }\href@noop {} {\bibfield  {journal} {\bibinfo  {journal} {PoS}\ }\textbf
  {\bibinfo {volume} {ICRC2015}},\ \bibinfo {pages} {1208} (\bibinfo {year}
  {2016})},\ \Eprint {http://arxiv.org/abs/1509.04123} {arXiv:1509.04123
  [astro-ph.HE]} \BibitemShut {NoStop}%
%%CITATION = ARXIV:1509.04123;%%
\bibitem [{\citenamefont {Profumo}\ \emph {et~al.}(2006)\citenamefont
  {Profumo}, \citenamefont {Sigurdson},\ and\ \citenamefont
  {Kamionkowski}}]{Profumo:2006}%
  \BibitemOpen
  \bibfield  {author} {\bibinfo {author} {\bibfnamefont {S.}~\bibnamefont
  {Profumo}}, \bibinfo {author} {\bibfnamefont {K.}~\bibnamefont {Sigurdson}},
  \ and\ \bibinfo {author} {\bibfnamefont {M.}~\bibnamefont {Kamionkowski}},\
  }\href {\doibase 10.1103/PhysRevLett.97.031301} {\bibfield  {journal}
  {\bibinfo  {journal} {Phys. Rev. Lett.}\ }\textbf {\bibinfo {volume} {97}},\
  \bibinfo {pages} {031301} (\bibinfo {year} {2006})},\ \Eprint
  {http://arxiv.org/abs/astro-ph/0603373} {arXiv:astro-ph/0603373 [astro-ph]}
  \BibitemShut {NoStop}%
%%CITATION = ASTRO-PH/0603373;%%
\bibitem [{\citenamefont {Cornell}\ \emph {et~al.}(2013)\citenamefont
  {Cornell}, \citenamefont {Profumo},\ and\ \citenamefont
  {Shepherd}}]{Cornell:2013rza}%
  \BibitemOpen
  \bibfield  {author} {\bibinfo {author} {\bibfnamefont {J.~M.}\ \bibnamefont
  {Cornell}}, \bibinfo {author} {\bibfnamefont {S.}~\bibnamefont {Profumo}}, \
  and\ \bibinfo {author} {\bibfnamefont {W.}~\bibnamefont {Shepherd}},\ }\href
  {\doibase 10.1103/PhysRevD.88.015027} {\bibfield  {journal} {\bibinfo
  {journal} {Phys. Rev. D}\ }\textbf {\bibinfo {volume} {88}},\ \bibinfo
  {pages} {015027} (\bibinfo {year} {2013})},\ \Eprint
  {http://arxiv.org/abs/1305.4676} {arXiv:1305.4676 [hep-ph]} \BibitemShut
  {NoStop}%
\bibitem [{\citenamefont {Hu}\ and\ \citenamefont
  {Sugiyama}(1996)}]{Hu:1995Meszaros}%
  \BibitemOpen
  \bibfield  {author} {\bibinfo {author} {\bibfnamefont {W.}~\bibnamefont
  {Hu}}\ and\ \bibinfo {author} {\bibfnamefont {N.}~\bibnamefont {Sugiyama}},\
  }\href {\doibase 10.1086/177989} {\bibfield  {journal} {\bibinfo  {journal}
  {Astrophys. J.}\ }\textbf {\bibinfo {volume} {471}},\ \bibinfo {pages} {542}
  (\bibinfo {year} {1996})},\ \Eprint {http://arxiv.org/abs/astro-ph/9510117}
  {arXiv:astro-ph/9510117 [astro-ph]} \BibitemShut {NoStop}%
%%CITATION = ASTRO-PH/9510117;%%
\bibitem [{\citenamefont {Meszaros}(1974)}]{Meszaros:1974}%
  \BibitemOpen
  \bibfield  {author} {\bibinfo {author} {\bibfnamefont {P.}~\bibnamefont
  {Meszaros}},\ }\href@noop {} {\bibfield  {journal} {\bibinfo  {journal}
  {Astron. Astrophys.}\ }\textbf {\bibinfo {volume} {37}},\ \bibinfo {pages}
  {225} (\bibinfo {year} {1974})}\BibitemShut {NoStop}%
%%CITATION = AAEJA,37,225;%%
\bibitem [{\citenamefont {Lewis}\ and\ \citenamefont
  {Challinor}(2007)}]{Lewis:2007CAMB}%
  \BibitemOpen
  \bibfield  {author} {\bibinfo {author} {\bibfnamefont {A.}~\bibnamefont
  {Lewis}}\ and\ \bibinfo {author} {\bibfnamefont {A.}~\bibnamefont
  {Challinor}},\ }\href {\doibase 10.1103/PhysRevD.76.083005} {\bibfield
  {journal} {\bibinfo  {journal} {Phys. Rev.}\ }\textbf {\bibinfo {volume}
  {D76}},\ \bibinfo {pages} {083005} (\bibinfo {year} {2007})},\ \Eprint
  {http://arxiv.org/abs/astro-ph/0702600} {arXiv:astro-ph/0702600 [ASTRO-PH]}
  \BibitemShut {NoStop}%
%%CITATION = ASTRO-PH/0702600;%%
\bibitem [{\citenamefont {Eisenstein}\ and\ \citenamefont
  {Hu}(1998)}]{Eisenstein:1997}%
  \BibitemOpen
  \bibfield  {author} {\bibinfo {author} {\bibfnamefont {D.~J.}\ \bibnamefont
  {Eisenstein}}\ and\ \bibinfo {author} {\bibfnamefont {W.}~\bibnamefont
  {Hu}},\ }\href {\doibase 10.1086/305424} {\bibfield  {journal} {\bibinfo
  {journal} {Astrophys. J.}\ }\textbf {\bibinfo {volume} {496}},\ \bibinfo
  {pages} {605} (\bibinfo {year} {1998})},\ \Eprint
  {http://arxiv.org/abs/astro-ph/9709112} {arXiv:astro-ph/9709112 [astro-ph]}
  \BibitemShut {NoStop}%
%%CITATION = ASTRO-PH/9709112;%%
\bibitem [{\citenamefont {Aghanim}\ \emph {et~al.}(2020)\citenamefont {Aghanim}
  \emph {et~al.}}]{Planck:2018vyg}%
  \BibitemOpen
  \bibfield  {author} {\bibinfo {author} {\bibfnamefont {N.}~\bibnamefont
  {Aghanim}} \emph {et~al.} (\bibinfo {collaboration} {Planck}),\ }\href
  {\doibase 10.1051/0004-6361/201833910} {\bibfield  {journal} {\bibinfo
  {journal} {Astron. Astrophys.}\ }\textbf {\bibinfo {volume} {641}},\ \bibinfo
  {pages} {A6} (\bibinfo {year} {2020})},\ \bibinfo {note} {[Erratum:
  Astron.Astrophys. 652, C4 (2021)]},\ \Eprint
  {http://arxiv.org/abs/1807.06209} {arXiv:1807.06209 [astro-ph.CO]}
  \BibitemShut {NoStop}%
\bibitem [{\citenamefont {Bertschinger}(2006)}]{Bertschinger:2006}%
  \BibitemOpen
  \bibfield  {author} {\bibinfo {author} {\bibfnamefont {E.}~\bibnamefont
  {Bertschinger}},\ }\href {\doibase 10.1103/PhysRevD.74.063509} {\bibfield
  {journal} {\bibinfo  {journal} {Phys. Rev.}\ }\textbf {\bibinfo {volume}
  {D74}},\ \bibinfo {pages} {063509} (\bibinfo {year} {2006})},\ \Eprint
  {http://arxiv.org/abs/astro-ph/0607319} {arXiv:astro-ph/0607319 [astro-ph]}
  \BibitemShut {NoStop}%
%%CITATION = ASTRO-PH/0607319;%%
\bibitem [{\citenamefont {Green}\ \emph {et~al.}(2005)\citenamefont {Green},
  \citenamefont {Hofmann},\ and\ \citenamefont {Schwarz}}]{Green:2005fa}%
  \BibitemOpen
  \bibfield  {author} {\bibinfo {author} {\bibfnamefont {A.~M.}\ \bibnamefont
  {Green}}, \bibinfo {author} {\bibfnamefont {S.}~\bibnamefont {Hofmann}}, \
  and\ \bibinfo {author} {\bibfnamefont {D.~J.}\ \bibnamefont {Schwarz}},\
  }\href {\doibase 10.1088/1475-7516/2005/08/003} {\bibfield  {journal}
  {\bibinfo  {journal} {J. Cosmol. Astropart. Phys.}\ }\textbf {\bibinfo
  {volume} {08}},\ \bibinfo {pages} {003} (\bibinfo {year} {2005})},\ \Eprint
  {http://arxiv.org/abs/astro-ph/0503387} {arXiv:astro-ph/0503387} \BibitemShut
  {NoStop}%
\bibitem [{\citenamefont {Loeb}\ and\ \citenamefont
  {Zaldarriaga}(2005)}]{Loeb:2005pm}%
  \BibitemOpen
  \bibfield  {author} {\bibinfo {author} {\bibfnamefont {A.}~\bibnamefont
  {Loeb}}\ and\ \bibinfo {author} {\bibfnamefont {M.}~\bibnamefont
  {Zaldarriaga}},\ }\href {\doibase 10.1103/PhysRevD.71.103520} {\bibfield
  {journal} {\bibinfo  {journal} {Phys. Rev. D}\ }\textbf {\bibinfo {volume}
  {71}},\ \bibinfo {pages} {103520} (\bibinfo {year} {2005})},\ \Eprint
  {http://arxiv.org/abs/astro-ph/0504112} {arXiv:astro-ph/0504112} \BibitemShut
  {NoStop}%
\bibitem [{\citenamefont {Bringmann}\ and\ \citenamefont
  {Hofmann}(2007)}]{Bringmann:2006mu}%
  \BibitemOpen
  \bibfield  {author} {\bibinfo {author} {\bibfnamefont {T.}~\bibnamefont
  {Bringmann}}\ and\ \bibinfo {author} {\bibfnamefont {S.}~\bibnamefont
  {Hofmann}},\ }\href {\doibase 10.1088/1475-7516/2007/04/016} {\bibfield
  {journal} {\bibinfo  {journal} {J. Cosmol. Astropart. Phys.}\ }\textbf
  {\bibinfo {volume} {04}},\ \bibinfo {pages} {016} (\bibinfo {year} {2007})},\
  \bibinfo {note} {[Erratum: JCAP 03, E02 (2016)]},\ \Eprint
  {http://arxiv.org/abs/hep-ph/0612238} {arXiv:hep-ph/0612238} \BibitemShut
  {NoStop}%
\bibitem [{\citenamefont {Gelmini}\ and\ \citenamefont
  {Gondolo}(2008)}]{Gelmini:2008KD}%
  \BibitemOpen
  \bibfield  {author} {\bibinfo {author} {\bibfnamefont {G.~B.}\ \bibnamefont
  {Gelmini}}\ and\ \bibinfo {author} {\bibfnamefont {P.}~\bibnamefont
  {Gondolo}},\ }\href {\doibase 10.1088/1475-7516/2008/10/002} {\bibfield
  {journal} {\bibinfo  {journal} {J. Cosmol. Astropart. Phys.}\ }\textbf
  {\bibinfo {volume} {0810}},\ \bibinfo {pages} {002} (\bibinfo {year}
  {2008})},\ \Eprint {http://arxiv.org/abs/0803.2349} {arXiv:0803.2349
  [astro-ph]} \BibitemShut {NoStop}%
%%CITATION = ARXIV:0803.2349;%%
\bibitem [{\citenamefont {Kolb}\ \emph {et~al.}(1985)\citenamefont {Kolb},
  \citenamefont {Seckel},\ and\ \citenamefont {Turner}}]{KST85}%
  \BibitemOpen
  \bibfield  {author} {\bibinfo {author} {\bibfnamefont {E.~W.}\ \bibnamefont
  {Kolb}}, \bibinfo {author} {\bibfnamefont {D.}~\bibnamefont {Seckel}}, \ and\
  \bibinfo {author} {\bibfnamefont {M.~S.}\ \bibnamefont {Turner}},\ }\href
  {\doibase 10.1038/314415a0} {\bibfield  {journal} {\bibinfo  {journal}
  {Nature}\ }\textbf {\bibinfo {volume} {314}},\ \bibinfo {pages} {415}
  (\bibinfo {year} {1985})}\BibitemShut {NoStop}%
\bibitem [{\citenamefont {Hodges}(1993)}]{PhysRevD.47.456}%
  \BibitemOpen
  \bibfield  {author} {\bibinfo {author} {\bibfnamefont {H.~M.}\ \bibnamefont
  {Hodges}},\ }\href {\doibase 10.1103/PhysRevD.47.456} {\bibfield  {journal}
  {\bibinfo  {journal} {Phys. Rev. D}\ }\textbf {\bibinfo {volume} {47}},\
  \bibinfo {pages} {456} (\bibinfo {year} {1993})}\BibitemShut {NoStop}%
\bibitem [{\citenamefont {Chen}\ and\ \citenamefont {Tye}(2006)}]{Chen:2006ni}%
  \BibitemOpen
  \bibfield  {author} {\bibinfo {author} {\bibfnamefont {X.}~\bibnamefont
  {Chen}}\ and\ \bibinfo {author} {\bibfnamefont {S.~H.~H.}\ \bibnamefont
  {Tye}},\ }\href {\doibase 10.1088/1475-7516/2006/06/011} {\bibfield
  {journal} {\bibinfo  {journal} {J. Cosmol. Astropart. Phys.}\ }\textbf
  {\bibinfo {volume} {06}},\ \bibinfo {pages} {011} (\bibinfo {year} {2006})},\
  \Eprint {http://arxiv.org/abs/hep-th/0602136} {arXiv:hep-th/0602136}
  \BibitemShut {NoStop}%
\bibitem [{\citenamefont {Feng}\ \emph {et~al.}(2008)\citenamefont {Feng},
  \citenamefont {Tu},\ and\ \citenamefont {Yu}}]{Feng:2008mu}%
  \BibitemOpen
  \bibfield  {author} {\bibinfo {author} {\bibfnamefont {J.~L.}\ \bibnamefont
  {Feng}}, \bibinfo {author} {\bibfnamefont {H.}~\bibnamefont {Tu}}, \ and\
  \bibinfo {author} {\bibfnamefont {H.-B.}\ \bibnamefont {Yu}},\ }\href
  {\doibase 10.1088/1475-7516/2008/10/043} {\bibfield  {journal} {\bibinfo
  {journal} {J. Cosmol. Astropart. Phys.}\ }\textbf {\bibinfo {volume} {10}},\
  \bibinfo {pages} {043} (\bibinfo {year} {2008})},\ \Eprint
  {http://arxiv.org/abs/0808.2318} {arXiv:0808.2318 [hep-ph]} \BibitemShut
  {NoStop}%
\bibitem [{\citenamefont {Berlin}\ \emph
  {et~al.}(2016{\natexlab{a}})\citenamefont {Berlin}, \citenamefont {Hooper},\
  and\ \citenamefont {Krnjaic}}]{Berlin:2016First}%
  \BibitemOpen
  \bibfield  {author} {\bibinfo {author} {\bibfnamefont {A.}~\bibnamefont
  {Berlin}}, \bibinfo {author} {\bibfnamefont {D.}~\bibnamefont {Hooper}}, \
  and\ \bibinfo {author} {\bibfnamefont {G.}~\bibnamefont {Krnjaic}},\ }\href
  {\doibase 10.1016/j.physletb.2016.06.037} {\bibfield  {journal} {\bibinfo
  {journal} {Phys. Lett.}\ }\textbf {\bibinfo {volume} {B760}},\ \bibinfo
  {pages} {106} (\bibinfo {year} {2016}{\natexlab{a}})},\ \Eprint
  {http://arxiv.org/abs/1602.08490} {arXiv:1602.08490 [hep-ph]} \BibitemShut
  {NoStop}%
%%CITATION = ARXIV:1602.08490;%%
\bibitem [{\citenamefont {Berlin}\ \emph
  {et~al.}(2016{\natexlab{b}})\citenamefont {Berlin}, \citenamefont {Hooper},\
  and\ \citenamefont {Krnjaic}}]{Berlin:2016}%
  \BibitemOpen
  \bibfield  {author} {\bibinfo {author} {\bibfnamefont {A.}~\bibnamefont
  {Berlin}}, \bibinfo {author} {\bibfnamefont {D.}~\bibnamefont {Hooper}}, \
  and\ \bibinfo {author} {\bibfnamefont {G.}~\bibnamefont {Krnjaic}},\ }\href
  {\doibase 10.1103/PhysRevD.94.095019} {\bibfield  {journal} {\bibinfo
  {journal} {Phys. Rev.}\ }\textbf {\bibinfo {volume} {D94}},\ \bibinfo {pages}
  {095019} (\bibinfo {year} {2016}{\natexlab{b}})},\ \Eprint
  {http://arxiv.org/abs/1609.02555} {arXiv:1609.02555 [hep-ph]} \BibitemShut
  {NoStop}%
%%CITATION = ARXIV:1609.02555;%%
\bibitem [{\citenamefont {Adshead}\ \emph {et~al.}(2016)\citenamefont
  {Adshead}, \citenamefont {Cui},\ and\ \citenamefont
  {Shelton}}]{Adshead:2016xxj}%
  \BibitemOpen
  \bibfield  {author} {\bibinfo {author} {\bibfnamefont {P.}~\bibnamefont
  {Adshead}}, \bibinfo {author} {\bibfnamefont {Y.}~\bibnamefont {Cui}}, \ and\
  \bibinfo {author} {\bibfnamefont {J.}~\bibnamefont {Shelton}},\ }\href
  {\doibase 10.1007/JHEP06(2016)016} {\bibfield  {journal} {\bibinfo  {journal}
  {JHEP}\ }\textbf {\bibinfo {volume} {06}},\ \bibinfo {pages} {016} (\bibinfo
  {year} {2016})},\ \Eprint {http://arxiv.org/abs/1604.02458} {arXiv:1604.02458
  [hep-ph]} \BibitemShut {NoStop}%
\bibitem [{\citenamefont {Press}\ and\ \citenamefont
  {Schechter}(1974)}]{Press:1973}%
  \BibitemOpen
  \bibfield  {author} {\bibinfo {author} {\bibfnamefont {W.~H.}\ \bibnamefont
  {Press}}\ and\ \bibinfo {author} {\bibfnamefont {P.}~\bibnamefont
  {Schechter}},\ }\href {\doibase 10.1086/152650} {\bibfield  {journal}
  {\bibinfo  {journal} {Astrophys. J.}\ }\textbf {\bibinfo {volume} {187}},\
  \bibinfo {pages} {425} (\bibinfo {year} {1974})}\BibitemShut {NoStop}%
%%CITATION = ASJOA,187,425;%%
\bibitem [{\citenamefont {Benson}\ \emph {et~al.}(2013)\citenamefont {Benson},
  \citenamefont {Farahi}, \citenamefont {Cole}, \citenamefont {Moustakas},
  \citenamefont {Jenkins}, \citenamefont {Lovell}, \citenamefont {Kennedy},
  \citenamefont {Helly},\ and\ \citenamefont {Frenk}}]{Benson:2012}%
  \BibitemOpen
  \bibfield  {author} {\bibinfo {author} {\bibfnamefont {A.~J.}\ \bibnamefont
  {Benson}}, \bibinfo {author} {\bibfnamefont {A.}~\bibnamefont {Farahi}},
  \bibinfo {author} {\bibfnamefont {S.}~\bibnamefont {Cole}}, \bibinfo {author}
  {\bibfnamefont {L.~A.}\ \bibnamefont {Moustakas}}, \bibinfo {author}
  {\bibfnamefont {A.}~\bibnamefont {Jenkins}}, \bibinfo {author} {\bibfnamefont
  {M.}~\bibnamefont {Lovell}}, \bibinfo {author} {\bibfnamefont
  {R.}~\bibnamefont {Kennedy}}, \bibinfo {author} {\bibfnamefont
  {J.}~\bibnamefont {Helly}}, \ and\ \bibinfo {author} {\bibfnamefont
  {C.}~\bibnamefont {Frenk}},\ }\href {\doibase 10.1093/mnras/sts159}
  {\bibfield  {journal} {\bibinfo  {journal} {Mon. Not. R. Astron. Soc.}\
  }\textbf {\bibinfo {volume} {428}},\ \bibinfo {pages} {1774} (\bibinfo {year}
  {2013})},\ \Eprint {http://arxiv.org/abs/1209.3018} {arXiv:1209.3018
  [astro-ph.CO]} \BibitemShut {NoStop}%
%%CITATION = ARXIV:1209.3018;%%
\bibitem [{\citenamefont {Blanco}\ and\ \citenamefont
  {Hooper}(2019)}]{Blanco:2018esa}%
  \BibitemOpen
  \bibfield  {author} {\bibinfo {author} {\bibfnamefont {C.}~\bibnamefont
  {Blanco}}\ and\ \bibinfo {author} {\bibfnamefont {D.}~\bibnamefont
  {Hooper}},\ }\href {\doibase 10.1088/1475-7516/2019/03/019} {\bibfield
  {journal} {\bibinfo  {journal} {J. Cosmol. Astropart. Phys.}\ }\textbf
  {\bibinfo {volume} {03}},\ \bibinfo {pages} {019} (\bibinfo {year} {2019})},\
  \Eprint {http://arxiv.org/abs/1811.05988} {arXiv:1811.05988 [astro-ph.HE]}
  \BibitemShut {NoStop}%
\bibitem [{\citenamefont {Ackermann}\ \emph
  {et~al.}(2015{\natexlab{b}})\citenamefont {Ackermann} \emph
  {et~al.}}]{Fermi-LAT:2014ryh}%
  \BibitemOpen
  \bibfield  {author} {\bibinfo {author} {\bibfnamefont {M.}~\bibnamefont
  {Ackermann}} \emph {et~al.} (\bibinfo {collaboration} {Fermi-LAT}),\ }\href
  {\doibase 10.1088/0004-637X/799/1/86} {\bibfield  {journal} {\bibinfo
  {journal} {Astrophys. J.}\ }\textbf {\bibinfo {volume} {799}},\ \bibinfo
  {pages} {86} (\bibinfo {year} {2015}{\natexlab{b}})},\ \Eprint
  {http://arxiv.org/abs/1410.3696} {arXiv:1410.3696 [astro-ph.HE]} \BibitemShut
  {NoStop}%
\bibitem [{\citenamefont {Ackermann}\ \emph
  {et~al.}(2015{\natexlab{c}})\citenamefont {Ackermann} \emph
  {et~al.}}]{Fermi-LAT:2015qzw}%
  \BibitemOpen
  \bibfield  {author} {\bibinfo {author} {\bibfnamefont {M.}~\bibnamefont
  {Ackermann}} \emph {et~al.} (\bibinfo {collaboration} {Fermi-LAT}),\ }\href
  {\doibase 10.1088/1475-7516/2015/09/008} {\bibfield  {journal} {\bibinfo
  {journal} {J. Cosmol. Astropart. Phys.}\ }\textbf {\bibinfo {volume} {09}},\
  \bibinfo {pages} {008} (\bibinfo {year} {2015}{\natexlab{c}})},\ \Eprint
  {http://arxiv.org/abs/1501.05464} {arXiv:1501.05464 [astro-ph.CO]}
  \BibitemShut {NoStop}%
\bibitem [{\citenamefont {Delos}\ \emph
  {et~al.}(2019{\natexlab{b}})\citenamefont {Delos}, \citenamefont {Bruff},\
  and\ \citenamefont {Erickcek}}]{Delos:2019mxl}%
  \BibitemOpen
  \bibfield  {author} {\bibinfo {author} {\bibfnamefont {M.~S.}\ \bibnamefont
  {Delos}}, \bibinfo {author} {\bibfnamefont {M.}~\bibnamefont {Bruff}}, \ and\
  \bibinfo {author} {\bibfnamefont {A.~L.}\ \bibnamefont {Erickcek}},\ }\href
  {\doibase 10.1103/PhysRevD.100.023523} {\bibfield  {journal} {\bibinfo
  {journal} {Phys. Rev. D}\ }\textbf {\bibinfo {volume} {100}},\ \bibinfo
  {pages} {023523} (\bibinfo {year} {2019}{\natexlab{b}})},\ \Eprint
  {http://arxiv.org/abs/1905.05766} {arXiv:1905.05766 [astro-ph.CO]}
  \BibitemShut {NoStop}%
\bibitem [{\citenamefont {Delos}\ and\ \citenamefont
  {White}(2022{\natexlab{a}})}]{Delos:2022yhn}%
  \BibitemOpen
  \bibfield  {author} {\bibinfo {author} {\bibfnamefont {M.~S.}\ \bibnamefont
  {Delos}}\ and\ \bibinfo {author} {\bibfnamefont {S.~D.~M.}\ \bibnamefont
  {White}},\ }\href {\doibase 10.1093/mnras/stac3373} {\bibfield  {journal}
  {\bibinfo  {journal} {Mon. Not. R. Astron. Soc.}\ }\textbf {\bibinfo {volume}
  {518}},\ \bibinfo {pages} {3509} (\bibinfo {year} {2022}{\natexlab{a}})},\
  \Eprint {http://arxiv.org/abs/2207.05082} {arXiv:2207.05082 [astro-ph.CO]}
  \BibitemShut {NoStop}%
\bibitem [{\citenamefont {Delos}\ and\ \citenamefont
  {Linden}(2022)}]{Delos:2021rqs}%
  \BibitemOpen
  \bibfield  {author} {\bibinfo {author} {\bibfnamefont {M.~S.}\ \bibnamefont
  {Delos}}\ and\ \bibinfo {author} {\bibfnamefont {T.}~\bibnamefont {Linden}},\
  }\href {\doibase 10.1103/PhysRevD.105.123514} {\bibfield  {journal} {\bibinfo
   {journal} {Phys. Rev. D}\ }\textbf {\bibinfo {volume} {105}},\ \bibinfo
  {pages} {123514} (\bibinfo {year} {2022})},\ \Eprint
  {http://arxiv.org/abs/2109.03240} {arXiv:2109.03240 [astro-ph.CO]}
  \BibitemShut {NoStop}%
\bibitem [{\citenamefont {White}(2022)}]{White:2022yoc}%
  \BibitemOpen
  \bibfield  {author} {\bibinfo {author} {\bibfnamefont {S.~D.~M.}\
  \bibnamefont {White}},\ }\href {\doibase 10.1093/mnrasl/slac107} {\bibfield
  {journal} {\bibinfo  {journal} {Mon. Not. Roy. Astron. Soc.}\ }\textbf
  {\bibinfo {volume} {517}},\ \bibinfo {pages} {L46} (\bibinfo {year}
  {2022})},\ \Eprint {http://arxiv.org/abs/2207.13565} {arXiv:2207.13565
  [astro-ph.CO]} \BibitemShut {NoStop}%
\bibitem [{\citenamefont {Delos}\ and\ \citenamefont
  {White}(2022{\natexlab{b}})}]{Delos:2022bhp}%
  \BibitemOpen
  \bibfield  {author} {\bibinfo {author} {\bibfnamefont {M.~S.}\ \bibnamefont
  {Delos}}\ and\ \bibinfo {author} {\bibfnamefont {S.~D.~M.}\ \bibnamefont
  {White}},\ }\href@noop {} {\  (\bibinfo {year} {2022}{\natexlab{b}})},\
  \Eprint {http://arxiv.org/abs/2209.11237} {arXiv:2209.11237 [astro-ph.CO]}
  \BibitemShut {NoStop}%
\bibitem [{\citenamefont {{Navarro}}\ \emph {et~al.}(1996)\citenamefont
  {{Navarro}}, \citenamefont {{Frenk}},\ and\ \citenamefont
  {{White}}}]{1996ApJ...462..563N}%
  \BibitemOpen
  \bibfield  {author} {\bibinfo {author} {\bibfnamefont {J.~F.}\ \bibnamefont
  {{Navarro}}}, \bibinfo {author} {\bibfnamefont {C.~S.}\ \bibnamefont
  {{Frenk}}}, \ and\ \bibinfo {author} {\bibfnamefont {S.~D.~M.}\ \bibnamefont
  {{White}}},\ }\href {\doibase 10.1086/177173} {\bibfield  {journal} {\bibinfo
   {journal} {\apj}\ }\textbf {\bibinfo {volume} {462}},\ \bibinfo {pages}
  {563} (\bibinfo {year} {1996})},\ \Eprint
  {http://arxiv.org/abs/astro-ph/9508025} {arXiv:astro-ph/9508025 [astro-ph]}
  \BibitemShut {NoStop}%
\bibitem [{\citenamefont {{Navarro}}\ \emph {et~al.}(1997)\citenamefont
  {{Navarro}}, \citenamefont {{Frenk}},\ and\ \citenamefont
  {{White}}}]{1997ApJ...490..493N}%
  \BibitemOpen
  \bibfield  {author} {\bibinfo {author} {\bibfnamefont {J.~F.}\ \bibnamefont
  {{Navarro}}}, \bibinfo {author} {\bibfnamefont {C.~S.}\ \bibnamefont
  {{Frenk}}}, \ and\ \bibinfo {author} {\bibfnamefont {S.~D.~M.}\ \bibnamefont
  {{White}}},\ }\href {\doibase 10.1086/304888} {\bibfield  {journal} {\bibinfo
   {journal} {\apj}\ }\textbf {\bibinfo {volume} {490}},\ \bibinfo {pages}
  {493} (\bibinfo {year} {1997})},\ \Eprint
  {http://arxiv.org/abs/astro-ph/9611107} {arXiv:astro-ph/9611107 [astro-ph]}
  \BibitemShut {NoStop}%
\bibitem [{\citenamefont {{Ogiya}}\ \emph {et~al.}(2016)\citenamefont
  {{Ogiya}}, \citenamefont {{Nagai}},\ and\ \citenamefont
  {{Ishiyama}}}]{2016MNRAS.461.3385O}%
  \BibitemOpen
  \bibfield  {author} {\bibinfo {author} {\bibfnamefont {G.}~\bibnamefont
  {{Ogiya}}}, \bibinfo {author} {\bibfnamefont {D.}~\bibnamefont {{Nagai}}}, \
  and\ \bibinfo {author} {\bibfnamefont {T.}~\bibnamefont {{Ishiyama}}},\
  }\href {\doibase 10.1093/mnras/stw1551} {\bibfield  {journal} {\bibinfo
  {journal} {\mnras}\ }\textbf {\bibinfo {volume} {461}},\ \bibinfo {pages}
  {3385} (\bibinfo {year} {2016})},\ \Eprint {http://arxiv.org/abs/1604.02866}
  {arXiv:1604.02866 [astro-ph.CO]} \BibitemShut {NoStop}%
\bibitem [{\citenamefont {{Gosenca}}\ \emph {et~al.}(2017)\citenamefont
  {{Gosenca}}, \citenamefont {{Adamek}}, \citenamefont {{Byrnes}},\ and\
  \citenamefont {{Hotchkiss}}}]{2017PhRvD..96l3519G}%
  \BibitemOpen
  \bibfield  {author} {\bibinfo {author} {\bibfnamefont {M.}~\bibnamefont
  {{Gosenca}}}, \bibinfo {author} {\bibfnamefont {J.}~\bibnamefont {{Adamek}}},
  \bibinfo {author} {\bibfnamefont {C.~T.}\ \bibnamefont {{Byrnes}}}, \ and\
  \bibinfo {author} {\bibfnamefont {S.}~\bibnamefont {{Hotchkiss}}},\ }\href
  {\doibase 10.1103/PhysRevD.96.123519} {\bibfield  {journal} {\bibinfo
  {journal} {\prd}\ }\textbf {\bibinfo {volume} {96}},\ \bibinfo {eid} {123519}
  (\bibinfo {year} {2017})},\ \Eprint {http://arxiv.org/abs/1710.02055}
  {arXiv:1710.02055 [astro-ph.CO]} \BibitemShut {NoStop}%
\bibitem [{\citenamefont {{Angulo}}\ \emph {et~al.}(2017)\citenamefont
  {{Angulo}}, \citenamefont {{Hahn}}, \citenamefont {{Ludlow}},\ and\
  \citenamefont {{Bonoli}}}]{2017MNRAS.471.4687A}%
  \BibitemOpen
  \bibfield  {author} {\bibinfo {author} {\bibfnamefont {R.~E.}\ \bibnamefont
  {{Angulo}}}, \bibinfo {author} {\bibfnamefont {O.}~\bibnamefont {{Hahn}}},
  \bibinfo {author} {\bibfnamefont {A.~D.}\ \bibnamefont {{Ludlow}}}, \ and\
  \bibinfo {author} {\bibfnamefont {S.}~\bibnamefont {{Bonoli}}},\ }\href
  {\doibase 10.1093/mnras/stx1658} {\bibfield  {journal} {\bibinfo  {journal}
  {\mnras}\ }\textbf {\bibinfo {volume} {471}},\ \bibinfo {pages} {4687}
  (\bibinfo {year} {2017})},\ \Eprint {http://arxiv.org/abs/1604.03131}
  {arXiv:1604.03131 [astro-ph.CO]} \BibitemShut {NoStop}%
\bibitem [{\citenamefont {{Ogiya}}\ and\ \citenamefont
  {{Hahn}}(2018)}]{2018MNRAS.473.4339O}%
  \BibitemOpen
  \bibfield  {author} {\bibinfo {author} {\bibfnamefont {G.}~\bibnamefont
  {{Ogiya}}}\ and\ \bibinfo {author} {\bibfnamefont {O.}~\bibnamefont
  {{Hahn}}},\ }\href {\doibase 10.1093/mnras/stx2639} {\bibfield  {journal}
  {\bibinfo  {journal} {\mnras}\ }\textbf {\bibinfo {volume} {473}},\ \bibinfo
  {pages} {4339} (\bibinfo {year} {2018})},\ \Eprint
  {http://arxiv.org/abs/1707.07693} {arXiv:1707.07693 [astro-ph.CO]}
  \BibitemShut {NoStop}%
\bibitem [{\citenamefont {{Ishiyama}}\ and\ \citenamefont
  {{Ando}}(2020)}]{2020MNRAS.492.3662I}%
  \BibitemOpen
  \bibfield  {author} {\bibinfo {author} {\bibfnamefont {T.}~\bibnamefont
  {{Ishiyama}}}\ and\ \bibinfo {author} {\bibfnamefont {S.}~\bibnamefont
  {{Ando}}},\ }\href {\doibase 10.1093/mnras/staa069} {\bibfield  {journal}
  {\bibinfo  {journal} {\mnras}\ }\textbf {\bibinfo {volume} {492}},\ \bibinfo
  {pages} {3662} (\bibinfo {year} {2020})},\ \Eprint
  {http://arxiv.org/abs/1907.03642} {arXiv:1907.03642 [astro-ph.CO]}
  \BibitemShut {NoStop}%
\bibitem [{\citenamefont {S\'anchez-Conde}\ and\ \citenamefont
  {Prada}(2014)}]{Sanchez-Conde:2013yxa}%
  \BibitemOpen
  \bibfield  {author} {\bibinfo {author} {\bibfnamefont {M.~A.}\ \bibnamefont
  {S\'anchez-Conde}}\ and\ \bibinfo {author} {\bibfnamefont {F.}~\bibnamefont
  {Prada}},\ }\href {\doibase 10.1093/mnras/stu1014} {\bibfield  {journal}
  {\bibinfo  {journal} {Mon. Not. R. Astron. Soc.}\ }\textbf {\bibinfo {volume}
  {442}},\ \bibinfo {pages} {2271} (\bibinfo {year} {2014})},\ \Eprint
  {http://arxiv.org/abs/1312.1729} {arXiv:1312.1729 [astro-ph.CO]} \BibitemShut
  {NoStop}%
\bibitem [{\citenamefont {Drakos}\ \emph {et~al.}(2019)\citenamefont {Drakos},
  \citenamefont {Taylor}, \citenamefont {Berrouet}, \citenamefont {Robotham},\
  and\ \citenamefont {Power}}]{Drakos:2018fgl}%
  \BibitemOpen
  \bibfield  {author} {\bibinfo {author} {\bibfnamefont {N.~E.}\ \bibnamefont
  {Drakos}}, \bibinfo {author} {\bibfnamefont {J.~E.}\ \bibnamefont {Taylor}},
  \bibinfo {author} {\bibfnamefont {A.}~\bibnamefont {Berrouet}}, \bibinfo
  {author} {\bibfnamefont {A.~S.~G.}\ \bibnamefont {Robotham}}, \ and\ \bibinfo
  {author} {\bibfnamefont {C.}~\bibnamefont {Power}},\ }\href {\doibase
  10.1093/mnras/stz1307} {\bibfield  {journal} {\bibinfo  {journal} {Mon. Not.
  R. Astron. Soc.}\ }\textbf {\bibinfo {volume} {487}},\ \bibinfo {pages}
  {1008} (\bibinfo {year} {2019})},\ \Eprint {http://arxiv.org/abs/1811.12844}
  {arXiv:1811.12844 [astro-ph.GA]} \BibitemShut {NoStop}%
\bibitem [{\citenamefont {{Bardeen}}\ \emph {et~al.}(1986)\citenamefont
  {{Bardeen}}, \citenamefont {{Bond}}, \citenamefont {{Kaiser}},\ and\
  \citenamefont {{Szalay}}}]{1986ApJ...304...15B}%
  \BibitemOpen
  \bibfield  {author} {\bibinfo {author} {\bibfnamefont {J.~M.}\ \bibnamefont
  {{Bardeen}}}, \bibinfo {author} {\bibfnamefont {J.~R.}\ \bibnamefont
  {{Bond}}}, \bibinfo {author} {\bibfnamefont {N.}~\bibnamefont {{Kaiser}}}, \
  and\ \bibinfo {author} {\bibfnamefont {A.~S.}\ \bibnamefont {{Szalay}}},\
  }\href {\doibase 10.1086/164143} {\bibfield  {journal} {\bibinfo  {journal}
  {\apj}\ }\textbf {\bibinfo {volume} {304}},\ \bibinfo {pages} {15} (\bibinfo
  {year} {1986})}\BibitemShut {NoStop}%
\bibitem [{\citenamefont {Delos}(2019)}]{Delos:2019lik}%
  \BibitemOpen
  \bibfield  {author} {\bibinfo {author} {\bibfnamefont {M.~S.}\ \bibnamefont
  {Delos}},\ }\href {\doibase 10.1103/PhysRevD.100.063505} {\bibfield
  {journal} {\bibinfo  {journal} {Phys. Rev. D}\ }\textbf {\bibinfo {volume}
  {100}},\ \bibinfo {pages} {063505} (\bibinfo {year} {2019})},\ \Eprint
  {http://arxiv.org/abs/1906.10690} {arXiv:1906.10690 [astro-ph.CO]}
  \BibitemShut {NoStop}%
\bibitem [{\citenamefont {Griest}\ and\ \citenamefont
  {Kamionkowski}(1990)}]{Griest:1989wd}%
  \BibitemOpen
  \bibfield  {author} {\bibinfo {author} {\bibfnamefont {K.}~\bibnamefont
  {Griest}}\ and\ \bibinfo {author} {\bibfnamefont {M.}~\bibnamefont
  {Kamionkowski}},\ }\href {\doibase 10.1103/PhysRevLett.64.615} {\bibfield
  {journal} {\bibinfo  {journal} {Phys. Rev. Lett.}\ }\textbf {\bibinfo
  {volume} {64}},\ \bibinfo {pages} {615} (\bibinfo {year} {1990})}\BibitemShut
  {NoStop}%
\bibitem [{\citenamefont {Abdallah}\ \emph {et~al.}(2016)\citenamefont
  {Abdallah} \emph {et~al.}}]{HESS:2016mib}%
  \BibitemOpen
  \bibfield  {author} {\bibinfo {author} {\bibfnamefont {H.}~\bibnamefont
  {Abdallah}} \emph {et~al.} (\bibinfo {collaboration} {H.E.S.S.}),\ }\href
  {\doibase 10.1103/PhysRevLett.117.111301} {\bibfield  {journal} {\bibinfo
  {journal} {Phys. Rev. Lett.}\ }\textbf {\bibinfo {volume} {117}},\ \bibinfo
  {pages} {111301} (\bibinfo {year} {2016})},\ \Eprint
  {http://arxiv.org/abs/1607.08142} {arXiv:1607.08142 [astro-ph.HE]}
  \BibitemShut {NoStop}%
\bibitem [{\citenamefont {Hall}\ \emph {et~al.}(2010)\citenamefont {Hall},
  \citenamefont {Jedamzik}, \citenamefont {March-Russell},\ and\ \citenamefont
  {West}}]{Hall:2009bx}%
  \BibitemOpen
  \bibfield  {author} {\bibinfo {author} {\bibfnamefont {L.~J.}\ \bibnamefont
  {Hall}}, \bibinfo {author} {\bibfnamefont {K.}~\bibnamefont {Jedamzik}},
  \bibinfo {author} {\bibfnamefont {J.}~\bibnamefont {March-Russell}}, \ and\
  \bibinfo {author} {\bibfnamefont {S.~M.}\ \bibnamefont {West}},\ }\href
  {\doibase 10.1007/JHEP03(2010)080} {\bibfield  {journal} {\bibinfo  {journal}
  {JHEP}\ }\textbf {\bibinfo {volume} {03}},\ \bibinfo {pages} {080} (\bibinfo
  {year} {2010})},\ \Eprint {http://arxiv.org/abs/0911.1120} {arXiv:0911.1120
  [hep-ph]} \BibitemShut {NoStop}%
\bibitem [{\citenamefont {Chung}\ \emph {et~al.}(2001)\citenamefont {Chung},
  \citenamefont {Crotty}, \citenamefont {Kolb},\ and\ \citenamefont
  {Riotto}}]{Chung:2001cb}%
  \BibitemOpen
  \bibfield  {author} {\bibinfo {author} {\bibfnamefont {D.~J.~H.}\
  \bibnamefont {Chung}}, \bibinfo {author} {\bibfnamefont {P.}~\bibnamefont
  {Crotty}}, \bibinfo {author} {\bibfnamefont {E.~W.}\ \bibnamefont {Kolb}}, \
  and\ \bibinfo {author} {\bibfnamefont {A.}~\bibnamefont {Riotto}},\ }\href
  {\doibase 10.1103/PhysRevD.64.043503} {\bibfield  {journal} {\bibinfo
  {journal} {Phys. Rev. D}\ }\textbf {\bibinfo {volume} {64}},\ \bibinfo
  {pages} {043503} (\bibinfo {year} {2001})},\ \Eprint
  {http://arxiv.org/abs/hep-ph/0104100} {arXiv:hep-ph/0104100} \BibitemShut
  {NoStop}%
\bibitem [{\citenamefont {Dai}\ and\ \citenamefont
  {Miralda-Escud\'e}(2020)}]{Dai:2019lud}%
  \BibitemOpen
  \bibfield  {author} {\bibinfo {author} {\bibfnamefont {L.}~\bibnamefont
  {Dai}}\ and\ \bibinfo {author} {\bibfnamefont {J.}~\bibnamefont
  {Miralda-Escud\'e}},\ }\href {\doibase 10.3847/1538-3881/ab5e83} {\bibfield
  {journal} {\bibinfo  {journal} {Astron. J.}\ }\textbf {\bibinfo {volume}
  {159}},\ \bibinfo {pages} {49} (\bibinfo {year} {2020})},\ \Eprint
  {http://arxiv.org/abs/1908.01773} {arXiv:1908.01773 [astro-ph.CO]}
  \BibitemShut {NoStop}%
\bibitem [{\citenamefont {Blinov}\ \emph {et~al.}(2021)\citenamefont {Blinov},
  \citenamefont {Dolan}, \citenamefont {Draper},\ and\ \citenamefont
  {Shelton}}]{Blinov:2021axd}%
  \BibitemOpen
  \bibfield  {author} {\bibinfo {author} {\bibfnamefont {N.}~\bibnamefont
  {Blinov}}, \bibinfo {author} {\bibfnamefont {M.~J.}\ \bibnamefont {Dolan}},
  \bibinfo {author} {\bibfnamefont {P.}~\bibnamefont {Draper}}, \ and\ \bibinfo
  {author} {\bibfnamefont {J.}~\bibnamefont {Shelton}},\ }\href {\doibase
  10.1103/PhysRevD.103.103514} {\bibfield  {journal} {\bibinfo  {journal}
  {Phys. Rev. D}\ }\textbf {\bibinfo {volume} {103}},\ \bibinfo {pages}
  {103514} (\bibinfo {year} {2021})},\ \Eprint
  {http://arxiv.org/abs/2102.05070} {arXiv:2102.05070 [astro-ph.CO]}
  \BibitemShut {NoStop}%
\bibitem [{\citenamefont {Dror}\ \emph {et~al.}(2019)\citenamefont {Dror},
  \citenamefont {Ramani}, \citenamefont {Trickle},\ and\ \citenamefont
  {Zurek}}]{Dror:2019twh}%
  \BibitemOpen
  \bibfield  {author} {\bibinfo {author} {\bibfnamefont {J.~A.}\ \bibnamefont
  {Dror}}, \bibinfo {author} {\bibfnamefont {H.}~\bibnamefont {Ramani}},
  \bibinfo {author} {\bibfnamefont {T.}~\bibnamefont {Trickle}}, \ and\
  \bibinfo {author} {\bibfnamefont {K.~M.}\ \bibnamefont {Zurek}},\ }\href
  {\doibase 10.1103/PhysRevD.100.023003} {\bibfield  {journal} {\bibinfo
  {journal} {Phys. Rev. D}\ }\textbf {\bibinfo {volume} {100}},\ \bibinfo
  {pages} {023003} (\bibinfo {year} {2019})},\ \Eprint
  {http://arxiv.org/abs/1901.04490} {arXiv:1901.04490 [astro-ph.CO]}
  \BibitemShut {NoStop}%
\bibitem [{\citenamefont {Ramani}\ \emph {et~al.}(2020)\citenamefont {Ramani},
  \citenamefont {Trickle},\ and\ \citenamefont {Zurek}}]{Ramani:2020hdo}%
  \BibitemOpen
  \bibfield  {author} {\bibinfo {author} {\bibfnamefont {H.}~\bibnamefont
  {Ramani}}, \bibinfo {author} {\bibfnamefont {T.}~\bibnamefont {Trickle}}, \
  and\ \bibinfo {author} {\bibfnamefont {K.~M.}\ \bibnamefont {Zurek}},\ }\href
  {\doibase 10.1088/1475-7516/2020/12/033} {\bibfield  {journal} {\bibinfo
  {journal} {J. Cosmol. Astropart. Phys.}\ }\textbf {\bibinfo {volume} {12}},\
  \bibinfo {pages} {033} (\bibinfo {year} {2020})},\ \Eprint
  {http://arxiv.org/abs/2005.03030} {arXiv:2005.03030 [astro-ph.CO]}
  \BibitemShut {NoStop}%
\bibitem [{\citenamefont {Lee}\ \emph {et~al.}(2021)\citenamefont {Lee},
  \citenamefont {Mitridate}, \citenamefont {Trickle},\ and\ \citenamefont
  {Zurek}}]{Lee:2020wfn}%
  \BibitemOpen
  \bibfield  {author} {\bibinfo {author} {\bibfnamefont {V.~S.~H.}\
  \bibnamefont {Lee}}, \bibinfo {author} {\bibfnamefont {A.}~\bibnamefont
  {Mitridate}}, \bibinfo {author} {\bibfnamefont {T.}~\bibnamefont {Trickle}},
  \ and\ \bibinfo {author} {\bibfnamefont {K.~M.}\ \bibnamefont {Zurek}},\
  }\href {\doibase 10.1007/JHEP06(2021)028} {\bibfield  {journal} {\bibinfo
  {journal} {JHEP}\ }\textbf {\bibinfo {volume} {06}},\ \bibinfo {pages} {028}
  (\bibinfo {year} {2021})},\ \Eprint {http://arxiv.org/abs/2012.09857}
  {arXiv:2012.09857 [astro-ph.CO]} \BibitemShut {NoStop}%
\bibitem [{\citenamefont {{Berezinsky}}\ \emph {et~al.}(2010)\citenamefont
  {{Berezinsky}}, \citenamefont {{Dokuchaev}}, \citenamefont {{Eroshenko}},
  \citenamefont {{Kachelrie{\ss}}},\ and\ \citenamefont
  {{Solberg}}}]{2010PhRvD..81j3529B}%
  \BibitemOpen
  \bibfield  {author} {\bibinfo {author} {\bibfnamefont {V.}~\bibnamefont
  {{Berezinsky}}}, \bibinfo {author} {\bibfnamefont {V.}~\bibnamefont
  {{Dokuchaev}}}, \bibinfo {author} {\bibfnamefont {Y.}~\bibnamefont
  {{Eroshenko}}}, \bibinfo {author} {\bibfnamefont {M.}~\bibnamefont
  {{Kachelrie{\ss}}}}, \ and\ \bibinfo {author} {\bibfnamefont {M.~A.}\
  \bibnamefont {{Solberg}}},\ }\href {\doibase 10.1103/PhysRevD.81.103529}
  {\bibfield  {journal} {\bibinfo  {journal} {\prd}\ }\textbf {\bibinfo
  {volume} {81}},\ \bibinfo {eid} {103529} (\bibinfo {year} {2010})},\ \Eprint
  {http://arxiv.org/abs/1002.3444} {arXiv:1002.3444 [astro-ph.CO]} \BibitemShut
  {NoStop}%
\bibitem [{\citenamefont {Berezinsky}\ \emph {et~al.}(2013)\citenamefont
  {Berezinsky}, \citenamefont {Dokuchaev},\ and\ \citenamefont
  {Eroshenko}}]{Berezinsky:2013fxa}%
  \BibitemOpen
  \bibfield  {author} {\bibinfo {author} {\bibfnamefont {V.~S.}\ \bibnamefont
  {Berezinsky}}, \bibinfo {author} {\bibfnamefont {V.~I.}\ \bibnamefont
  {Dokuchaev}}, \ and\ \bibinfo {author} {\bibfnamefont {Y.~N.}\ \bibnamefont
  {Eroshenko}},\ }\href {\doibase 10.1088/1475-7516/2013/11/059} {\bibfield
  {journal} {\bibinfo  {journal} {J. Cosmol. Astropart. Phys.}\ }\textbf
  {\bibinfo {volume} {11}},\ \bibinfo {pages} {059} (\bibinfo {year} {2013})},\
  \Eprint {http://arxiv.org/abs/1308.6742} {arXiv:1308.6742 [astro-ph.CO]}
  \BibitemShut {NoStop}%
\bibitem [{\citenamefont {Delos}\ and\ \citenamefont
  {Silk}(2023)}]{StenDelos:2022jld}%
  \BibitemOpen
  \bibfield  {author} {\bibinfo {author} {\bibfnamefont {M.~S.}\ \bibnamefont
  {Delos}}\ and\ \bibinfo {author} {\bibfnamefont {J.}~\bibnamefont {Silk}},\
  }\href {\doibase 10.1093/mnras/stad356} {\bibfield  {journal} {\bibinfo
  {journal} {Mon. Not. R. Astron. Soc.}\ }\textbf {\bibinfo {volume} {520}},\
  \bibinfo {pages} {4370} (\bibinfo {year} {2023})},\ \Eprint
  {http://arxiv.org/abs/2210.04904} {arXiv:2210.04904 [astro-ph.CO]}
  \BibitemShut {NoStop}%
\bibitem [{\citenamefont {Delos}\ and\ \citenamefont
  {Franciolini}(2023)}]{Delos:2023fpm}%
  \BibitemOpen
  \bibfield  {author} {\bibinfo {author} {\bibfnamefont {M.~S.}\ \bibnamefont
  {Delos}}\ and\ \bibinfo {author} {\bibfnamefont {G.}~\bibnamefont
  {Franciolini}},\ }\href {\doibase 10.1103/PhysRevD.107.083505} {\bibfield
  {journal} {\bibinfo  {journal} {Phys. Rev. D}\ }\textbf {\bibinfo {volume}
  {107}},\ \bibinfo {pages} {083505} (\bibinfo {year} {2023})},\ \Eprint
  {http://arxiv.org/abs/2301.13171} {arXiv:2301.13171 [astro-ph.CO]}
  \BibitemShut {NoStop}%
\bibitem [{\citenamefont {Watson}\ \emph {et~al.}(2013)\citenamefont {Watson},
  \citenamefont {Iliev}, \citenamefont {D'Aloisio}, \citenamefont {Knebe},
  \citenamefont {Shapiro},\ and\ \citenamefont {Yepes}}]{Watson:2012mt}%
  \BibitemOpen
  \bibfield  {author} {\bibinfo {author} {\bibfnamefont {W.~A.}\ \bibnamefont
  {Watson}}, \bibinfo {author} {\bibfnamefont {I.~T.}\ \bibnamefont {Iliev}},
  \bibinfo {author} {\bibfnamefont {A.}~\bibnamefont {D'Aloisio}}, \bibinfo
  {author} {\bibfnamefont {A.}~\bibnamefont {Knebe}}, \bibinfo {author}
  {\bibfnamefont {P.~R.}\ \bibnamefont {Shapiro}}, \ and\ \bibinfo {author}
  {\bibfnamefont {G.}~\bibnamefont {Yepes}},\ }\href {\doibase
  10.1093/mnras/stt791} {\bibfield  {journal} {\bibinfo  {journal} {Mon. Not.
  R. Astron. Soc.}\ }\textbf {\bibinfo {volume} {433}},\ \bibinfo {pages}
  {1230} (\bibinfo {year} {2013})},\ \Eprint {http://arxiv.org/abs/1212.0095}
  {arXiv:1212.0095 [astro-ph.CO]} \BibitemShut {NoStop}%
\bibitem [{\citenamefont {Madau}\ \emph {et~al.}(2008)\citenamefont {Madau},
  \citenamefont {Diemand},\ and\ \citenamefont {Kuhlen}}]{Madau:2008fr}%
  \BibitemOpen
  \bibfield  {author} {\bibinfo {author} {\bibfnamefont {P.}~\bibnamefont
  {Madau}}, \bibinfo {author} {\bibfnamefont {J.}~\bibnamefont {Diemand}}, \
  and\ \bibinfo {author} {\bibfnamefont {M.}~\bibnamefont {Kuhlen}},\ }\href
  {\doibase 10.1086/587545} {\bibfield  {journal} {\bibinfo  {journal}
  {Astrophys. J.}\ }\textbf {\bibinfo {volume} {679}},\ \bibinfo {pages} {1260}
  (\bibinfo {year} {2008})},\ \Eprint {http://arxiv.org/abs/0802.2265}
  {arXiv:0802.2265 [astro-ph]} \BibitemShut {NoStop}%
\bibitem [{\citenamefont {Springel}\ \emph {et~al.}(2008)\citenamefont
  {Springel}, \citenamefont {Wang}, \citenamefont {Vogelsberger}, \citenamefont
  {Ludlow}, \citenamefont {Jenkins}, \citenamefont {Helmi}, \citenamefont
  {Navarro}, \citenamefont {Frenk},\ and\ \citenamefont
  {White}}]{Springel:2008cc}%
  \BibitemOpen
  \bibfield  {author} {\bibinfo {author} {\bibfnamefont {V.}~\bibnamefont
  {Springel}}, \bibinfo {author} {\bibfnamefont {J.}~\bibnamefont {Wang}},
  \bibinfo {author} {\bibfnamefont {M.}~\bibnamefont {Vogelsberger}}, \bibinfo
  {author} {\bibfnamefont {A.}~\bibnamefont {Ludlow}}, \bibinfo {author}
  {\bibfnamefont {A.}~\bibnamefont {Jenkins}}, \bibinfo {author} {\bibfnamefont
  {A.}~\bibnamefont {Helmi}}, \bibinfo {author} {\bibfnamefont {J.~F.}\
  \bibnamefont {Navarro}}, \bibinfo {author} {\bibfnamefont {C.~S.}\
  \bibnamefont {Frenk}}, \ and\ \bibinfo {author} {\bibfnamefont {S.~D.~M.}\
  \bibnamefont {White}},\ }\href {\doibase 10.1111/j.1365-2966.2008.14066.x}
  {\bibfield  {journal} {\bibinfo  {journal} {Mon. Not. R. Astron. Soc.}\
  }\textbf {\bibinfo {volume} {391}},\ \bibinfo {pages} {1685} (\bibinfo {year}
  {2008})},\ \Eprint {http://arxiv.org/abs/0809.0898} {arXiv:0809.0898
  [astro-ph]} \BibitemShut {NoStop}%
\bibitem [{\citenamefont {Diemer}\ and\ \citenamefont
  {Joyce}(2019)}]{Diemer:2018vmz}%
  \BibitemOpen
  \bibfield  {author} {\bibinfo {author} {\bibfnamefont {B.}~\bibnamefont
  {Diemer}}\ and\ \bibinfo {author} {\bibfnamefont {M.}~\bibnamefont {Joyce}},\
  }\href {\doibase 10.3847/1538-4357/aafad6} {\bibfield  {journal} {\bibinfo
  {journal} {Astrophys. J.}\ }\textbf {\bibinfo {volume} {871}},\ \bibinfo
  {pages} {168} (\bibinfo {year} {2019})},\ \Eprint
  {http://arxiv.org/abs/1809.07326} {arXiv:1809.07326 [astro-ph.CO]}
  \BibitemShut {NoStop}%
\bibitem [{\citenamefont {{Widrow}}(2000)}]{2000ApJS..131...39W}%
  \BibitemOpen
  \bibfield  {author} {\bibinfo {author} {\bibfnamefont {L.~M.}\ \bibnamefont
  {{Widrow}}},\ }\href {\doibase 10.1086/317367} {\bibfield  {journal}
  {\bibinfo  {journal} {\apjs}\ }\textbf {\bibinfo {volume} {131}},\ \bibinfo
  {pages} {39} (\bibinfo {year} {2000})}\BibitemShut {NoStop}%
\end{thebibliography}%

\end{document}